\documentclass[twocolumn, twocolappendix]{aastex7}

\shorttitle{AASTeX v6.31}
\shortauthors{Huang et al.}
\graphicspath{{./}{figures/}}
\usepackage{times}
\usepackage{mathptmx}
\usepackage{graphicx}
\usepackage{natbib}
\usepackage{amsmath}
\hypersetup
{
	colorlinks=true,
	linkcolor=blue,
	filecolor=blue,
	urlcolor=blue,
	citecolor=blue,
}

\begin{document}

\title{A Morpho-kinematic Study of Galactic High-ADF PNe Based on the VLT/UVES Deep Spectroscopy\footnote{Based on the observations obtained with the Very Large Telescope (VLT) through program ID \#69.D-0174A.}} 

\author[0009-0004-3407-0848]{Haomiao Huang}
\affiliation{CAS Key Laboratory of Optical Astronomy, National Astronomical Observatories, Chinese Academy of Sciences (NAOC), Beijing 100101, P.~R.\ China}
\affiliation{School of Astronomy and Space Sciences, University of Chinese Academy of Science (UCAS), Beijing 100049, P.~R.\ China.}
\email{huanghm@bao.ac.cn}

\author[0000-0003-1286-2743]{Xuan Fang}
\affiliation{CAS Key Laboratory of Optical Astronomy, National Astronomical Observatories, Chinese Academy of Sciences (NAOC), Beijing 100101, P.~R.\ China}
\affiliation{School of Astronomy and Space Sciences, University of Chinese Academy of Science (UCAS), Beijing 100049, P.~R.\ China.}
\affiliation{Xinjiang Astronomical Observatory, Chinese Academy of Sciences, 150 Science 1-Street, Urumqi, Xinjiang, 830011, P.~R.\ China}
\affiliation{Laboratory for Space Research, Faculty of Science, The University of Hong Kong, Pokfulam Road, Hong Kong, P.~R.\ China}
\email[show]{fangx@nao.cas.cn}

\author[0000-0002-6138-1869]{Jorge, Garc{\'\i}a-Rojas}
\affiliation{Instituto de Astrof{\'\i}sica de Canarias, 38205 La Laguna, Tenerife, Spain}
\affiliation{Departamento de Astrof{\'\i}sica, Universidad de La Laguna, 38206 La Laguna, Tenerife, Spain}
\email{jogarcia@iac.es}

\author[0009-0000-7976-7383]{Zhijun Tu}
\affiliation{CAS Key Laboratory of Optical Astronomy, National Astronomical Observatories, Chinese Academy of Sciences (NAOC), Beijing 100101, P.~R.\ China}
\email{zjtu@bao.ac.cn}

\author[0000-0002-2874-2706]{Jifeng Liu}
\affiliation{CAS Key Laboratory of Optical Astronomy, National Astronomical Observatories, Chinese Academy of Sciences (NAOC), Beijing 100101, P.~R.\ China}
\affiliation{School of Astronomy and Space Sciences, University of Chinese Academy of Science (UCAS), Beijing 100049, P.~R.\ China.}
\affiliation{Institute for Frontiers in Astronomy and Astrophysics, Beijing Normal University, Beijing 102206, P.~R.\ China}
\affiliation{New Cornerstone Science Laboratory, National Astronomical Observatories, Chinese Academy of Sciences, Beijing 100101, P.~R.\ China}
\email{jfliu@nao.cas.cn}

\author[0000-0003-1295-2909]{Xiaowei Liu}
\affiliation{South-Western Institute for Astronomy Research, Yunnan University, Kunming 650500, Yunnan, P.~R.\ China}
\email{x.liu@ynu.edu.cn}

\correspondingauthor{Xuan Fang}

\begin{abstract} 
We report detailed analyses of deep, high-resolution spectra of three Galactic planetary nebulae (PNe) with high abundance discrepancy factors (ADFs), Hf\,2-2, M\,1-42 and NGC\,6153, obtained with the Ultraviolet and Visual Echelle Spectrograph (UVES) on the 8.2\,m Very Large Telescope (VLT).  These spectra were carefully reduced, including rigorous absolute flux calibration, yielding detection of $\sim$410--800 emission lines in each PN.  Plasma diagnostics and abundance calculations were performed using nebular lines.  In all three PNe, the electron temperatures derived using the collisionally excited lines (CELs) are higher than that yielded by the H\,{\sc i} Balmer and Paschen jumps, while the temperatures yielded by the O\,{\sc ii} and N\,{\sc ii} optical recombination lines (ORLs) are very low, $\lesssim$2000\,K, indicating that the heavy-element ORLs probe cold nebular regions.  The ORL abundances of N, O and Ne are systematically higher than the corresponding CEL values, confirming high ADFs in the three objects.  Position-velocity (PV) diagrams were created, and spatio-kinematical studies show that CELs come from the outer nebular regions, while the ORL-emitting regions are close to nebular center.  Additionally, the velocity indicated by CEL line-splitting decreases with ionization potential, which was not obvious in ORLs.  These spatial and kinematic differences support two distinct components of ionized gas: a cold, metal-rich component and a warmer component with normal metallicity.  Heavy elements are strongly enriched in the cold gas, while its H$^{+}$ fraction is low but still produces significant H\,{\sc i} emission, affecting CEL abundance estimates. 
\end{abstract}



\section{Introduction} 
\label{sec:intro}

Planetary nebulae (PNe), the ionized gaseous remnants of low- to intermediate-mass stars (1–8 M$_{\odot}$), represent a critical phase of stellar evolution \citep{2000oepn.book.....K}. As stars exhaust their nuclear fuel, they expel their outer layers through stellar winds, and the interactions of the stellar wind form intricate nebular structures illuminated by the ultraviolet radiation of the central white dwarf \citep{1978ApJ...219L.125K, 2011AJ....142...91R}. These objects serve as laboratories for studying nucleosynthesis processes, mass-loss mechanisms, and the chemical evolution of galaxies \citep[e.g.][]{2018ApJ...853...50F, 2010ApJ...714.1096S}.  The formation and evolution of PNe are shaped by the interplay between stellar dynamics, radiation fields, and the surrounding interstellar medium (ISM), making them key targets for understanding the life cycles of stars and their galactic ecosystems.

Spectroscopic studies of PNe have long been central to determining their physical and chemical properties. For distant and faint PNe, only optical collisionally excited lines (CELs) can be detected with high quality in short exposure times.  Surveys or case observations of these sources \citep[e.g.][]{2004MNRAS.349.1291E, 2015ApJ...815...69F, 2018ApJ...853...50F} have therefore generally presented results derived only from CELs.  As the spectral quality improves, faint optical recombination lines (ORLs) emitted by heavy-element ions can be detected.  Because the emissivities of ORLs are generally less temperature-sensitive than those of optical CELs, analyses based on ORLs are less affected by temperature fluctuations \citep[e.g.][]{2006IAUS..234..219L}.  However, the measurement accuracy for heavy-element ORLs is generally low (compared to that of optical CELs) due to their faintness ($\lesssim$10$^{-4}$--10$^{-2}$ of H$\beta$ flux). 

Elemental abundances of photoionized nebulae can be obtained using both ORLs and CELs; however, spectroscopic analysis of a large number of Galactic PNe and H\,{\sc ii} regions found that for the same element (C, N, O, and Ne), the ORL abundance is always higher than the CEL abundance.  This is the renowned yet unresolved ``abundance discrepancy'' problem, which was first identified by \citet{1942ApJ....95..356W} and later confirmed by many other observations.  This discrepancy, quantified as the abundance discrepancy factor (ADF), is defined as the ratio of abundance derived from ORLs to that from CELs.  Several deep surveys of ORLs revealed that abundance discrepancy is universal among Galactic PNe \citep[e.g.][]{2004MNRAS.353..953T, 2004MNRAS.353.1251L, 2005MNRAS.362..424W, 2007MNRAS.381..669W}. The ADF values of PNe cover a broad range, with a median value of $\sim$2, but exceeding 100 in A30 \citep{2003MNRAS.340..253W} and A46 \citep{2015ApJ...803...99C}. 

Understanding and resolving the abundance discrepancy problem is important for the accurate determination of elemental abundances in ionized gas, especially in galaxies where only optical CELs can be detected and where a precise determination of chemical composition helps to constrain the chemical evolution of the Universe.  A number of hypotheses have been proposed to address the abundance discrepancy, including temperature fluctuations \citep[e.g.][]{1967ApJ...150..825P}, non-Maxwellian distribution of electron energies \citep[e.g., $\kappa$ distribution,][]{2012ApJ...752..148N,2013ApJS..207...21N}, fluorescence contamination in ORLs \citep[e.g.][]{1968MNRAS.139..129S,2012MNRAS.426.2318E} and chemical inhomogeneities \citep[e.g.][]{1990A&A...233..540T,2000MNRAS.312..585L}.  The temperature fluctuation scenario well explains abundance discrepancy in H\,{\sc ii} regions and the low-ADF PNe, but failed to explain the high-ADF PNe because the required mean-square temperature fluctuation, $t^{2}$, is too large \citep{2000MNRAS.312..585L}.  The chemical inhomogeneities hypothesis suggests that the abundance discrepancy arises from the presence of plasmas with different compositions.  In fact, in high-ADF PNe, low-temperature, high-metallicity plasmas dominate the heavy-element ORL emission \citep[e.g.][]{2011MNRAS.411.1035Y}. 

With the advances in observational techniques, chemical inhomogeneities have been shown to exist in some PNe.  Imaging observations using tunable filters mounted on the 10.4-m Gran Telescopio Canarias (GTC) have revealed different spatial distributions of O\,{\sc ii} ORLs and [O\,{\sc iii}] CELs in the Galactic PN NGC\,6778 \citep{2016ApJ...824L..27G}.  This phenomenon has also been confirmed through integral-field unit (IFU) spectroscopy of other high-ADF PNe \citep{2008MNRAS.386...22T, 2019MNRAS.484.3251A, 2022MNRAS.510.5444G, 2024AA...689A.228G}.  Recently, observations also found a link between high ADF and binarity of PN central stars \citep{2015ApJ...803...99C, 2016MNRAS.455.3263J, 2018MNRAS.480.4589W}.  The most extreme abundance discrepancies occur in PNe with close binary central stars which have undergone a common envelope (CE) phase.  However, how these H-deficient materials are ejected during the CE phase -- or whether they originate from other processes -- remain unclear. 

The wavelengths of ORLs and CELs from the same ion are usually very different.  IFU spectroscopy typical struggles to provide both broad wavelength coverage and high spectral resolution simultaneously.  Studying the differences in CEL and ORL kinematics, and thus hypothesizing the origin of the metal-rich gas, requires deep high-resolution spectroscopy.  Previous studies based on high-dispersion spectra have revealed that the kinematics of heavy element ORLs and CELs are different, with the former exhibiting smaller expansion velocities \citep{2004ApJ...615..323S, 2017MNRAS.472.1182P, 2013ApJ...773..133R, 2017AJ....153..140R, 2022AJ....164..243R}.  The differences of ORL and CEL kinematics in a larger sample of PNe need to be studied to assess their generality as well as to provide enough data to probe the origin of metal-rich regions.

Hf\,2-2, M\,1-42, and NGC\,6153 are three high-ADF PNe that have been extensively studied through various spectroscopic techniques \citep[e.g.][]{2000MNRAS.312..585L, 2001MNRAS.327..141L, 2006MNRAS.368.1959L,  2022MNRAS.510.5444G, 2024AA...689A.228G}.  The VLT/UVES spectra of these sources were obtained more than two decades ago, but their resolution and quality remain excellent compared to most of the observations of Galactic PNe nowadays.  Several studies based on these data have been published \citep{2016MNRAS.461.2818M, 2022AJ....164..243R}.  However, the previous analysis using these data \citep{2016MNRAS.461.2818M} had some problems, which we will revisit here and thoroughly analyze to provide improved and more reliable results.  Beyond the abundance discrepancy, high-resolution deep spectra also enable the detection of rare $s$-process elements, e.g., Kr \citep{2015MNRAS.452.2606G, 2008ApJS..174..158S}, offering new avenues to probe nucleosynthetic pathways.

\begin{figure*}[ht!]
\begin{center}
\includegraphics[width=17cm,angle=0]{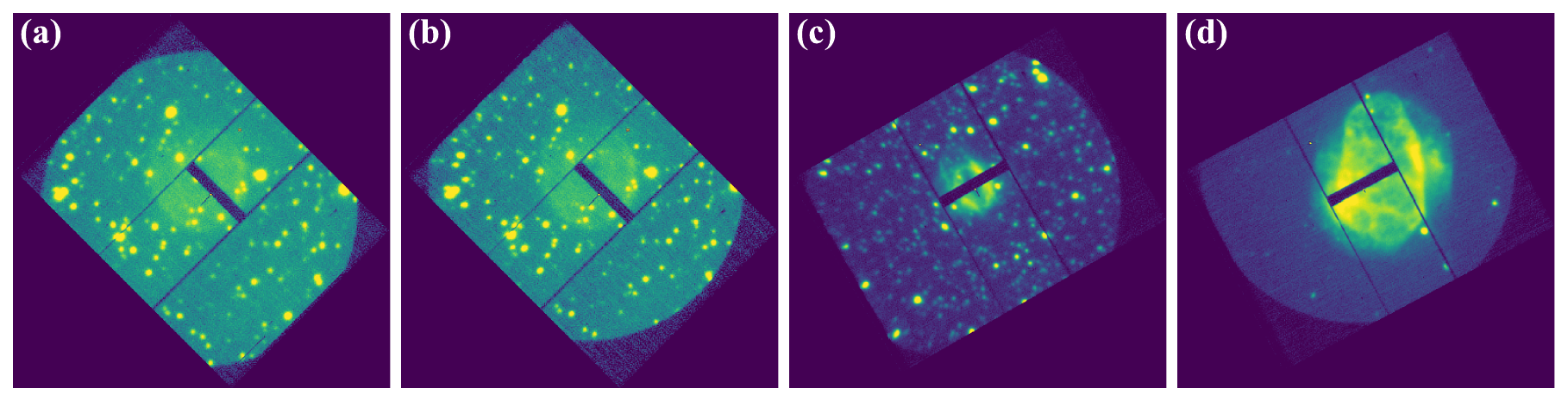}
\caption{VLT UVES slit-view images of Hf\,2-2 (panels~a and b), M\,1-42 (panel~c), and NGC\,6153 (panel~d), showing the red-arm slit positions during spectroscopic observations with Dichroic\,2.  Panels~(a) and (b) show the slit positions for the first and second science exposures of Hf\,2-2, respectively; note that there is a slight offset in slit position (perpendicular to the direction of slit length) between the two exposures.  Slit width is 2\arcsec; north is up and east to the left.} 
\label{fig:slitview}
\end{center}
\end{figure*}

This paper presents a comprehensive reanalysis of VLT/UVES echelle spectra for three high-ADF Galactic PNe: Hf\,2-2, M\,1-42, and NGC\,6153.  Our study carried out meticulous data reduction and flux calibrations, to revisit plasma diagnostics and abundance calculations.  We also employ position-velocity (PV) diagrams to probe the spatial and dynamical segregation of CEL- and ORL-emitting regions.  The paper is structured as follows: Section\,\ref{sec:data} describes the data reduction; Section\,\ref{sec:1D_lines} presents the line measurements and identifications; plasma diagnostics and abundance calculation are presented in Sections\,\ref{sec:diagnostics} and \ref{sec:abundance}, respectively.  In Section\,\ref{sec:2d}, we explore the kinematics of ORLs and CELs mainly based on the PV maps derived from 2D spectra.  Since \citet{2022AJ....164..243R} already conducted a similar analysis of NGC\,6153 based on the same dataset, we focus the analysis of Hf\,2-2 and M\,1-42.  Discussion is presented in Section\,\ref{sec:discussion}, and our conclusions are summarized in Section\,\ref{sec:summary}.

\section{The VLT/UVES Spectroscopy and Data Reduction} \label{sec:data}

\subsection{Observations, and Previous Relevant Studies} \label{observations}

The data reported in this article were obtained on 2002 June 8, using the Ultraviolet and Visual Echelle Spectrograph \citep[UVES,][]{2000SPIE.4008..534D} mounted on the 8.2\,m Very Large Telescope (VLT) Kueyen (UT2) at the European Southern Observatory (ESO), under program ID: 69.D-0174A (P.I.: I.~J. Danziger).  UVES is a cross-dispersed echelle spectrograph that operates through dual-arm configurationp: The blue arm employs an EEV 44-82 CCD with a 2048$\times$4096 array of 15-$\mu$m pixels (only 2048$\times$3000 pixels were utilized during observations), while the red arm features a mosaic CCD combining an EEV 44-82 and MIT-LL CCID-20 detector, both also consisting with 2048 $\times$ 4096 15$\mu$m pixels. 

To optimize detection, all detectors operated in 2$\times$2 binning mode during the observations.  Four cross-dispersed settings (CD\#1–-CD\#4) were engaged to cover the broad wavelength range of 3043–-10655\,{\AA}, with Dichroic\,1 pairing CD\#1/CD\#3 and Dichroic\,2 combining CD\#2/CD\#4.  This configuration introduces gaps in the spectra obtained by the latter three cross-dispersers, as well as between the last few orders of CD\#4.  Additionally, the mosaic CCD in the red arm introduces further gaps in both the CD\#3 and CD\#4 spectra.  To facilitate reference, the spectra on either side of the gaps for CD\#3 and CD\#4 are systematically labeled as CD\#3b/CD\#3r and CD\#4b/CD\#4r for the short- and long-wavelength segments, respectively.  Spatial sampling along the slit differed between arms, with the blue spectra achieving 0$\farcs$492 per pixel resolution versus 0$\farcs$362 per pixel for the red.  The entrance slit configuration consisted of a slit length of 10$^{\prime\prime}$ in the blue arm and 13\arcsec--14\arcsec\ in the CD\#3/\#4 spectra.  A slit width of 2$\arcsec$ was used in the spectroscopy, producing a spectral resolution of $R\approx$20,000, corresponding to $\sim$15 km\,s$^{-1}$ in velocity resolution. 

All targets were observed using multiple exposures with repeated long integrations.  Among the three PNe, Hf\,2-2 is the faintest and was observed with two 1800\,s exposures using a 2$^{\prime\prime}$ slit width and one 900\,s exposure using a 10$^{\prime\prime}$ slit width.  As its strong emission lines remained unsaturated even in 30\,min integrations, no short exposure observations with a 2$^{\prime\prime}$ slit width were conducted. In contrast, NGC\,6153 and M\,1-42 required additional short exposures (60 -- 120\,s) to mitigate saturation from bright emission lines in their 2$^{\prime\prime}$ slit width configurations.  Specifically, NGC\,6153 was observed with three 1200\,s long exposures, 60\,s short exposures for CD\#1/CD\#3, and 120\,s exposures for CD\#2/CD\#4.  M\,1-42 was observed with two 1800\,s exposures plus a 60\,s short exposure through a 2$^{\prime\prime}$ slit width, complemented by a 900\,s exposure with a 10$^{\prime\prime}$ slit width.  Observations of the three PNe are summarized in \citet[][Table\,1 therein]{2016MNRAS.461.2818M}, and the spatial coverage of the slits is illustrated in Figure\,\ref{fig:slitview}.

\begin{figure}[ht!]
\begin{center}
\includegraphics[width=8.5 cm,angle=0]{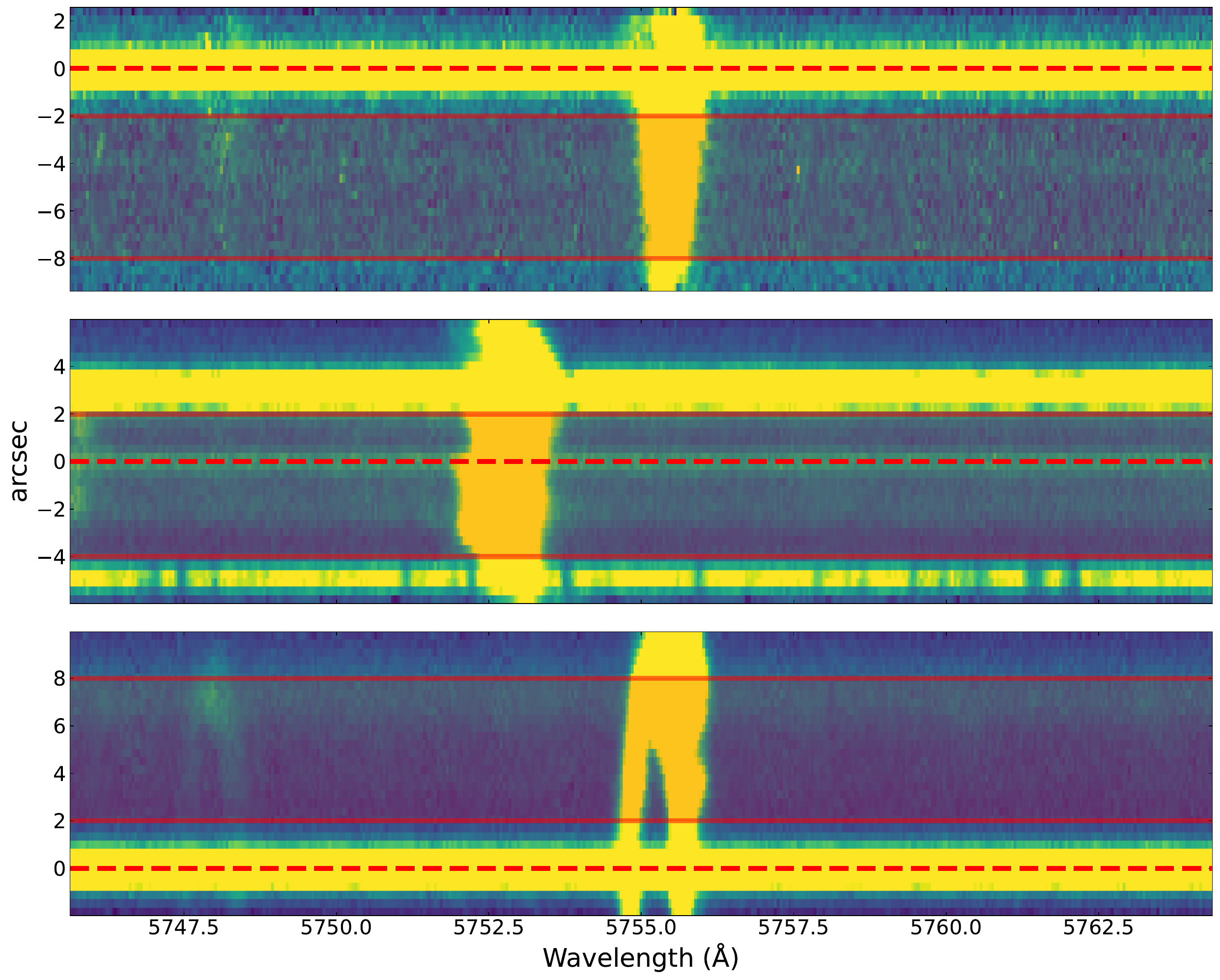}
\caption{Central star positions (dashed lines) and extraction regions (shaded area between two translucent lines) of Hf\,2-2 (upper panel), M\,1-42 (middle panel) and NGC\,6153 (lower panel). Spectra of field stars near M\,1-42 are largely ruled out.} 
\label{fig:extraction}
\end{center}
\end{figure}

These datasets have supported multiple investigations. \citet{2016MNRAS.461.2818M} conducted basic plasma diagnostics and abundance determinations using their emission line measurements; however, their 1D spectral extraction did not account for the different red/blue arm slit lengths, and their extractions of the 1D spectra from different orders were not co-spatial.  \citet{2022AJ....164..243R} revealed kinematic discrepancy in NGC\,6153: ORLs from ions with different ionization potentials exhibit expansion velocities that diverge from those of CELs, indicating coexisting plasma components –- a nebular shell and a denser, cooler phase.  While this dual-component framework helps to advance the interpretation of NGC\,6153, the kinematics of the remaining two PNe, particularly those traced by permitted lines require a systematic exploration to obtain a comprehensive understanding of their chemical and kinematical properties.

\subsection{Data Reduction} \label{reduction}

Although \citet{2022AJ....164..243R} successfully calibrated the spectra of NGC\,6153 using the same dataset with interstellar reddening references, some wavelength intervals lack a sufficient number of lines for robust extinction determination \citep[e.g., CD\#3b with only He\,{\sc i} $\lambda\lambda$5016,5876, and CD\#3r with He\,{\sc i} $\lambda$6678 and H$\alpha$, see][Tables\,3-4 therein]{2022AJ....164..243R}.  Notably, the temperature-sensitive He\,{\sc i} lines used in this calibration introduce uncertainties, as their relative intensities vary with electron temperature.  As a result, flux scales estimated solely from extinction may therefore be inaccurate, and \emph{absolute} flux calibration is essential to ensure the accuracy of spectral analysis. 

The ESO pipeline-processed 2D spectra retrieved from the ESO archive\footnote{\url{https://doi.org/10.18727/archive/50} (The UVES reduced data processed by standard ESO pipeline)} lack accurate absolute flux calibration, necessitating a complete re-processing of these data.  The red and blue bias frames were combined separately and subtracted from corresponding CCD images.  For flat fields, images of sources and standards, inter-order scattered light was quantified and removed.  Since the standard flat field used for the CD\#1 spectra has a low signal-to-noise ratio (SNR) at short wavelengths, we used a hybrid flat field combining the deuterium flat (DFLAT) and the normal flat, with DFLAT being used at short wavelengths and the normal flat being used at long wavelengths to avoid contamination of the DFLAT emission lines.  All other images were divided by flat fields with matching slit width and wavelength range to eliminate pixel-to-pixel variations, correct for the spectrograph's blaze function, and remove fringe effects in the far-red region. 

All long exposures were combined, and cosmic rays removed via sigma-clipping.  Single short exposures and those obtained with the 10$^{\prime\prime}$ slit width were cosmic-ray removed using the \texttt{cosmicray\char`_lacosmic} algorithm \citep{2001PASP..113.1420V} in \texttt{ccdproc}. 2D spectra were extracted using order definition flats and slit lengths.  Due to nebular angular sizes comparable to/exceeding slit lengths, sky subtraction was omitted. Standard stars (LTT\,3864, LTT\,9293, CD\,-32$^{\circ}$\,9927 and Feige\,56) underwent identical processing with sky subtraction. Wavelength calibration utilized ThAr arc lamp spectra and the ThAr atlases for UVES to establish wavelength-to-pixel solutions.  Theoretical spectra of the standard star \citep[50\,{\AA} sampling;][]{1992PASP..104..533H, 1994PASP..106..566H} required merging the spectra orders within each CCD into single continuous spectra to derive the response curve for each wavelength range.  Individual response curves were generated using each standard star, and their average was employed for the final absolute flux calibration.

\section{Measurements of Emission Lines} \label{sec:1D_lines}

\subsection{1D spectra extraction} \label{sec:extraction}

\begin{table*}
\begin{center}
\caption{Atomic Data}
\label{tab:atomic_data}
\begin{tabular}{lcc}
\hline\hline
Ion & \multicolumn{2}{c}{CELs} \\
 & Transition Probabilities & Collision Strengths \\
\hline
$[$C\,{\sc i}$]$ & \citet{1985PhyS...32..181F} & \citet{\detokenize{1976A&A....50..141P}} \\ 
$[$N\,{\sc i}$]$ & \citet{\detokenize{1996atpc.book.....W}} & \citet{\detokenize{1976A&A....50..141P}} \\
$[$N\,{\sc ii}$]$ & \citet{2004ADNDT..87....1F} & \citet{2011ApJS..195...12T} \\
$[$O\,{\sc i}$]$ & \citet{\detokenize{1996atpc.book.....W}} & \citet{1995ApJS...96..325B} \\
$[$O\,{\sc ii}$]$ & \citet{1982MNRAS.198..111Z} & \citet{2009MNRAS.397..903K} \\
$[$O\,{\sc iii}$]$ & \citet{2000MNRAS.312..813S} & \citet{2014MNRAS.441.3028S} \\
 & \citet{2004ADNDT..87....1F} &  \\
$[$Ne~{\sc iii}$]$ & \citet{\detokenize{1997A&AS..123..159G}} & \citet{2000JPhB...33..597M} \\
$[$Ne~{\sc iv}$]$ & \citet{1984JPhB...17..681G} & \citet{1981MNRAS.195P..63G} \\
$[$Ne~{\sc v}$]$ & \citet{\detokenize{1997A&AS..123..159G}} & \citet{2013MNRAS.435.1576D} \\
$[$S~{\sc ii}$]$ & \citet{\detokenize{2019A&A...623A.155R}} & \citet{2010ApJS..188...32T} \\
$[$S~{\sc iii}$]$ & \citet{2006ADNDT..92..607F} & \citet{1999ApJ...526..544T} \\
$[$Cl~{\sc ii}$]$ & \citet{1983MNRAS.202..981M} & \citet{2004\detokenize{A&A...418..363T}} \\
$[$Cl~{\sc iii}$]$ & \citet{\detokenize{2019A&A...623A.155R}} & \citet{1989\detokenize{A&A...208..337B}} \\
$[$Cl~{\sc iv}$]$ & \citet{1986JPCRD..15..321K} & \citet{\detokenize{1995A&AS..111..347G}} \\
$[$Ar~{\sc iii}$]$ & \citet{\detokenize{2009A&A...500.1253M}} & \citet{\detokenize{2009A&A...500.1253M}} \\
$[$Ar~{\sc iv}$]$ & \citet{\detokenize{2019A&A...623A.155R}} & \citet{1997ADNDT..66...65R} \\
$[$Ar~{\sc v}$]$ & \citet{1986JPCRD..15..321K} & \citet{\detokenize{1995A&AS..111..347G}} \\
$[$K~{\sc iv}$]$ & \citet{1986JPCRD..15..321K} & \citet{\detokenize{1995A&AS..111..347G}} \\
$[$Kr~{\sc iv}$]$ & \citet{1986PhyS...34..116B} & \citet{\detokenize{1997A&AS..122..277S}} \\
\hline
Ion & \multicolumn{2}{c}{ORLs} \\
 & Effective recombination coefficients & Case\\
\hline
H\,{\sc i} & \citet{1995MNRAS.272...41S} & B \\
He\,{\sc i} & \citet{2012MNRAS.425L..28P, 2013MNRAS.433L..89P} & B \\
 & \citet{2022MNRAS.513.1198D} & B \\
He~{\sc ii} & \citet{1995MNRAS.272...41S} & B \\
C\,{\sc ii} & \citet{\detokenize{2000A&AS..142...85D}} & A, B (see text for details) \\
C~{\sc iii} & \citet{\detokenize{1991A&A...251..680P}} & A \\
C~{\sc iv} & \citet{\detokenize{1991A&A...251..680P}} & A \\
N\,{\sc ii} & \citet{\detokenize{2011A&A...530A..18F, 2013A&A...550C...2F}} & B \\
N\,{\sc iii} & \citet{\detokenize{1991A&A...251..680P}} & A \\
O\,{\sc i} & \citet{\detokenize{1991A&A...251..680P}} & A \\
O\,{\sc ii} & \citet{2017MNRAS.470..379S} & B \\
O\,{\sc iii} & \citet{\detokenize{1991A&A...251..680P}} & B \\
Ne~{\sc ii} & \citet{\detokenize{1998A&AS..133..257K}} & B \\
 & Storey (unpublished; private communication) &  \\
\hline
\end{tabular}
\end{center}
\end{table*}

As discussed in Section\,\ref{observations}, differences in the slit lengths for the red and blue spectra introduce spatial coverage discrepancies.  Since the angular diameters of all the three PNe exceed 10$^{\prime\prime}$, integrating the entire slit length for 1D spectra would overestimate fluxes in longer wavelength. To ensure consistency, 2D spectra were aligned spatially using the central star as a reference. The central star positions were determined via Gaussian fitting along the spatial direction. Atmospheric dispersion caused wavelength-dependent positional shifts, which were modified by measuring the central star’s location at multiple wavelengths and modeling these shifts with low-order polynomial functions. These positional functions for central stars anchored both the 1D spectral extraction and subsequent 2D spectral analysis to a common spatial framework. 

The extraction regions were tailored to each object. In Hf\,2-2 and NGC\,6153, the bright central stars lie near the slit edges in blue spectra, risking contamination.  Thus the spectra were extracted $2^{\prime\prime}$--$8^{\prime\prime}$ from the central stars to avoid stellar flux loss in the edge.  However, the central star of M\,1-42 is much fainter, and is located near the central part of 2D frame, therefore the region from 2$^{\prime\prime}$ “above” to 4$^{\prime\prime}$ “below” the central star was used to extract 1D spectra.  The extraction regions are marked in Figure\,\ref{fig:extraction}.

\subsection{Measurements and Identification of Emission Lines} \label{sec:line}

\begin{table*}
\begin{center}
\caption{Plasma Diagnostics}
\label{temden}
\begin{tabular}{llccc}
\hline\hline
ID & Diagnostic Ratio & Hf\,2-2 & M\,1-42 & NGC\,6153 \\
\hline
 & & \multicolumn{3}{c}{$T_{\rm e}$ (K)} \\
1 & $[$O\,{\sc iii}$]$ $\lambda4363/(\lambda4959+\lambda5007)$\tablenotemark{\rm{\scriptsize a}}  & 9300$\pm$120 & 9300$\pm$100 & 9400$\pm$60 \\
 & $[$O\,{\sc ii}$]$ $(\lambda3727+)/(\lambda7325+)$\tablenotemark{\rm{\scriptsize b}} & 34,000$\pm$4000 & 17,500$\pm$1500 & 17,200$\pm$400 \\
2 & $[$S~{\sc iii}$]$ $\lambda6312/(\lambda9069+\lambda9531)$ & 7400$\pm$250 & 9100$\pm$100 & 9100$\pm$100 \\
3 & $[$N\,{\sc ii}$]$ $\lambda5755/(\lambda6548+\lambda6583)$\tablenotemark{\rm{\scriptsize c}} & 10,100$\pm$400 & 9300$\pm$100 & 8350$\pm$100 \\
4 & $[$Ar~{\sc iii}$]$ $\lambda5192/(\lambda7751+\lambda7136)$ & $\cdots$ & 9400$\pm$400 & 8800$\pm$200\\
 & $[$C\,{\sc i}$]$ $\lambda8727/(\lambda9824+\lambda9850)$ & 10,350$^{+11000}_{-2100}$ & $>$20,000 & 9300$^{+1600}_{-900}$\\
5 & $[$O\,{\sc i}$]$ $\lambda5577/(\lambda6300+\lambda6363)$ &$\cdots$ & 8500$\pm$300 & 8300$\pm$1400 \\
 & He\,{\sc i} $\lambda5876/\lambda4471$ & 1500$\pm$150 & 1500$\pm$60 & 1500$\pm$100 \\
 & He\,{\sc i} $\lambda6678/\lambda4471$ & 1300$\pm$150 & 1400$\pm$100 & 1300$\pm$90 \\
 & He\,{\sc i} $\lambda7281/\lambda6678$ & 500$\pm$200 & 2100$\pm$120 & 3700$\pm$200\\
 & O\,{\sc ii} $\lambda4649/\lambda4089$ & 400$^{+600}_{-:}$ & 1370$^{+570}_{-400}$ & 2240$^{+640}_{-430}$ \\
 & N\,{\sc ii} $\lambda5679/\lambda4041$\tablenotemark{\rm{\scriptsize d}} & 2500$^{+2500}_{-1300}$ & 2150$^{+730}_{-260}$ & 2500$^{+700}_{-500}$\\
 & BJ/H11 & 840$\pm$60 & 3440$\pm$170 & 5750$\pm$300 \\
 & PJ/P11 & 610$\pm$110 & 2370$\pm$140 & 5150$\pm$250 \\
\hline
 & & \multicolumn{3}{c}{$N_{\rm e}$ (cm$^{-3}$)} \\
6 & $[$S~{\sc ii}$]$ $\lambda6731/\lambda6716$ & 400$\pm$50 & 1800$\pm$100 & 5550$\pm$100\\
7 & $[$O\,{\sc ii}$]$ $\lambda3726/\lambda3729$\tablenotemark{\rm{\scriptsize a}} & 1300$\pm$150 & 1700$\pm300$ & 4000$\pm$200\\
8 & $[$Ar~{\sc iv}$]$ $\lambda4740/\lambda4711$ & 510$\pm$: & 800$\pm$100 & 4000$\pm$100\\
9 & $[$Cl~{\sc iii}$]$ $\lambda5537/\lambda5517$ & 440$\pm$: & 1640$\pm120$ & 3900$\pm$150\\
10 & $[$N\,{\sc i}$]$ $\lambda5198/\lambda5200$ & 260$\pm$60 & 640$\pm50$ & 1700$\pm$700\\
 & O\,{\sc ii} $\lambda4649/\lambda4661$ & 800$^{+200}_{-170}$  & 5600$^{+2100}_{-1200}$ & 11,000$^{+1600}_{-1000}$\\
 & N\,{\sc ii} $\lambda5679/\lambda5666$ & 1800$^{+2700}_{-1100}$  & 13,000$^{+:}_{-5700}$ & $>$20000\\
 & Balmer decrement & $\sim 10^3$ & $10^3 \sim 10^4$ & $10^3 \sim 10^4$ \\
 & Paschen decrement & $10^2 \sim 10^3$ & $\sim 10^3$ & $10^3 \sim 10^4$ \\
\hline
\end{tabular}
\begin{description}
NOTE. -- Symbol ``:” indicates high uncertainty, and ``$\cdots$" means that the values cannot be derived because the auroral lines of these diagnostics were not detected .\\
\tablenotemark{\rm{\scriptsize a}} The contributions of recombination excitation in [O\,{\sc iii}] $\lambda$4363 were corrected based on O\,{\sc iii} $\lambda\lambda$3260,3265 with atomic data from \citet{1991A&A...251..680P} at 10000 K. The contributions of charge exchange were removed using O\,{\sc iii} $\lambda$5592 based on the transition probabilities of singlet cascade \citep{1982ApJ...257L..87D}. \\
\tablenotemark{\rm{\scriptsize b}} The recombination contributions were not corrected, see text for more details. \\
\tablenotemark{\rm{\scriptsize c}} The recombination contributions were corrected based on N\,{\sc ii} $\lambda$5679 with atomic data \citep{2011A&A...530A..18F} at N\,{\sc ii} temperatures and densities. \\
\tablenotemark{\rm{\scriptsize d}} The N\,{\sc ii} temperature of Hf\,2-2 were derived based on N\,{\sc ii} $\lambda5005/\lambda4041$ ratio.\\
\end{description}
\end{center}
\end{table*}

Emission lines of PNe were analyzed using 1D spectra from designated regions mentioned in Section\,\ref{sec:extraction}, supplemented by the 2D spectra to distinguish nebular emission features from telluric contamination.  After continuum subtraction, line fluxes were integrated over the line profiles and normalized to $F$(H$\beta$)=100, with uncertainties estimated based on the standard deviation of nearby continuum.  Wavelengths were determined via flux-weighted averaging of line profiles. A total of 417, 674, and 773 emission lines were detected in the UVES spectra of Hf\,2-2, M\,1-42, and NGC\,6153, respectively; the faintest lines detected in the three PNe have fluxes of $\sim2\times10^{-4}\,F(\mathrm{H}\beta)$ in Hf\,2-2 and $\sim1\times10^{-4}\,F(\mathrm{H}\beta)$ in M\,1-42 and NGC\,6153. 

Emission lines were identified utilizing the NIST Atomic Spectra Database\footnote{\url{https://www.nist.gov/pml/atomic-spectra-database}} and the online database of Atomic Line List\footnote{\url{https://linelist.pa.uky.edu/newpage/}} \citep{2018Galax...6...63V}.  We first performed manual identification, and then used the newly developed spectral-line identification code PyEMILI \citep{2025ApJS..277...13T} to identify those lines that were not recognized in the manual work; the final ID for each emission line was decided by empirical yet rigorous analysis followed by more rigorous software ranking.  Identifications (and measurements) of the emission lines detected in the VLT/UVES spectra of the three PNe are summarized in Tables\,\ref{Hf22_linelist}, \ref{M142_linelist} and \ref{NGC_linelist}.  Those emission lines that still cannot be identified are marked with ``??''.  Radial velocities were calculated from wavelength offsets relative to the laboratory values.  Notably, some emission lines of low-ionization species (e.g., [O\,{\sc i}] in Hf\,2-2) exhibit kinematic deviations from dominant nebular components (see Section\,6.2), and thus their radial velocities deviate significantly from the average value.  Other lines with velocities inconsistent with nebular averages are flagged with ``?'', indicating uncertain identification.

\subsection{Extinction and Flux Scale Correction} \label{sec:correction}

\begin{figure*}[ht!]
\begin{center}
\includegraphics[width=18 cm,angle=0]{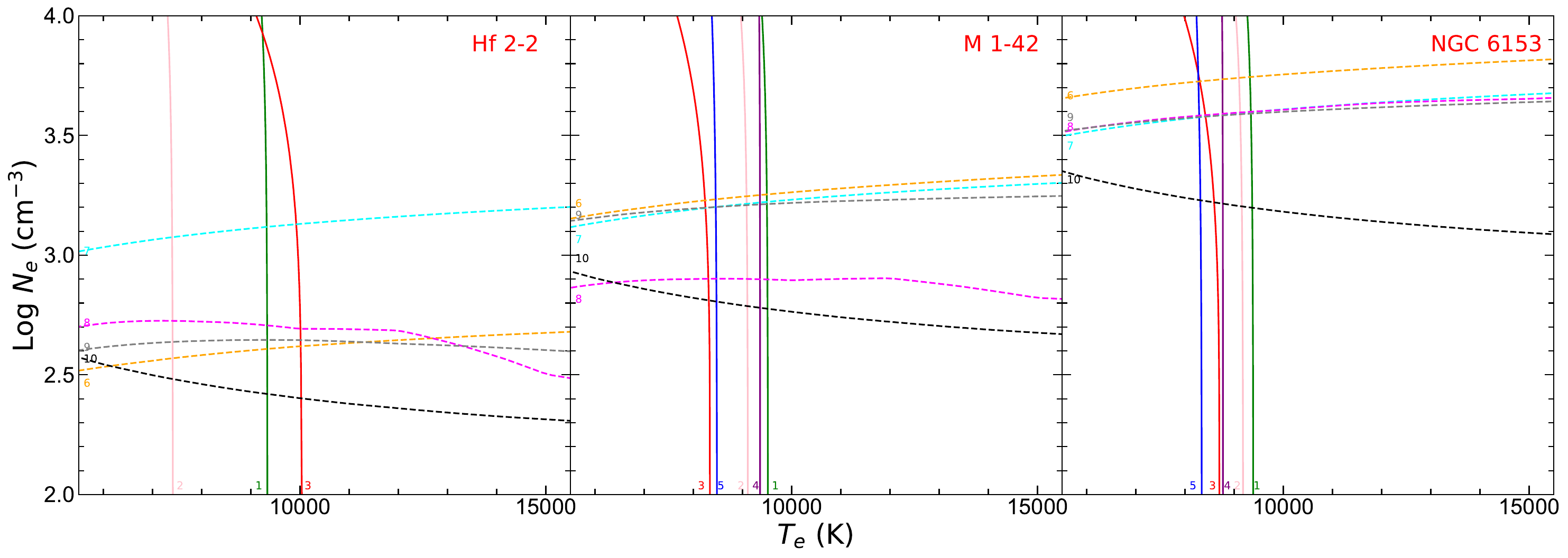}
\caption{CELs plasma diagnostic diagram of Hf\,2-2 (left), M\,1-42 (middle) and NGC\,6153 (right).  Numbers of the same colour labeled near curves represent the diagnostic line ratios with the IDs in Table \ref{temden}. The solid and dashed lines are the temperature and density diagnostics. The auroral lines used in the [N\,{\sc ii}] and [O\,{\sc iii}] temperature diagnostics have been corrected to remove the recombination excitation contributions.} 
\label{fig:celdiag}
\end{center}
\end{figure*}

The logarithmic extinction parameter, $c$(H$\beta$), for each nebula was determined using flux ratios of Balmer and Paschen lines relative to H$\beta$, with theoretical ratios from \citet{1995MNRAS.272...41S} (for $T_{\rm e}=10^4$ K and $n_{\rm e}=10^4$ cm$^{-3}$). Dereddened fluxes were calculated via:
\begin{equation}
I ({\rm{\lambda}})=10^{c({\rm{H}} \beta)[1+f(\lambda)]} F(\lambda),
\end{equation} 
where $f$($\lambda$) follows the \citet{1989ApJ...345..245C} extinction curve with a total-to-selective extinction ratio $R_V=3.1$, and all dereddened fluxes were normalized to $I$(H$\beta$) = 100. The c(H$\beta$) and dereddened fluxes are presented in the line table in supplementary material. The c(H$\beta$) we derived for NGC\,6153 is about 1.13, larger than the that obtained by \citet{2022AJ....164..243R} and close to the results of other previous studies \citep[e.g.,][]{2000MNRAS.312..585L, 2003A&A...409..599P, 2008MNRAS.386...22T, 2024AA...689A.228G}. 

Flux calibration inconsistencies between different wavelength intervals and short/long exposures were reported by \citet{2022AJ....164..243R} for the NGC\,6153 spectra and also confirmed in our data.  No such problems were found in the Hf\,2-2 or M\,1-42 spectra, which were reduced following the same procedures.  To ensure a reliable analysis, the flux scales of the NGC\,6153 spectra need to be corrected.  We applied corrections using unsaturated lines to match the flux levels between short and long exposures.  This step was unnecessary for the CD\#1 and CD\#4b spectra because both wavelength ranges have no saturated lines and all analysis were based on long exposures.  The flux scale factors divided between short and long exposures are consistent with those reported by \citet{2022AJ....164..243R}, with short/long ratios equal to 1.40 and 1.34 for CD\#3b and CD\#3r, respectively, while the ratios for CD\#2 and CD\#4r are very close to 1. 

Flux scale corrections between wavelength intervals of NGC\,6153 spectra were addressed through multiple approaches.  For overlapping regions between CD\#1 and CD\#2, the flux scale factor of CD\#1/CD\#2\,$\approx$1.20 was derived by comparing line flux differences in their shared spectral range.  Since there is a gap between the CD\#2 and CD\#3b wavelength coverages, the theoretical intensity ratio of [O\,{\sc iii}] $\lambda$5007/$\lambda$4959 = 2.98 was utilized to determine a CD\#3b/CD\#2 scaling factor of 1.086. Corrections for CD\#4b and CD\#4r relative to CD\#2 employed consistency between extinction coefficients derived from Paschen and Balmer lines, yielding CD\#4b/CD\#2 and CD\#4r/CD\#2 $\approx$ 1.09. The CD\#3r/CD\#2 factor 1.08 was constrained by matching H$\alpha$/H$\beta$ ratios to values from \citet{2000MNRAS.312..585L}, whose spectra covered both lines in the same order. This result aligns with other red-spectra corrections, prompting adoption of a unified 1.085 scaling factor for all red spectra relative to CD\#2. 

\begin{figure*}[ht!]
\begin{center}
\includegraphics[width=18 cm,angle=0]{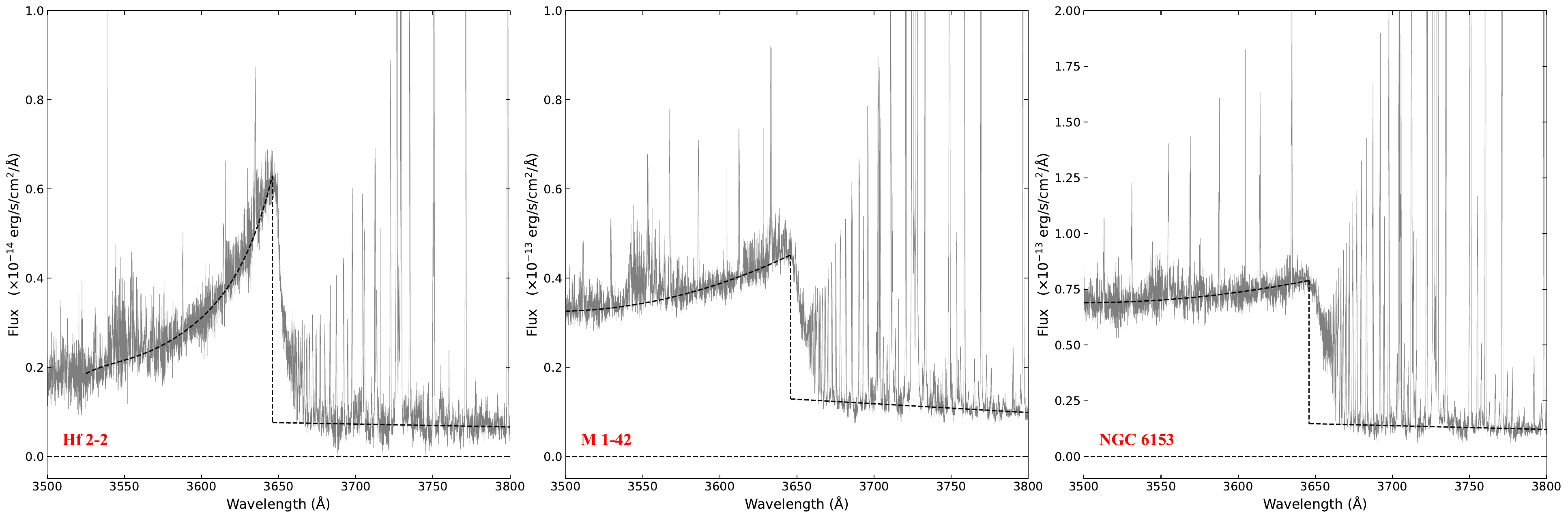}
\caption{The VLT UVES spectra (grey lines) of Hf\,2-2 (left),M\,1-42 (middle) and NGC\,6153 (right) in the vicinity of Balmer jump/discontinuity, with radial velocity and extinction corrected.  The black dashed lines are fittings of continua on both sides the Balmer jump and extrapolated towards 3643\,\AA.} 
\label{fig:BJ}
\end{center}
\end{figure*}

\begin{figure*}[ht!]
\begin{center}
\includegraphics[width=18 cm,angle=0]{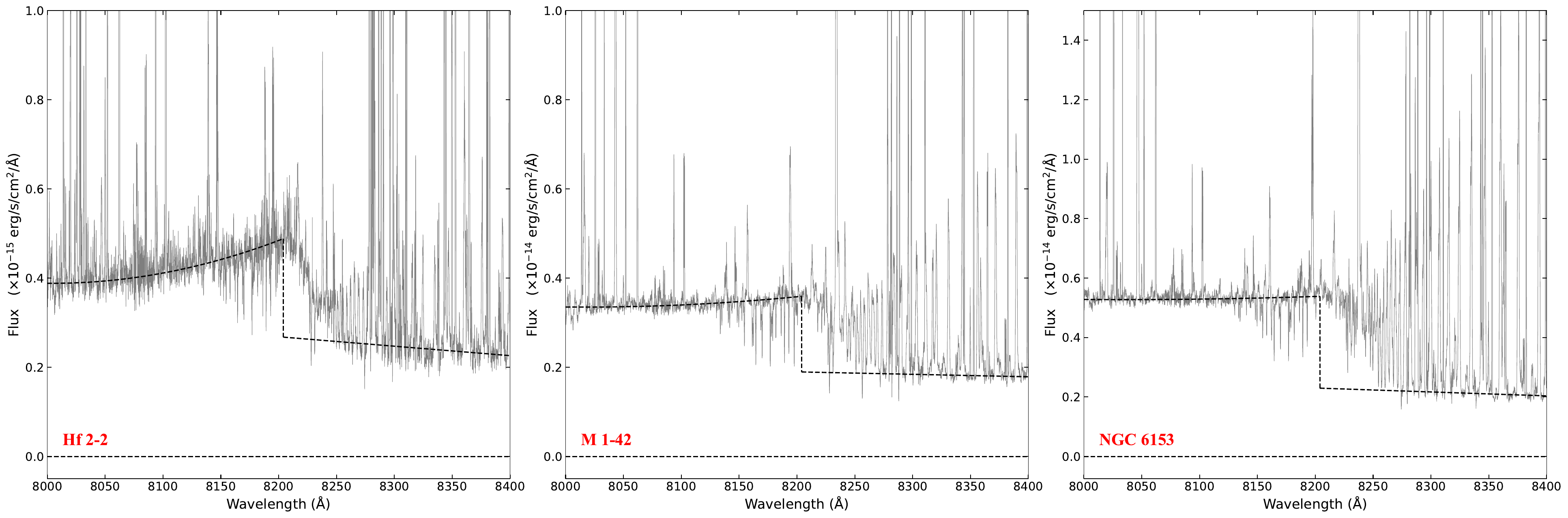}
\caption{Same as Figure\,\ref{fig:BJ}, but the UVES spectra and continuum fittings near the Paschen discontinuity at 8204\,{\AA}.} 
\label{fig:PJ}
\end{center}
\end{figure*}

During the second exposure of Hf\,2-2 with Dichroic\,2, the UVES slit exhibits a slight offset perpendicular to its spatial direction (see Figure\,\ref{fig:slitview}, panels (a) and (b)).  Fortunately, flux comparisons between overlapping regions of the CD\#1 and CD\#2 spectra revealed that the majority of emission lines showed flux differences smaller than their measurement uncertainties, likely attributable to Hf\,2-2's uniform surface brightness.  The sole exception was [Ne~{\sc iii}] $\lambda$3869, for which we measured in CD\#2 a flux that was 6\% higher than in CD\#1.  Given this line's position near the edge of both the CD\#1 spectral coverage and the CCD detector, the quality of data processing in that region may not be optimal.  Therefore, we decided no to apply a flux scale correction.  However, the difference in the slit position introduced slight variations in the spectral profile and the position-velocity (PV) map.  Subsequent analyses of PV diagrams prioritized contemporaneously observed lines to mitigate these inconsistencies.  The measurements of emission lines (including laboratory/observed wavelengths, fluxes, identifications) detected in the spectra of the three PNe are tabulated in Tables\,\ref{Hf22_linelist}, \ref{M142_linelist} and \ref{NGC_linelist} (in Appendix\,\ref{appendix:linelist}), where the asterisk``$\ast$" indicates that the line is blended with the adjacent one right above.  

\section{Plasma diagnostics} \label{sec:diagnostics}

\subsection{Plasma Diagnostics with CELs} 
\label{sec:celdiagnostics}

Electron temperatures and densities of the three PNe were determined based on the CEL ratios presented in Table\,\ref{temden} using the code \texttt{PyNeb} v1.1.28 \citep[][]{2015A&A...573A..42L,2020Atoms...8...66M}.  Fluxes of the [O\,{\sc ii}] and [N\,{\sc ii}] auroral lines can be affected by recombination excitation and require corrections to avoid temperature overestimates \citep[e.g.][]{1986ApJ...309..334R,2000MNRAS.312..585L}.  Recombination contributions to these forbidden lines were quantified using the equations: 
\begin{equation}\label{eq1} 
I_r (5755)={j(5755)\over j(5679)}\times I(5679)
\end{equation} 
and
\begin{equation} 
I_r (7325+)={j(7325+)\over j(4649)}\times I(4649),
\end{equation} 
where $j$($\lambda$) denotes recombination emissivity for a specific line at wavelength $\lambda$, and ``7325$+$'' represents a sum of the four O\,{\sc ii} $\lambda$$\lambda$7319,7320,7330,7331 auroral lines.  References to the atomic data used in plasma diagnostics and ionic abundance determinations are summarized in Table\,\ref{tab:atomic_data}.  
The electron temperatures and densities derived using the O\,{\sc ii} and N\,{\sc ii} ORL ratios (in Section\,\ref{sec:orl_diagnostics}) were then used to estimate the contribution of recombination excitation to the fluxes of the [O\,{\sc ii}] and [N\,{\sc ii}] auroral lines, respectively.

\begin{figure*}[ht!]
\begin{center}
\includegraphics[width=18 cm,angle=0]{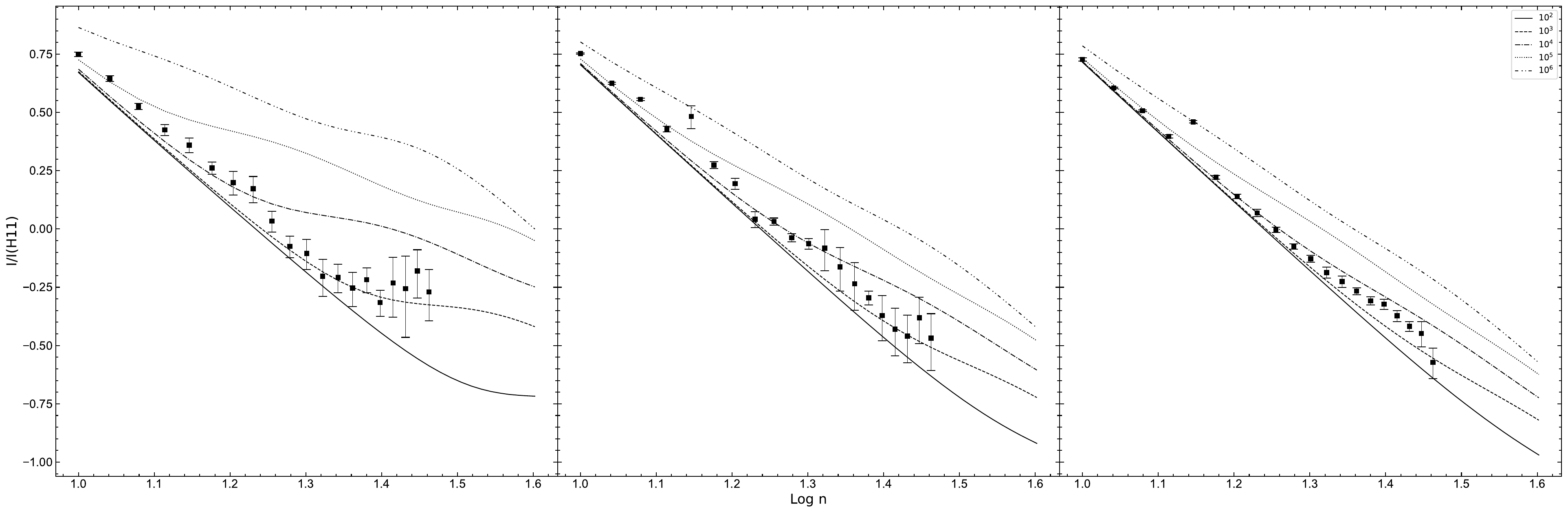}
\caption{Intensities (normalized to H$\beta$ =100) of high-order (10$\leqslant\,n\,<$30) Balmer lines as a function of the principal quantum number $n$ for Hf\,2-2 (left), M\,1-42 (middle) and NGC\,6153 (right).  The curves are the theoretical intensities of high-order Balmer lines, calculated using the effective recombination coefficients from \citet{1995MNRAS.272...41S}, as a function of $n$ at various density cases (10$^2$--10$^6$ cm$^{-3}$), at the Balmer jump temperature of each PN given in Table\,\ref{temden}.  The intensity of H14 may be overestimated possibly due to a blend with [S\,{\sc iii}] $\lambda$3722.} 
\label{fig:BD}
\end{center}
\end{figure*}

\begin{figure*}[ht!]
\begin{center}
\includegraphics[width=18cm,angle=0]{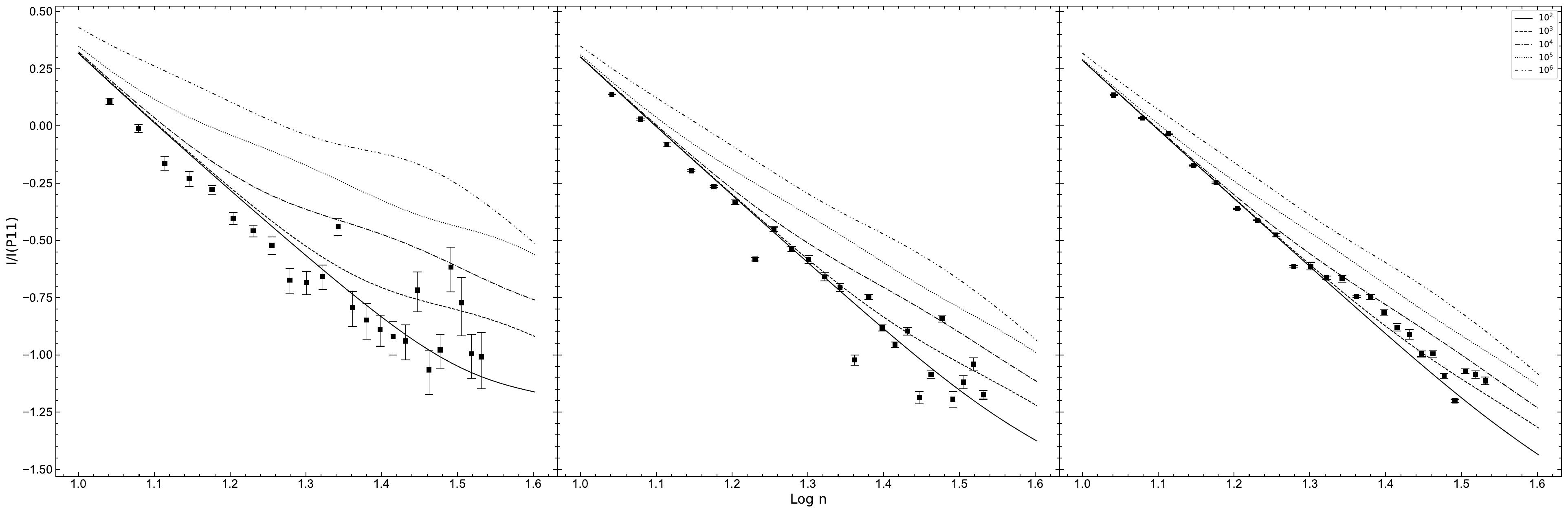}
\caption{Same as Figure\,\ref{fig:BD} but for the H\,{\sc i} Paschen decrements (10$<\,n\,<$35).  The curves are the theoretical intensities of high-order Paschen lines, calculated using the effective recombination coefficients from \citet{1995MNRAS.272...41S}, as a function of $n$ at different density cases, at the Balmer jump temperature of each PN.} 
\label{fig:PD}
\end{center}
\end{figure*}

The nebular lines are theoretically susceptible to recombination excitation as well. However, the nebular lines are usually strong and the recombination contribution is negligible, except for the [O\,{\sc ii}] $\lambda\lambda$3726,3729 nebular lines.  However, their low critical densities ($n_{\rm crit}\sim$10$^{3}$ cm$^{-3}$ for $\lambda$3726, and $n_{\rm crit}\sim$3$\times$10$^{3}$ cm$^{-3}$ for $\lambda$3729) enable collisional de-excitation of recombination-populated upper levels, causing calculated recombination contributions to exceed observed fluxes in all samples. Besides, the recombination contribution to the [O\,{\sc ii}] auroral line flux matches the observed flux.  Given the considerable uncertainty in the recombination contribution, we decided not to apply these corrections and present the uncorrected (and certainly inaccurate) [O\,{\sc ii}] temperatures/densities in Table\,\ref{temden}.  

The recombination corrections for the [O\,{\sc iii}] $\lambda$4363 auroral line were addressed differently across targets.  For Hf\,2-2, which is a low-excitation nebula with undetected pure O\,{\sc iii} recombination lines. We derived the O$^{3+}$/H$^{+}$ abundance ratio based on the differences of oxygen elemental abundance and O$^{+}$/H$^+$ + O$^{2+}$/H$^+$, then computed recombination effects. This yielded negligible corrections due to the low ionization nature of Hf\,2-2. For M\,1-42 and NGC\,6153, detected the O\,{\sc iii} $\lambda\lambda$3260,3265 pure recombination lines and the [O\,{\sc iii}] $\lambda$5592 charge-exchange line enabled direct corrections.  Recombination contributions used temperature-dependent emissivities from [O\,{\sc iii}] diagnostics, while charge-exchange effects employed transition probabilities from \citet{1982ApJ...257L..87D}.  Both processes were insignificant for [O\,{\sc iii}] nebular lines due to their high intrinsic strengths, so that no correction was made for these two lines.

\begin{figure}[ht!]
\begin{center}
\includegraphics[width=8.5 cm,angle=0]{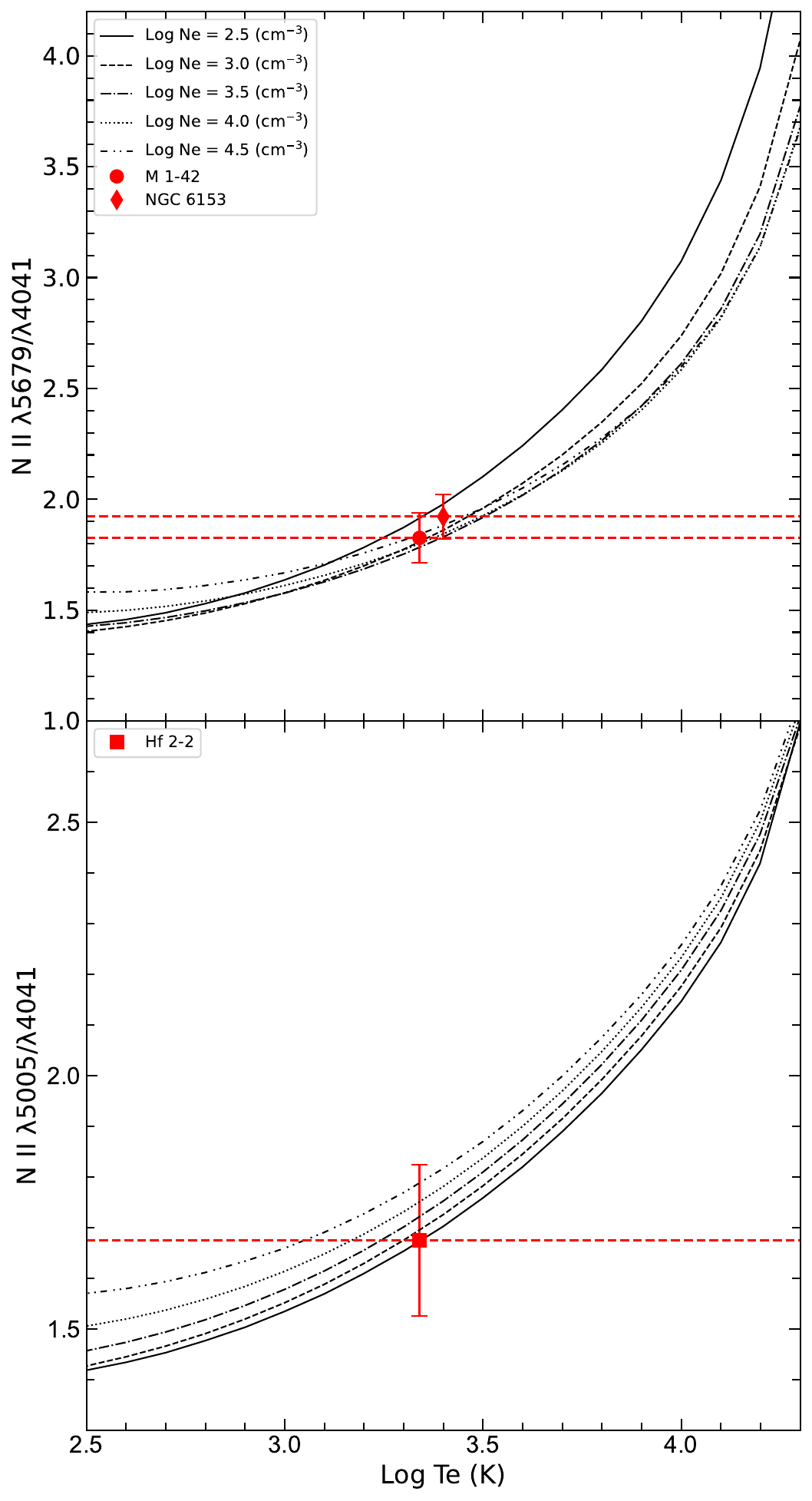}
\caption{Top: The theoretical N\,{\sc ii} $\lambda$5679/$\lambda$4041 line ratio as a function of electron temperature at various density cases; overplotted are the observed line ratios in M\,1-42 and NGC\,6153.  The theoretical N\,{\sc ii} $\lambda$5679/$\lambda$4041 line ratio observed in Hf\,2-2 is out of the diagnostic range.  Bottom: Same as the top panel but for the N\,{\sc ii} $\lambda$5005/$\lambda$4041 line ratio; overplotted is the observed data in Hf\,2-2.  The theoretical curves are based on the effective recombination coefficients for the N\,{\sc ii} nebular recombination lines calculated by \citet{2011A&A...530A..18F,2013A&A...550C...2F}.} 
\label{fig:NIITe}
\end{center}
\end{figure}

The CELs of neutral atoms were detected and employed to probe the photodissociation regions (PDRs).  As the [C\,{\sc i}] diagnostic ratio is not available in \texttt{PyNeb}, we adopted the methodology of \citet{1995MNRAS.273...47L}.  The anomalously high [C\,{\sc i}] temperature derived for M\,1-42 (Table\,\ref{temden}) may indicate possible recombination contribution to these lines, which needs careful investigations. 

The CEL diagnostic diagrams for the three PNe are presented in Figure\,\ref{fig:celdiag}.  Most Temperature-sensitive and density-sensitive diagnostic curves show convergence across all three PNe. Despite differing excitation classes, all nebulae exhibit characteristic PN electron temperatures near 10,000 K. Density variations are pronounced. Among the three nebulae, Hf\,2-2 has the lowest density, about a few hundred cm$^{-3}$, while M\,1-42 and NGC\,6153 show densities of $\sim$2000 and $\sim$6000\,cm$^{-3}$, respectively.  

\subsection{Plasma Diagnostics with the H\,{\sc i} Recombination Spectrum} \label{sec:HI_diagnostics}

The recombination spectra of H\,{\sc i} provided additional nebular physical diagnostics.  The Balmer jump temperature was derived from the H11 line intensity and continuum difference across the jump (i.e., height of the jump), $[I_c(\lambda3643)-I_c(\lambda3681)]$, as measured via local continuum fitting and extrapolation.  We first mask all emission features in the spectra using a sigma-clipping procedure (iteratively rejecting the $>3\sigma$ outliers).  We then fit the spectral continua on the blue and red sides of the Balmer jump separately, using low-order polynomials, and extrapolate towards the centre of the jump; subsequently, we measured the height of Balmer jump from the extrapolated continua on both edges.  Uncertainties were estimated through propagation from the polynomial-fit covariances and the root-mean-square values of the pre-fit continuum.  Calculations followed \citet[][Equation\,3 therein]{2001MNRAS.327..141L}, incorporating the He$^+$/H$^{+}$ and He$^{2+}$/H$^{+}$ abundance ratios in Table\,\ref{tab:cel_ionic}.  The Paschen jump temperatures were similarly determined, using the methodology of \citet[][Equation\,7 therein]{2011MNRAS.415..181F}.  Both jump temperatures show consistency with each other but remain systematically lower than the CEL-derived temperatures.  The continuum fits for the Balmer and Paschen jumps/discontinuities are illustrated in Figures\,\ref{fig:BJ} and \ref{fig:PJ}, respectively.

\begin{figure}[ht!]
\begin{center}
\includegraphics[width=8.5 cm,angle=0]{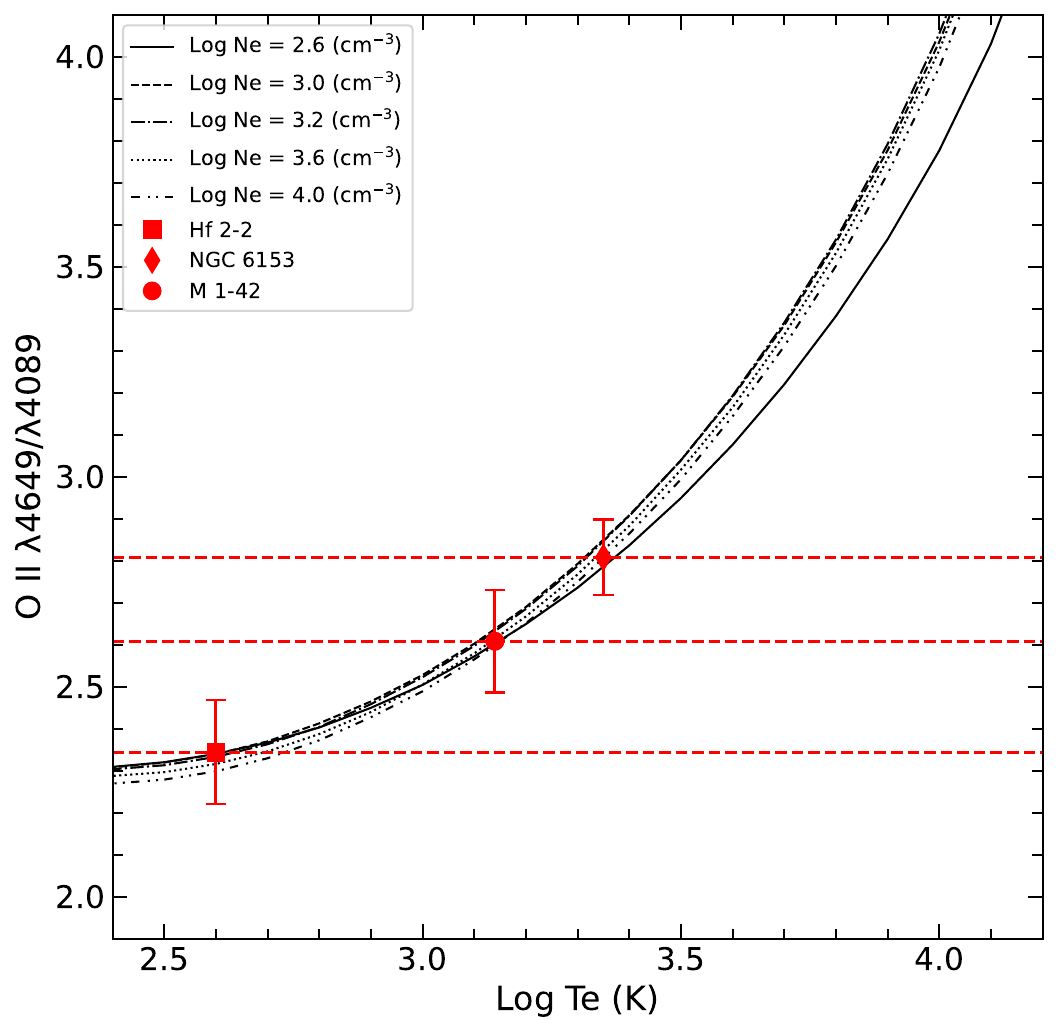}
\caption{Same as Figure\,\ref{fig:NIITe} but for the O\,{\sc ii} $\lambda$4649/$\lambda$4089 line ratio.  The theoretical curves are based on the effective recombination coefficients of the O\,{\sc ii} nebular recombination lines calculated by \citet{2017MNRAS.472.1182P}.} 
\label{fig:OIITe}
\end{center}
\end{figure}

High-order Balmer and Paschen line fluxes were used to estimate electron density. The continuum close to the discontinuity is difficult to fit and the hydrogen lines there are weak. Thus only Balmer lines with principal quantum numbers $n<30$ and Paschen lines with principal quantum numbers $n<35$ ($n<30$ in Hf\,2-2) were adopted. Partial blending between faint hydrogen lines as well as with other emission lines introduces large measurement uncertainties, permitting only approximate density determinations. The comparisons of theoretical and observed intensities of these transitions are presented in Figure \ref{fig:BD} and \ref{fig:PD}.

\subsection{Plasma Diagnostics with the He\,{\sc i} and Heavy-element ORLs} \label{sec:orl_diagnostics}

Electron temperatures were derived using the He\,{\sc i} $\lambda$5876/$\lambda$4471, $\lambda$6678/$\lambda$4471 and $\lambda$7281/$\lambda$6678 line ratios, based on the updated He\,{\sc i} atomic data from \citet[][in Case\,B assumption]{2022MNRAS.513.1198D}.  In each PN, the He\,{\sc i} $\lambda$5876/$\lambda$4471 and $\lambda$6678/$\lambda$4471 ratios yield consistent temperatures, which are different from that yielded by the $\lambda$7281/$\lambda$6678 ratio (Table\,\ref{temden}).  He\,{\sc i} $\lambda$5876, $\lambda$6678 and $\lambda$7281 are transitions among the triplet states of helium, and may potentially be affected by the optical depth effect.  Following the suggestion of \citet{2005MNRAS.358..457Z}, who demonstrated that the He\,{\sc i} $\lambda$7281/$\lambda$6678 ratio has superior reliability in temperature determination, we adopted $T_{\rm e}$(He\,{\sc i} $\lambda$7281/$\lambda$6678) in subsequent ionic abundance calculations.

\begin{figure}[ht!]
\begin{center}
\includegraphics[width=8.5 cm,angle=0]{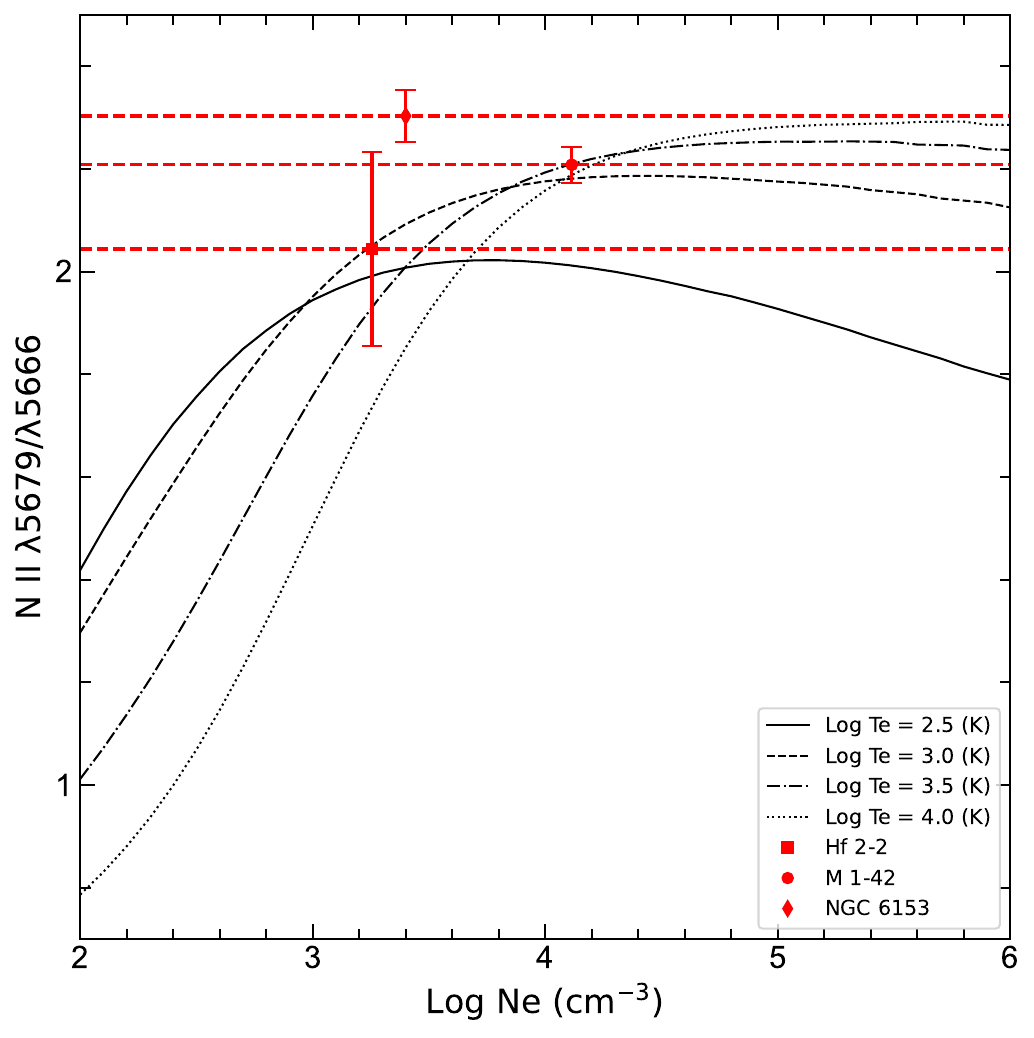}
\caption{The theoretical N\,{\sc ii} $\lambda$5679/$\lambda$5666 line ratio as a function of electron density at different temperature cases; overplotted are the observed line ratios in the three PNe.  The observed ratio in NGC\,6153 is above the upper limit of theoretical calculations.  The curves are based on the effective recombination coefficients for the N\,{\sc ii} nebular recombination lines calculated by \citet{2011A&A...530A..18F,2013A&A...550C...2F}.} 
\label{fig:NIINe}
\end{center}
\end{figure}

The He\,{\sc i} temperatures in the three PNe are systematically lower than those yielded by the CELs (e.g., $T_{\rm e}$([O\,{\sc iii}]), see Table\,\ref{temden}).  The $T_{\rm e}$(He\,{\sc i})-$T_{\rm e}$([O\,{\sc iii}]) difference has recently been investigated by \citet{2025ApJ...986...74M} through spectral analysis of large samples of H\,{\sc ii} regions and PNe.  Deviations from Case\,B in the recombination of He\,{\sc i} and temperature inhomogeneities in nebulae were suggested to be the possible culprit of such temperature discrepancy \citep{2025ApJ...986...74M}. 

Emissivities of heavy-element ORLs have differential temperature dependence across the multiplets with different orbital angular momentum quantum numbers, $l$, enabling temperature determination with flux ratios of the lines with different $l$ \citep[e.g.][]{2011A&A...530A..18F,2017MNRAS.472.1182P}.  The upper panel of Figure\,\ref{fig:NIITe} shows the N\,{\sc ii} $\lambda$5679/$\lambda$4041 line ratio as a function of electron temperature at different density cases.  Comparisons with theoretical curves yielded N\,{\sc ii} ORL temperatures of about 2150 K for M\,1-42 and 2500 K for NGC\,6153. For Hf\,2-2, where the $\lambda$5679/$\lambda$4041 ratio fell outside the theoretical range, the $\lambda$5005/$\lambda$4041 ratio was employed instead, yielding a N\,{\sc ii} temperature of 2500 K as shown in the lower panel of Figure \ref{fig:NIITe}.

\begin{figure}[ht!]
\begin{center}
\includegraphics[width=8.5 cm,angle=0]{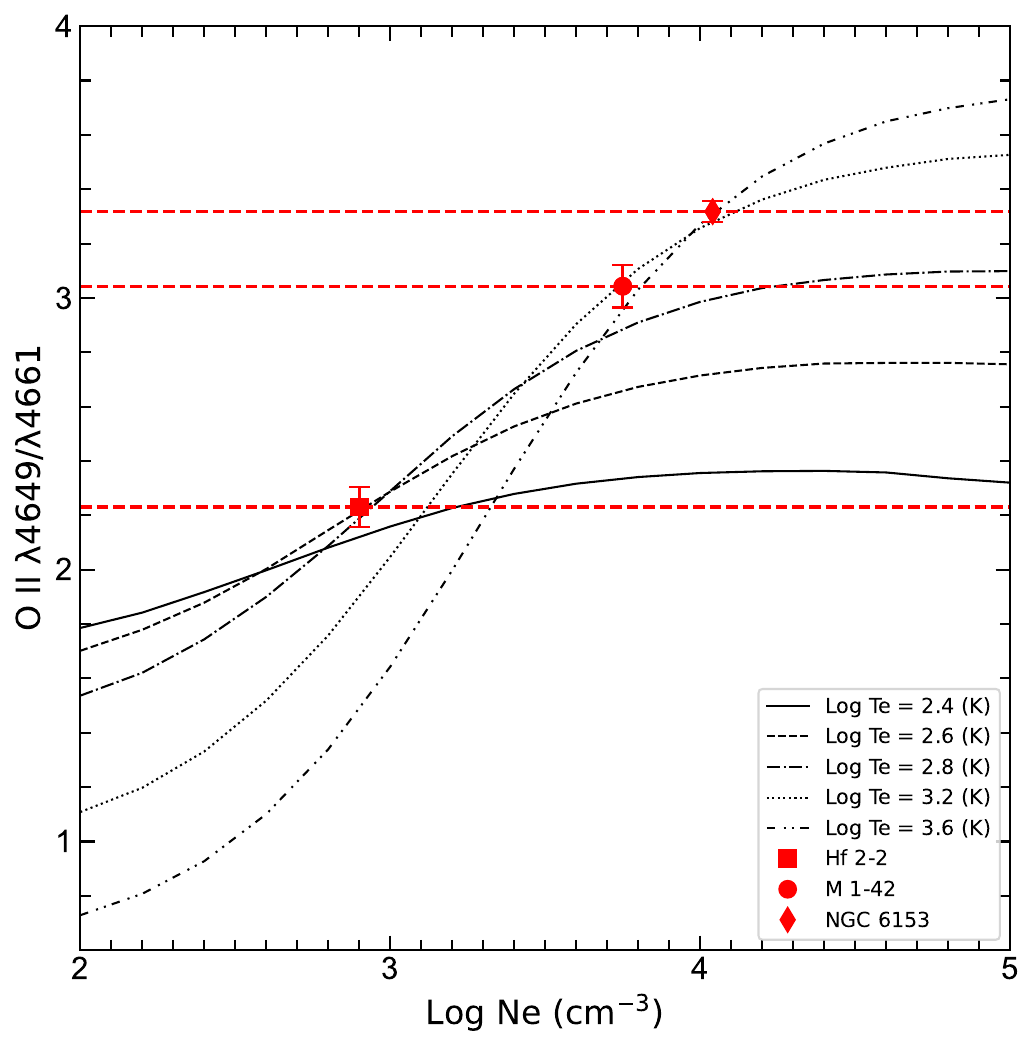}
\caption{Same as Figure\,\ref{fig:NIINe} but for the O\,{\sc ii} $\lambda$4649/$\lambda$4661 density diagnostic.} 
\label{fig:OIINe} 
\end{center}
\end{figure}

The temperature relations of O\,{\sc ii} $\lambda$4649/$\lambda$4089 ratio as a function of electron temperature are presented in Figure\,\ref{fig:OIITe}.  We obtained O\,{\sc ii} temperatures of about 400, 1370 and 2240 K for Hf\,2-2, M\,1-42 and NGC\,6153, respectively.  These O\,{\sc ii} temperatures are systematically lower than N\,{\sc ii} temperature across all three PNe, with the discrepancy most pronounced in Hf\,2-2.  The observed O\,{\sc ii} and N\,{\sc ii} line ratios lie near the extremes/limits of the diagnostic curves, resulting in the large uncertainties of temperatures despite relatively small measurement errors of emission lines. Collectively, heavy-element ORL temperatures remain lower than those from CELs, H\,{\sc i} and He\,{\sc i} recombination diagnostics.

In low-density PNe, electron density influences the populations of ions' ground-term fine-structure levels, which can be used to determine electron density by comparing the intensities of ORLs from the same multiplet \citep{2011A&A...530A..18F}. Line ratios N\,{\sc ii} $\lambda$5679/$\lambda$5666 and O\,{\sc ii} $\lambda$4649/$\lambda$4661 were utilized for density estimations, which are shown in Figure\,\ref{fig:NIINe} and \ref{fig:OIINe}, respectively. The O\,{\sc ii} densities of Hf\,2-2, M\,1-42 and NGC\,6153 are about 800, 5600 and 11,000 cm$^{-3}$, but their N\,{\sc ii} densities are systemically higher, and the N\,{\sc ii} diagnostic ratio of NGC\,6153 exceed the range of theoretical curves. The N\,{\sc ii} line ratios of M\,1-42 is at the edge of the diagnostic region, and its temperature uncertainty is significant.

Heavy-element ORL densities are higher than those from CELs, together with the temperature results, suggesting these emission lines originate from distinct regions:  ORLs from high-density, low-temperature zones, while CELs come from high-temperature, low-density areas. Our results are consistent with those yielded by other observations of the three PNe \citep{2022MNRAS.510.5444G, 2024AA...689A.228G} as well as those of other high-ADF PNe, such as NGC\,6778 \citep[e.g.][]{2022MNRAS.510.5444G}.

\section{Abundance Determination} \label{sec:abundance}

\subsection{Ionic Abundances from CELs} \label{sec:cels_ionic}

\begin{table*}
\begin{center}
\caption{Ionic Abundances Derived Using CELs and He Lines}
\label{tab:cel_ionic}
\begin{tabular}{llccc}
\hline\hline
Ion & Line & \multicolumn{3}{c}{Abundance (X$^{i+}$/H$^+$)} \\
\cline{3-5}
 & (\AA) & Hf\,2-2 & M\,1-42 & NGC\,6153\\
\hline
He$^+$ & He~{\sc i} $\lambda$4711 & 0.140($\pm$0.002) & 0.132($\pm$0.002) & 0.109($\pm$0.001)\\
 & He~{\sc i} $\lambda$5876 & 0.137($\pm$0.001) & 0.133($\pm$0.001) & 0.109($\pm$0.001)\\
 & He~{\sc i} $\lambda$6678 & 0.142($\pm$0.001) & 0.134($\pm$0.001) & 0.111($\pm$0.001)\\
adopted\tablenotemark{\rm{\scriptsize a}} & & 0.137($\pm$0.001) & 0.133($\pm$0.001) & 0.109($\pm$0.001)\\
He$^{2+}$ & He~{\sc ii} $\lambda$4686 & 2.95($\pm$0.05)$\times10^{-3}$ & 1.77($\pm$0.01)$\times10^{-2}$ & 1.44($\pm$0.01)$\times10^{-2}$\\
C$^0$ & $[$C~{\sc i}$]$ $\lambda$8727 & 2.10($\pm$1.10)$\times10^{-7}$ & 5.90($\pm$0.70)$\times10^{-7}$ & 9.40($\pm$2.09)$\times10^{-8}$\\
 & $[$C~{\sc i}$]$ $\lambda$9824 & 2.12($\pm$0.98)$\times10^{-7}$ & 2.69($\pm$0.32)$\times10^{-7}$ & 5.41($\pm$0.54)$\times10^{-8}$\\
 & $[$C~{\sc i}$]$ $\lambda$9850 & 1.93($\pm$0.17)$\times10^{-7}$ & 2.36($\pm$0.10)$\times10^{-7}$ & 6.23($\pm$0.25)$\times10^{-8}$\\
adopted\tablenotemark{\rm{\scriptsize b}} & & 1.97($\pm$0.22)$\times10^{-7}$ & 2.44($\pm$0.11)$\times10^{-7}$ & 5.97($\pm$0.25)$\times10^{-8}$\\
N$^0$ & $[$N~{\sc i}$]$ $\lambda$5198 & 8.99($\pm$0.36)$\times10^{-7}$ & 3.95($\pm$0.05)$\times10^{-6}$ & 3.14($\pm$0.18)$\times10^{-7}$\\
 & $[$N~{\sc i}$]$ $\lambda$5200 & 1.02($\pm$0.04)$\times10^{-6}$ & 3.95($\pm$0.05)$\times10^{-6}$ & 3.30($\pm$0.30)$\times10^{-7}$\\
adopted\tablenotemark{\rm{\scriptsize b}} & & 9.56($\pm$0.27)$\times10^{-7}$ & 3.95($\pm$0.05)$\times10^{-6}$ & 3.20($\pm$0.16)$\times10^{-7}$\\
N$^+$ & $[$N~{\sc ii}$]$ $\lambda$5755 & 4.56($\pm$0.46)$\times10^{-6}$ & 8.42($\pm$0.06)$\times10^{-5}$ & 1.80($\pm$0.02)$\times10^{-5}$\\
 & $[$N~{\sc ii}$]$ $\lambda$6548 & 4.21($\pm$0.05)$\times10^{-6}$ & 8.13($\pm$0.04)$\times10^{-5}$ & 1.72($\pm$0.01)$\times10^{-5}$\\
 & $[$N~{\sc ii}$]$ $\lambda$6583 & 4.63($\pm$0.02)$\times10^{-6}$ & 8.59($\pm$0.02)$\times10^{-5}$ & 1.80($\pm$0.01)$\times10^{-5}$\\
adopted\tablenotemark{\rm{\scriptsize c}} & & 4.63($\pm$0.02)$\times10^{-6}$ & 8.59($\pm$0.02)$\times10^{-5}$ & 1.80($\pm$0.01)$\times10^{-5}$\\
O$^0$ & $[$O~{\sc i}$]$ $\lambda$5577 & $\cdots$ & 1.52($\pm$0.17)$\times10^{-5}$ & 2.51($\pm$1.26)$\times10^{-6}$\\
 & $[$O~{\sc i}$]$ $\lambda$6300 & 3.20($\pm$0.29)$\times10^{-7}$ & 1.52($\pm$0.01)$\times10^{-5}$ & 2.54($\pm$0.03)$\times10^{-6}$\\
 & $[$O~{\sc i}$]$ $\lambda$6363 & 2.87($\pm$0.97)$\times10^{-7}$ & 1.58($\pm$0.02)$\times10^{-5}$ & 2.61($\pm$0.05)$\times10^{-6}$\\
adopted\tablenotemark{\rm{\scriptsize b}} & & 3.12($\pm$0.32)$\times10^{-7}$ & 1.54($\pm$0.01)$\times10^{-5}$ & 2.57($\pm$0.04)$\times10^{-6}$\\
O$^+$ & $[$O~{\sc ii}$]$ $\lambda$3726 & 5.99($\pm$0.32)$\times10^{-6}$ & 3.30($\pm$0.14)$\times10^{-5}$ & 9.31($\pm$0.12)$\times10^{-6}$\\
O$^{2+}$ & $[$O~{\sc iii}$]$ $\lambda$4363 & 9.45($\pm$0.41)$\times10^{-5}$ & 2.65($\pm$0.02)$\times10^{-4}$ & 4.04($\pm$0.02)$\times10^{-4}$\\
 & $[$O~{\sc iii}$]$ $\lambda$4959 & 9.21($\pm$0.02)$\times10^{-5}$ & 2.56($\pm$0.01)$\times10^{-4}$ & 4.06($\pm$0.01)$\times10^{-4}$\\
 & $[$O~{\sc iii}$]$ $\lambda$5007 & 9.37($\pm$0.01)$\times10^{-5}$ & 2.64($\pm$0.01)$\times10^{-4}$ & 4.07($\pm$0.01)$\times10^{-4}$\\
adopted\tablenotemark{\rm{\scriptsize c}} & & 9.37($\pm$0.01)$\times10^{-5}$ & 2.64($\pm$0.01)$\times10^{-4}$ & 4.07($\pm$0.01)$\times10^{-4}$\\
Ne$^{2+}$ & $[$Ne~{\sc iii}$]$ $\lambda$3869 & 4.42($\pm$0.03)$\times10^{-5}$ & 1.13($\pm$0.01)$\times10^{-4}$ & 1.45($\pm$0.01)$\times10^{-4}$\\
 & $[$Ne~{\sc iii}$]$ $\lambda$3967 & 3.01($\pm$0.06)$\times10^{-5}$ & 1.05($\pm$0.01)$\times10^{-4}$ & 1.39($\pm$0.01)$\times10^{-4}$\\
adopted\tablenotemark{\rm{\scriptsize d}} & & 3.95($\pm$0.03)$\times10^{-5}$ & 1.09($\pm$0.01)$\times10^{-4}$ & 1.42($\pm$0.01)$\times10^{-4}$\\
Ne$^{3+}$ & $[$Ne~{\sc iv}$]$ $\lambda$4714 & $\cdots$ & 2.06($\pm$0.21)$\times10^{-4}$ & 1.34($\pm$0.15)$\times10^{-4}$\\
 & $[$Ne~{\sc iv}$]$ $\lambda$4724 & $\cdots$ & 1.20($\pm$0.16)$\times10^{-4}$ & 8.19($\pm$1.32)$\times10^{-5}$\\
 & $[$Ne~{\sc iv}$]$ $\lambda$4725 & $\cdots$ & 8.65($\pm$1.75)$\times10^{-5}$ & 4.74($\pm$1.48)$\times10^{-5}$\\
adopted\tablenotemark{\rm{\scriptsize d}} & & $\cdots$ & 1.33($\pm$0.11)$\times10^{-4}$ & 8.73($\pm$0.83)$\times10^{-5}$\\
Ne$^{4+}$ & $[$Ne~{\sc v}$]$ $\lambda$3426 & $\cdots$ & 2.41($\pm$0.87)$\times10^{-6}$ & $\cdots$\\
S$^+$ & $[$S~{\sc ii}$]$ $\lambda$4069 & 2.96($\pm$0.50)$\times10^{-7}$ & 1.52($\pm$0.03)$\times10^{-6}$ & 5.04($\pm$0.13)$\times10^{-7}$\\
 & $[$S~{\sc ii}$]$ $\lambda$4076 & $\cdots$ & 6.09($\pm$0.11)$\times10^{-6}$ & 3.72($\pm$0.24)$\times10^{-7}$\\
 & $[$S~{\sc ii}$]$ $\lambda$6716 & 1.86($\pm$0.03)$\times10^{-7}$ & 1.81($\pm$0.01)$\times10^{-6}$ & 6.15($\pm$0.01)$\times10^{-7}$\\
 & $[$S~{\sc ii}$]$ $\lambda$6731 & 1.88($\pm$0.04)$\times10^{-7}$ & 1.78($\pm$0.01)$\times10^{-6}$ & 6.26($\pm$0.02)$\times10^{-7}$\\
 & $[$S~{\sc ii}$]$ $\lambda$10287 & $\cdots$ & 1.49($\pm$0.10)$\times10^{-6}$ & 4.23($\pm$0.41)$\times10^{-7}$\\
 & $[$S~{\sc ii}$]$ $\lambda$10336 & $\cdots$ & 1.54($\pm$0.20)$\times10^{-6}$ & 5.11($\pm$0.56)$\times10^{-7}$\\
 & $[$S~{\sc ii}$]$ $\lambda$10287 & $\cdots$ & 1.70($\pm$0.38)$\times10^{-6}$ & 6.24($\pm$1.10)$\times10^{-7}$\\
adopted\tablenotemark{\rm{\scriptsize e}} & & 1.87($\pm$0.03)$\times10^{-7}$ & 1.80($\pm$0.01)$\times10^{-6}$ & 6.19($\pm$0.01)$\times10^{-7}$\\
S$^{2+}$ & $[$S~{\sc iii}$]$ $\lambda$6312 & 2.97($\pm$0.31)$\times10^{-6}$ & 4.54($\pm$0.04)$\times10^{-6}$ & 5.01($\pm$0.02)$\times10^{-6}$\\
 & $[$S~{\sc iii}$]$ $\lambda$9069 & 3.27($\pm$0.02)$\times10^{-6}$ & 6.75($\pm$0.05)$\times10^{-6}$ & 5.66($\pm$0.02)$\times10^{-6}$\\
 & $[$S~{\sc iii}$]$ $\lambda$9531 & 2.83($\pm$0.02)$\times10^{-6}$ & 4.04($\pm$0.03)$\times10^{-6}$ & 5.10($\pm$0.01)$\times10^{-6}$\\
adopted\tablenotemark{\rm{\scriptsize c}} & & 2.83($\pm$0.02)$\times10^{-6}$ & 4.04($\pm$0.03)$\times10^{-6}$ & 5.10($\pm$0.01)$\times10^{-6}$\\
Cl$^+$ & $[$Cl~{\sc ii}$]$ $\lambda$8579 & 2.60($\pm$0.96)$\times10^{-9}$ & 2.12($\pm$0.08)$\times10^{-8}$ & 5.90($\pm$0.15)$\times10^{-9}$\\
 & $[$Cl~{\sc ii}$]$ $\lambda$9124 & $\cdots$ & 1.91($\pm$0.11)$\times10^{-8}$ & 4.42($\pm$1.10)$\times10^{-9}$\\
adopted\tablenotemark{\rm{\scriptsize f}} & & 2.60($\pm$0.96)$\times10^{-9}$ & 2.12($\pm$0.08)$\times10^{-8}$ & 5.90($\pm$0.15)$\times10^{-9}$\\
Cl$^{2+}$ & $[$Cl~{\sc iii}$]$ $\lambda$5518 & 2.89($\pm$0.22)$\times10^{-8}$ & 1.08($\pm$0.01)$\times10^{-7}$ & 1.18($\pm$0.02)$\times10^{-7}$\\
\hline
\end{tabular}
\addtocounter{table}{-1}
\caption{(Continued)}
\end{center}
\end{table*}

\begin{table*}
\begin{center}
\addtocounter{table}{-1}
\caption{(Continued)}
\begin{tabular}{llccc}
\hline\hline
Ion & Line & \multicolumn{3}{c}{Abundance (X$^{i+}$/H$^+$)} \\
\cline{3-5}
 & (\AA) & Hf\,2-2 & M\,1-42 & NGC\,6153\\
\hline
 & $[$Cl~{\sc iii}$]$ $\lambda$5538 & 2.91($\pm$0.38)$\times10^{-8}$ & 1.08($\pm$0.02)$\times10^{-7}$ & 1.17($\pm$0.01)$\times10^{-7}$\\
 & $[$Cl~{\sc iii}$]$ $\lambda$8481 & $\cdots$ & 1.57($\pm$0.22)$\times10^{-7}$ & 1.36($\pm$0.10)$\times10^{-7}$\\
adopted\tablenotemark{\rm{\scriptsize g}} & & 2.90($\pm$0.20)$\times10^{-8}$ & 1.08($\pm$0.01)$\times10^{-7}$ & 1.17($\pm$0.01)$\times10^{-7}$\\
Cl$^{3+}$ & $[$Cl~{\sc iv}$]$ $\lambda$7531 & 1.48($\pm$0.18)$\times10^{-8}$ & 7.22($\pm$0.15)$\times10^{-8}$ & 7.18($\pm$0.06)$\times10^{-8}$\\
 & $[$Cl~{\sc iv}$]$ $\lambda$8046 & 9.08($\pm$1.64)$\times10^{-9}$ & 4.38($\pm$0.06)$\times10^{-8}$ & 4.97($\pm$0.04)$\times10^{-8}$\\
adopted\tablenotemark{\rm{\scriptsize h}} & & 9.08($\pm$1.64)$\times10^{-9}$ & 4.38($\pm$0.06)$\times10^{-8}$ & 4.97($\pm$0.04)$\times10^{-8}$\\
Ar$^{2+}$ & $[$Ar~{\sc iii}$]$ $\lambda$5192 & $\cdots$ & 1.93($\pm$0.23)$\times10^{-6}$ & 1.60($\pm$0.10)$\times10^{-6}$\\
 & $[$Ar~{\sc iii}$]$ $\lambda$7136 & 5.10($\pm$0.04)$\times10^{-7}$ & 1.91($\pm$0.02)$\times10^{-6}$ & 1.96($\pm$0.01)$\times10^{-6}$\\
 & $[$Ar~{\sc iii}$]$ $\lambda$7751 & 5.33($\pm$0.12)$\times10^{-7}$ & 1.82($\pm$0.01)$\times10^{-6}$ & 1.92($\pm$0.01)$\times10^{-6}$\\
adopted\tablenotemark{\rm{\scriptsize g}} & & 5.16($\pm$0.04)$\times10^{-7}$ & 1.85($\pm$0.02)$\times10^{-6}$ & 1.94($\pm$0.01)$\times10^{-6}$\\
Ar$^{3+}$ & $[$Ar~{\sc iv}$]$ $\lambda$4711 & 9.89($\pm$0.78)$\times10^{-8}$ & 8.62($\pm$0.04)$\times10^{-7}$ & 1.05($\pm$0.01)$\times10^{-6}$\\
 & $[$Ar~{\sc iv}$]$ $\lambda$4740 & 1.00($\pm$0.11)$\times10^{-7}$ & 8.61($\pm$0.05)$\times10^{-7}$ & 1.04($\pm$0.01)$\times10^{-6}$\\
 & $[$Ar~{\sc iv}$]$ $\lambda$7171 & $\cdots$ & 1.42($\pm$0.13)$\times10^{-6}$ & 1.71($\pm$0.12)$\times10^{-6}$\\
 & $[$Ar~{\sc iv}$]$ $\lambda$7263 & $\cdots$ & 1.18($\pm$0.16)$\times10^{-6}$ & 1.54($\pm$0.20)$\times10^{-6}$\\
adopted\tablenotemark{\rm{\scriptsize g}} & & 9.94($\pm$0.65)$\times10^{-8}$ & 8.62($\pm$0.03)$\times10^{-7}$ & 1.05($\pm$0.01)$\times10^{-6}$\\
Ar$^{4+}$ & $[$Ar~{\sc v}$]$ $\lambda$7006 & $\cdots$ & 2.21($\pm$0.19)$\times10^{-8}$ & 8.03($\pm$2.36)$\times10^{-9}$\\
K$^{3+}$ & $[$K~{\sc iv}$]$ $\lambda$6102 & $\cdots$ & 1.33($\pm$0.08)$\times10^{-8}$ & 9.95($\pm$0.64)$\times10^{-9}$\\
 & $[$K~{\sc iv}$]$ $\lambda$6795 & $\cdots$ & 1.07($\pm$0.15)$\times10^{-8}$ & 1.02($\pm$0.15)$\times10^{-8}$\\
adopted\tablenotemark{\rm{\scriptsize d}} & & $\cdots$ & 1.24($\pm$0.07)$\times10^{-8}$ & 1.00($\pm$0.07)$\times10^{-8}$\\
Kr$^{3+}$ & $[$Kr~{\sc iv}$]$ $\lambda$5346 & $\cdots$ & $\cdots$ & 2.01($\pm$0.46)$\times10^{-9}$\\
 & $[$Kr~{\sc iv}$]$ $\lambda$5868 & $\cdots$ & $\cdots$ & 2.05($\pm$0.19)$\times10^{-9}$\\
adopted\tablenotemark{\rm{\scriptsize d}} & & $\cdots$ & $\cdots$ & 2.04($\pm$0.20)$\times10^{-9}$\\
\hline
\end{tabular}
\begin{description}
NOTE. -- This table presents the ionic abundances (and their uncertainties) from all CELs and helium lines. Symbol  ``$\cdots$" means that the line does not exist in the spectra. \\
\tablenotemark{\rm{\scriptsize a}} The He$^+$/H$^+$ ratio derived from the He~{\sc i} $\lambda$5876 line is adopted.\\ 
\tablenotemark{\rm{\scriptsize b}} For neutral species, the weighted average values of abundances listed in this table are adopted. \\ 
\tablenotemark{\rm{\scriptsize c}} For N$^+$, O$^{2+}$ and S$^{2+}$, which have very intense CELs, the results from the strongest transition ([N~{\sc ii}] $\lambda$6583, [O~{\sc iii}] $\lambda$4959 and [S~{\sc iii}] $\lambda$9531) are adopted. \\ 
\tablenotemark{\rm{\scriptsize d}} The weighted averages of all listed results are adopted for this ion. \\
\tablenotemark{\rm{\scriptsize e}} A weighted average value of the S$^+$/H$^+$ ratio derived from the [S~{\sc ii}] $\lambda$6716 and $\lambda$6731 lines is adopted. \\
\tablenotemark{\rm{\scriptsize f}} The Cl$^{+}$ abundance derived from the [Cl~{\sc ii}] $\lambda$8579 line is adopted. \\
\tablenotemark{\rm{\scriptsize g}} The weighted average value of the results derived from the simultaneous detections lines in three PNe is adopted. \\
\tablenotemark{\rm{\scriptsize h}} The Cl$^{3+}$ abundance derived from the [Cl~{\sc iv}] $\lambda$8046 line is adopted, as the [Cl~{\sc iv}] $\lambda$7531 is probably contaminated by O~{\sc ii} $\lambda$7531 $3d\ ^4D_{7/2}-4p\ ^4D^{\rm o}_{7/2}$.\\
\end{description}
\end{center}
\end{table*}

\begin{table*}
\begin{center}
\caption{C$^{2+}$ abundances}
\label{tab:cii_ionic}
\begin{tabular}{llccc}
\hline\hline
Line & Case & \multicolumn{3}{c}{Abundance (X$^{i+}$/H$^+$)} \\
\cline{3-5}
(\AA) & & Hf\,2-2 & M\,1-42 & NGC\,6153\\
\hline
$\lambda$4267 & B & 4.78($\pm$0.05)$\times10^{-3}$ & 2.45($\pm$0.02)$\times10^{-3}$ & 2.05($\pm$0.02)$\times10^{-3}$\\
$\lambda$7235 & A & 1.50($\pm$0.05)$\times10^{-1}$ & 7.19($\pm$0.07)$\times10^{-2}$ & 8.04($\pm$0.08)$\times10^{-2}$\\
$\lambda$7235 & B & 2.15($\pm$0.07)$\times10^{-3}$ & 1.03($\pm$0.01)$\times10^{-3}$ & 1.15($\pm$0.01)$\times10^{-3}$\\
$\lambda$5342 & A & 4.74($\pm$0.47)$\times10^{-3}$ & 2.38($\pm$0.09)$\times10^{-3}$ & 2.20($\pm$0.07)$\times10^{-3}$\\
$\lambda$6151 & B & 4.09($\pm$0.65)$\times10^{-3}$ & 2.67($\pm$0.29)$\times10^{-3}$ & 2.26($\pm$0.15)$\times10^{-3}$\\
adopted & & 4.78($\pm$0.05)$\times10^{-3}$ & 2.45($\pm$0.02)$\times10^{-3}$ & 2.05($\pm$0.02)$\times10^{-3}$\\
\hline
\end{tabular}
\begin{description}
NOTE. -- The abundances derived from C~{\sc ii} $\lambda4267$ line were adopted. For C~{\sc ii} $\lambda$7235, abundances derived with Case A atomic data are significantly higher than the results from other lines 
\end{description}
\end{center}
\end{table*}

\begin{table*}
\begin{center}
\caption{N$^{2+}$ ORL abundances}
\label{tab:nii_ionic}
\begin{tabular}{llccc}
\hline\hline
Line & Case & \multicolumn{3}{c}{Abundance (X$^{i+}$/H$^+$)} \\
\cline{3-5}
(\AA) & & Hf\,2-2 & M\,1-42 & NGC\,6153\\
\hline
N~{\sc ii} $\lambda$5666.63 & M3 & 1.95($\pm$0.18)$\times10^{-3}$ & 3.88($\pm$0.06)$\times10^{-3}$ & 1.60($\pm$0.04)$\times10^{-3}$\\
N~{\sc ii} $\lambda$5676.02 & M3 & 1.99($\pm$0.15)$\times10^{-3}$ & 3.72($\pm$0.08)$\times10^{-3}$ & 1.54($\pm$0.05)$\times10^{-3}$\\
N~{\sc ii} $\lambda$5679.56 & M3 & 2.02($\pm$0.05)$\times10^{-3}$ & 3.95($\pm$0.02)$\times10^{-3}$ & 1.67($\pm$0.02)$\times10^{-3}$\\
N~{\sc ii} $\lambda$5686.21 & M3 & 2.04($\pm$0.63)$\times10^{-3}$ & 4.11($\pm$0.16)$\times10^{-3}$ & 1.70($\pm$0.11)$\times10^{-3}$\\
N~{\sc ii} $\lambda$5710.77 & M3 & 1.85($\pm$0.16)$\times10^{-3}$ & 3.27($\pm$0.10)$\times10^{-3}$ & 1.46($\pm$0.05)$\times10^{-3}$\\
N~{\sc ii} $\lambda$5730.66 & M3 & $\cdots$ & 3.12($\pm$1.30)$\times10^{-3}$ & $\cdots$\\
N~{\sc ii} Mult. 3 &  & 1.98($\pm$0.06)$\times10^{-3}$ & 3.84($\pm$0.03)$\times10^{-3}$ & 1.60($\pm$0.02)$\times10^{-3}$\\
N~{\sc ii} $\lambda$5010.62\tablenotemark{\rm{\scriptsize a}} & M4 & $\cdots$ & $\cdots$ & 3.15($\pm$0.14)$\times10^{-3}$\\
N~{\sc ii} $\lambda$5045.10 & M4 & 1.32($\pm$0.17)$\times10^{-3}$ & 2.90($\pm$0.13)$\times10^{-3}$ & 2.06($\pm$0.09)$\times10^{-4}$\tablenotemark{\rm{\scriptsize b}}\\
N~{\sc ii} $\lambda$4601.48 & M5 & $\cdots$ & 3.31($\pm$0.10)$\times10^{-3}$ & 1.17($\pm$0.10)$\times10^{-3}$\\
N~{\sc ii} $\lambda$4607.16 & M5 & 1.70($\pm$0.77)$\times10^{-3}$ & 2.59($\pm$0.17)$\times10^{-3}$ & 1.08($\pm$0.11)$\times10^{-3}$\\
N~{\sc ii} $\lambda$4613.87\tablenotemark{\rm{\scriptsize a}} & M5 & 3.28($\pm$0.47)$\times10^{-3}$ & $\cdots$ & 1.74($\pm$0.17)$\times10^{-3}$\\
N~{\sc ii} $\lambda$4621.39 & M5 & 1.88($\pm$0.37)$\times10^{-3}$ & 2.90($\pm$0.15)$\times10^{-3}$ & 1.28($\pm$0.16)$\times10^{-3}$\\
N~{\sc ii} $\lambda$4630.54 & M5 & 1.58($\pm$0.11)$\times10^{-3}$ & 2.76($\pm$0.06)$\times10^{-3}$ & 1.23($\pm$0.05)$\times10^{-3}$\\
N~{\sc ii} $\lambda$4643.09 & M5 & 1.47($\pm$0.23)$\times10^{-3}$ & 2.81($\pm$0.16)$\times10^{-3}$ & 1.28($\pm$0.15)$\times10^{-3}$\\
N~{\sc ii} Mult. 5 &  & 1.60($\pm$0.12)$\times10^{-3}$ & 2.89($\pm$0.05)$\times10^{-3}$ & 1.20($\pm$0.05)$\times10^{-3}$\\
N~{\sc ii} $\lambda$6482.05 & M8 & 2.30($\pm$0.29)$\times10^{-3}$ & 4.16($\pm$0.27)$\times10^{-3}$ & 1.86($\pm$0.18)$\times10^{-3}$\\
N~{\sc ii} $\lambda$3995.00 & M12 & 2.22($\pm$0.49)$\times10^{-3}$ & 3.86($\pm$0.21)$\times10^{-3}$ & 1.52($\pm$0.24)$\times10^{-3}$\\
N~{\sc ii} $\lambda$5001.48 & M19 & 2.02($\pm$0.06)$\times10^{-3}$ & 4.11($\pm$0.30)$\times10^{-3}$ & 1.42($\pm$0.02)$\times10^{-3}$\\
N~{\sc ii} $\lambda$5005.15\tablenotemark{\rm{\scriptsize a}} & M19 & 2.46($\pm$0.07)$\times10^{-3}$ & $\cdots$ & 2.72($\pm$0.05)$\times10^{-3}$\\
N~{\sc ii} $\lambda$5025.66 & M19 & 2.14($\pm$0.53)$\times10^{-3}$ & 3.13($\pm$0.33)$\times10^{-3}$ & 1.49($\pm$0.25)$\times10^{-3}$\\
N~{\sc ii} Mult. 19 &  & 2.03($\pm$0.07)$\times10^{-3}$ & 3.64($\pm$0.22)$\times10^{-3}$ & 1.43($\pm$0.03)$\times10^{-3}$\\
N~{\sc ii} $\lambda$4035.08\tablenotemark{\rm{\scriptsize a}} & 3d$-$4f & 2.33($\pm$0.26)$\times10^{-3}$ & 3.79($\pm$0.17)$\times10^{-3}$ & 1.77($\pm$0.14)$\times10^{-3}$\\
N~{\sc ii} $\lambda$4041.31\tablenotemark{\rm{\scriptsize a}} & 3d$-$4f & 2.12($\pm$0.18)$\times10^{-3}$ & 3.64($\pm$0.08)$\times10^{-3}$ & 1.58($\pm$0.09)$\times10^{-3}$\\
N~{\sc ii} $\lambda$4043.53 & 3d$-$4f & 1.89($\pm$0.33)$\times10^{-3}$ & 3.59($\pm$0.33)$\times10^{-3}$ & 1.43($\pm$0.20)$\times10^{-3}$\\
N~{\sc ii} $\lambda$4176.16 & 3d$-$4f & 1.89($\pm$0.37)$\times10^{-3}$ & 3.74($\pm$0.24)$\times10^{-3}$ & 1.44($\pm$0.20)$\times10^{-3}$\\
N~{\sc ii} $\lambda$4236.91 & 3d$-$4f & 2.12($\pm$0.17)$\times10^{-3}$ & 4.12($\pm$0.08)$\times10^{-3}$ & 1.58($\pm$0.06)$\times10^{-3}$\\
N~{\sc ii} $\lambda$4241.78\tablenotemark{\rm{\scriptsize a}} & 3d$-$4f & 2.39($\pm$0.14)$\times10^{-3}$ & 4.60($\pm$0.16)$\times10^{-3}$ & 1.70($\pm$0.04)$\times10^{-3}$\\
N~{\sc ii} $\lambda$4432.74 & 3d$-$4f & 1.79($\pm$0.23)$\times10^{-3}$ & 3.93($\pm$0.11)$\times10^{-3}$ & 1.38($\pm$0.07)$\times10^{-3}$\\
N~{\sc ii} $\lambda$4530.41 & 3d$-$4f & 1.76($\pm$0.17)$\times10^{-3}$ & 3.65($\pm$0.11)$\times10^{-3}$ & 1.56($\pm$0.07)$\times10^{-3}$\\
N~{\sc ii} $\lambda$4552.53\tablenotemark{\rm{\scriptsize a}} & 3d$-$4f & 2.34($\pm$0.33)$\times10^{-3}$ & 4.40($\pm$0.23)$\times10^{-3}$ & 2.10($\pm$0.19)$\times10^{-3}$\\
N~{\sc ii} $\lambda$3d – 4f &  & 1.95($\pm$0.10)$\times10^{-3}$ & 3.82($\pm$0.05)$\times10^{-3}$ & 1.51($\pm$0.04)$\times10^{-3}$\\
N~{\sc ii} adopted &  & 1.91($\pm$0.04)$\times10^{-3}$ & 3.56($\pm$0.03)$\times10^{-3}$ & 1.49($\pm$0.02)$\times10^{-3}$\\
\hline
\end{tabular}
\begin{description}
NOTE. -- The adopted N$^{2+}$ abundances are the weighted averages of the results from lines free of blending and bad-pixel contamination.\\
\tablenotemark{\rm{\scriptsize a}} Blended with other nebular emission line(s) that cannot be neglected. (If the blend consist of only N~{\sc ii} lines with effective recombination coefficients, the calculation was carried out using the coefficients of both lines without additional labeling.) \\
\tablenotemark{\rm{\scriptsize b}} Contaminated by bad pixels. \\
\end{description}
\end{center}
\end{table*}

\begin{table*}
\begin{center}
\caption{O$^{2+}$ ORL abundances}
\label{tab:oii_ionic}
\begin{tabular}{llccc}
\hline\hline
Line & Mult. & \multicolumn{3}{c}{Abundance (X$^{i+}$/H$^+$)} \\
\cline{3-5}
(\AA) & & Hf\,2-2 & M\,1-42 & NGC\,6153\\
\hline
O~{\sc ii} $\lambda$4638.85 & M1 & 4.49($\pm$0.15)$\times10^{-3}$ & 5.99($\pm$0.12)$\times10^{-3}$ & 4.65($\pm$0.11)$\times10^{-3}$\\
O~{\sc ii} $\lambda$4641.81\tablenotemark{\rm{\scriptsize a}} & M1 & 4.41($\pm$0.10)$\times10^{-3}$ & 7.22($\pm$0.04)$\times10^{-3}$ & 5.69($\pm$0.05)$\times10^{-3}$\\
O~{\sc ii} $\lambda$4649.13 & M1 & 4.38($\pm$0.11)$\times10^{-3}$ & 5.65($\pm$0.03)$\times10^{-3}$ & 4.26($\pm$0.02)$\times10^{-3}$\\
O~{\sc ii} $\lambda$4650.84 & M1 & 4.47($\pm$0.22)$\times10^{-3}$ & 6.16($\pm$0.06)$\times10^{-3}$\tablenotemark{\rm{\scriptsize a}} & 4.74($\pm$0.05)$\times10^{-3}$\\
O~{\sc ii} $\lambda$4661.63 & M1 & 4.49($\pm$0.10)$\times10^{-3}$ & 5.72($\pm$0.17)$\times10^{-3}$ & 4.26($\pm$0.05)$\times10^{-3}$\\
O~{\sc ii} $\lambda$4673.73 & M1 & 4.32($\pm$0.76)$\times10^{-3}$ & 5.04($\pm$0.30)$\times10^{-3}$ & 3.81($\pm$0.29)$\times10^{-3}$\\
O~{\sc ii} $\lambda$4676.23 & M1 & 4.35($\pm$0.33)$\times10^{-3}$ & 5.58($\pm$0.08)$\times10^{-3}$ & 4.11($\pm$0.08)$\times10^{-3}$\\
O~{\sc ii} $\lambda$4696.35\tablenotemark{\rm{\scriptsize a}} & M1 & 6.86($\pm$1.40)$\times10^{-3}$ & 5.06($\pm$1.02)$\times10^{-3}$ & 5.30($\pm$0.62)$\times10^{-3}$\\
O~{\sc ii} Mult. 1  &  & 4.43($\pm$0.06)$\times10^{-3}$ & 5.65($\pm$0.04)$\times10^{-3}$ & 4.34($\pm$0.03)$\times10^{-3}$\\
O~{\sc ii} $\lambda$4317.14\tablenotemark{\rm{\scriptsize a}} & M2 & 6.53($\pm$0.51)$\times10^{-3}$ & 5.66($\pm$0.17)$\times10^{-3}$ & 3.94($\pm$0.16)$\times10^{-3}$\\
O~{\sc ii} $\lambda$4319.63\tablenotemark{\rm{\scriptsize a}} & M2 & 5.22($\pm$0.56)$\times10^{-3}$ & 5.99($\pm$0.27)$\times10^{-3}$ & 4.35($\pm$0.20)$\times10^{-3}$\\
O~{\sc ii} $\lambda$4325.76 & M2 & 2.77($\pm$0.99)$\times10^{-3}$\tablenotemark{\rm{\scriptsize b}} & 2.80($\pm$1.20)$\times10^{-3}$\tablenotemark{\rm{\scriptsize b}} & 4.16($\pm$0.57)$\times10^{-3}$\tablenotemark{\rm{\scriptsize b}}\\
O~{\sc ii} $\lambda$4336.86 & M2 & 3.82($\pm$0.66)$\times10^{-3}$ & 4.79($\pm$0.40)$\times10^{-3}$ & 4.10($\pm$0.55)$\times10^{-3}$\\
O~{\sc ii} $\lambda$4345.56 & M2 & 4.28($\pm$0.28)$\times10^{-3}$ & 4.77($\pm$0.14)$\times10^{-3}$ & 4.28($\pm$0.19)$\times10^{-3}$\\
O~{\sc ii} $\lambda$4349.43\tablenotemark{\rm{\scriptsize a}} & M2 & 4.89($\pm$0.21)$\times10^{-3}$ & 5.80($\pm$0.16)$\times10^{-3}$ & 4.42($\pm$0.16)$\times10^{-3}$\\
O~{\sc ii} $\lambda$4366.89\tablenotemark{\rm{\scriptsize a}} & M2 & 5.76($\pm$0.96)$\times10^{-3}$ & 6.05($\pm$0.28)$\times10^{-3}$ & 4.33($\pm$0.17)$\times10^{-3}$\\
O~{\sc ii} Mult. 2  &  & 4.14($\pm$0.28)$\times10^{-3}$ & 4.78($\pm$0.15)$\times10^{-3}$ & 4.10($\pm$0.18)$\times10^{-3}$\\
O~{\sc ii} $\lambda$3727.32\tablenotemark{\rm{\scriptsize a}} & M3 & 9.53($\pm$0.65)$\times10^{-3}$ & 1.46($\pm$0.48)$\times10^{-2}$ & 7.40($\pm$0.37)$\times10^{-3}$\\
O~{\sc ii} $\lambda$3749.48\tablenotemark{\rm{\scriptsize a}} & M3 & 5.55($\pm$0.28)$\times10^{-3}$ & 2.56($\pm$0.16)$\times10^{-3}$ & 2.54($\pm$0.27)$\times10^{-3}$\\
O~{\sc ii} $\lambda$4414.90\tablenotemark{\rm{\scriptsize a}} & M5 & 3.44($\pm$0.68)$\times10^{-3}$\tablenotemark{\rm{\scriptsize c}} & 6.22($\pm$0.37)$\times10^{-3}$ & 4.44($\pm$0.18)$\times10^{-3}$\\
O~{\sc ii} $\lambda$4416.97\tablenotemark{\rm{\scriptsize a}} & M5 & 4.73($\pm$0.77)$\times10^{-3}$ & 7.74($\pm$0.57)$\times10^{-3}$ & 5.14($\pm$0.32)$\times10^{-3}$\\
O~{\sc ii} $\lambda$4452.38\tablenotemark{\rm{\scriptsize a}} & M5 & $\cdots$ & 1.24($\pm$0.20)$\times10^{-2}$ & 4.57($\pm$0.95)$\times10^{-3}$\\
O~{\sc ii} $\lambda$3973.26 & M6 & 4.04($\pm$0.86)$\times10^{-3}$ & 4.29($\pm$0.75)$\times10^{-3}$ & 5.04($\pm$0.65)$\times10^{-3}$\\
O~{\sc ii} $\lambda$4069.89 & M10 & 4.35($\pm$0.19)$\times10^{-3}$ & 5.91($\pm$0.09)$\times10^{-3}$ & 4.66($\pm$0.11)$\times10^{-3}$\\
O~{\sc ii} $\lambda$4072.16 & M10 & 4.10($\pm$0.20)$\times10^{-3}$ & 5.58($\pm$0.08)$\times10^{-3}$ & 4.32($\pm$0.10)$\times10^{-3}$\\
O~{\sc ii} $\lambda$4075.86\tablenotemark{\rm{\scriptsize a}} & M10 & 4.83($\pm$0.20)$\times10^{-3}$ & $\cdots$ & 4.30($\pm$0.08)$\times10^{-3}$\\
O~{\sc ii} $\lambda$4078.84 & M10 & 4.40($\pm$0.51)$\times10^{-3}$ & 4.29($\pm$0.34)$\times10^{-3}$ & 4.05($\pm$0.44)$\times10^{-3}$\\
O~{\sc ii} $\lambda$4085.11 & M10 & 4.87($\pm$0.40)$\times10^{-3}$ & 4.90($\pm$0.27)$\times10^{-3}$ & 4.64($\pm$0.41)$\times10^{-3}$\\
O~{\sc ii} $\lambda$4092.93 & M10 & 4.71($\pm$0.75)$\times10^{-3}$ & 5.00($\pm$0.45)$\times10^{-3}$ & 3.65($\pm$0.27)$\times10^{-3}$\\
O~{\sc ii} Mult. 10  &  & 4.39($\pm$0.14)$\times10^{-3}$ & 5.46($\pm$0.07)$\times10^{-3}$ & 4.40($\pm$0.24)$\times10^{-3}$\\
O~{\sc ii} $\lambda$3851.03 & M12 & 3.46($\pm$1.83)$\times10^{-3}$\tablenotemark{\rm{\scriptsize c}} & $\cdots$ & $\cdots$\\
O~{\sc ii} $\lambda$3856.13\tablenotemark{\rm{\scriptsize a}} & M12 & 1.15($\pm$0.28)$\times10^{-2}$ & 2.58($\pm$0.21)$\times10^{-2}$ & 1.80($\pm$0.14)$\times10^{-2}$\\
O~{\sc ii} $\lambda$3864.67\tablenotemark{\rm{\scriptsize a}} & M12 & $\cdots$ & 6.16($\pm$0.87)$\times10^{-3}$ & 5.96($\pm$1.12)$\times10^{-3}$\\
O~{\sc ii} $\lambda$3882.19 & M12 & 4.48($\pm$0.61)$\times10^{-3}$ & 5.00($\pm$0.55)$\times10^{-3}$ & 4.34($\pm$0.37)$\times10^{-3}$\\
O~{\sc ii} $\lambda$4121.46 & M19 & 4.35($\pm$0.19)$\times10^{-3}$\tablenotemark{\rm{\scriptsize c}} & $\cdots$ & 2.42($\pm$0.22)$\times10^{-3}$\tablenotemark{\rm{\scriptsize a}}\\
O~{\sc ii} $\lambda$4132.80 & M19 & 4.10($\pm$0.20)$\times10^{-3}$\tablenotemark{\rm{\scriptsize c}} & 3.92($\pm$0.23)$\times10^{-3}$ & 2.77($\pm$0.12)$\times10^{-3}$\\
O~{\sc ii} $\lambda$4153.30 & M19 & 4.83($\pm$0.20)$\times10^{-3}$ & 4.78($\pm$0.12)$\times10^{-3}$ & 3.44($\pm$0.12)$\times10^{-3}$\\
O~{\sc ii} $\lambda$4156.53\tablenotemark{\rm{\scriptsize a}} & M19 & 4.40($\pm$0.51)$\times10^{-3}$ & 1.62($\pm$0.12)$\times10^{-2}$ & 1.04($\pm$0.08)$\times10^{-2}$\\
O~{\sc ii} $\lambda$4169.22\tablenotemark{\rm{\scriptsize a}} & M19 & 4.87($\pm$0.40)$\times10^{-3}$ & 7.99($\pm$0.61)$\times10^{-3}$ & 5.62($\pm$0.62)$\times10^{-3}$\\
O~{\sc ii} Mult. 19  &  & 4.39($\pm$0.14)$\times10^{-3}$ & 4.78($\pm$0.12)$\times10^{-3}$ & 3.11($\pm$0.09)$\times10^{-3}$\\
O~{\sc ii} $\lambda$4097.26\tablenotemark{\rm{\scriptsize a}} & M20 & 6.87($\pm$0.25)$\times10^{-3}$ & $\cdots$ & $\cdots$\\
O~{\sc ii} $\lambda$4103.00\tablenotemark{\rm{\scriptsize a}} & M20 & 2.37($\pm$0.31)$\times10^{-2}$ & $\cdots$ & $\cdots$\\
O~{\sc ii} $\lambda$4104.99 & M20 & 3.07($\pm$0.42)$\times10^{-3}$ & 3.40($\pm$0.25)$\times10^{-3}$\tablenotemark{\rm{\scriptsize c}} & 2.89($\pm$0.36)$\times10^{-3}$\\
O~{\sc ii} $\lambda$4110.79 & M20 & 3.58($\pm$0.66)$\times10^{-3}$ & 3.70($\pm$0.45)$\times10^{-3}$\tablenotemark{\rm{\scriptsize c}} & 2.34($\pm$0.54)$\times10^{-3}$\\
O~{\sc ii} $\lambda$4119.22 & M20 & 4.01($\pm$0.34)$\times10^{-3}$ & 5.67($\pm$0.12)$\times10^{-3}$ & 4.20($\pm$0.10)$\times10^{-3}$\\
O~{\sc ii} $\lambda$4120.55\tablenotemark{\rm{\scriptsize a}} & M20 & 6.43($\pm$0.82)$\times10^{-3}$ & $\cdots$ & 2.82($\pm$0.17)$\times10^{-3}$\\
O~{\sc ii} Mult. 20  &  & 3.57($\pm$0.26)$\times10^{-3}$ & 5.67($\pm$0.12)$\times10^{-3}$ & 3.72($\pm$0.18)$\times10^{-3}$\\
\hline
\end{tabular}
\addtocounter{table}{-1}
\caption{(Continued)}
\end{center}
\end{table*}

\begin{table*}
\begin{center}
\addtocounter{table}{-1}
\caption{(Continued)}
\begin{tabular}{llccc}
\hline\hline
Line & Mult. & \multicolumn{3}{c}{Abundance (X$^{i+}$/H$^+$)} \\
\cline{3-5}
(\AA) & & Hf\,2-2 & M\,1-42 & NGC\,6153\\
\hline
O~{\sc ii} $\lambda$4083.89 & 3d$-$4f & 4.06($\pm$0.39)$\times10^{-3}$ & 5.01($\pm$0.31)$\times10^{-3}$ & 4.53($\pm$0.46)$\times10^{-3}$\\
O~{\sc ii} $\lambda$4087.15\tablenotemark{\rm{\scriptsize a}} & 3d$-$4f & 4.14($\pm$0.36)$\times10^{-3}$ & 5.87($\pm$0.18)$\times10^{-3}$ & 3.97($\pm$0.39)$\times10^{-3}$\\
O~{\sc ii} $\lambda$4089.29\tablenotemark{\rm{\scriptsize a}} & 3d$-$4f & 4.37($\pm$0.20)$\times10^{-3}$ & 5.71($\pm$0.27)$\times10^{-3}$ & 4.27($\pm$0.14)$\times10^{-3}$\\
O~{\sc ii} $\lambda$4095.64 & 3d$-$4f & 4.87($\pm$0.74)$\times10^{-3}$ & 5.72($\pm$0.58)$\times10^{-3}$ & 3.77($\pm$0.40)$\times10^{-3}$\\
O~{\sc ii} $\lambda$4098.24 & 3d$-$4f & 3.83($\pm$0.42)$\times10^{-3}$ & 5.97($\pm$0.58)$\times10^{-3}$\tablenotemark{\rm{\scriptsize c}} & 3.62($\pm$0.38)$\times10^{-3}$\\
O~{\sc ii} $\lambda$4107.09 & 3d$-$4f & 3.78($\pm$0.53)$\times10^{-3}$ & 5.31($\pm$0.51)$\times10^{-3}$ & 3.89($\pm$0.82)$\times10^{-3}$\\
O~{\sc ii} $\lambda$4275.55 & 3d$-$4f & 5.22($\pm$0.36)$\times10^{-3}$ & 5.91($\pm$0.11)$\times10^{-3}$ & 4.03($\pm$0.10)$\times10^{-3}$\\
O~{\sc ii} $\lambda$4276.74\tablenotemark{\rm{\scriptsize a}} & 3d$-$4f & 6.12($\pm$0.97)$\times10^{-3}$ & 1.50($\pm$0.04)$\times10^{-2}$ & 6.72($\pm$0.22)$\times10^{-3}$\\
O~{\sc ii} $\lambda$4285.69\tablenotemark{\rm{\scriptsize a}} & 3d$-$4f & 4.42($\pm$0.48)$\times10^{-3}$ & 5.59($\pm$0.28)$\times10^{-3}$ & 3.74($\pm$0.28)$\times10^{-3}$\\
O~{\sc ii} $\lambda$4291.25 & 3d$-$4f & 4.07($\pm$1.01)$\times10^{-3}$ & 4.91($\pm$0.38)$\times10^{-3}$ & 3.87($\pm$0.28)$\times10^{-3}$\\
O~{\sc ii} $\lambda$4294.78 & 3d$-$4f & 4.84($\pm$0.52)$\times10^{-3}$ & 5.35($\pm$0.23)$\times10^{-3}$ & 3.80($\pm$0.23)$\times10^{-3}$\\
O~{\sc ii} $\lambda$4303.82\tablenotemark{\rm{\scriptsize a}} & 3d$-$4f & 5.20($\pm$0.30)$\times10^{-3}$ & 6.72($\pm$0.19)$\times10^{-3}$ & 4.71($\pm$0.16)$\times10^{-3}$\\
O~{\sc ii} $\lambda$4342.00 & 3d$-$4f & 4.28($\pm$0.21)$\times10^{-3}$ & 9.98($\pm$0.14)$\times10^{-3}$\tablenotemark{\rm{\scriptsize c}} & 6.17($\pm$0.17)$\times10^{-3}$\\
O~{\sc ii} $\lambda$4491.23\tablenotemark{\rm{\scriptsize a}} & 3d$-$4f & 5.08($\pm$0.31)$\times10^{-3}$ & 6.98($\pm$0.24)$\times10^{-3}$ & 5.29($\pm$0.32)$\times10^{-3}$\\
O~{\sc ii} $\lambda$4602.13\tablenotemark{\rm{\scriptsize a}} & 3d$-$4f & 6.18($\pm$0.88)$\times10^{-3}$ & 5.37($\pm$0.24)$\times10^{-3}$ & 4.39($\pm$0.31)$\times10^{-3}$\\
O~{\sc ii} $\lambda$4609.44 & 3d$-$4f & 3.61($\pm$0.17)$\times10^{-3}$ & 5.77($\pm$0.19)$\times10^{-3}$ & 4.00($\pm$0.08)$\times10^{-3}$\\
O~{\sc ii} $\lambda$3d$-$4f  &  & 4.15($\pm$0.12)$\times10^{-3}$ & 5.38($\pm$0.13)$\times10^{-3}$ & 3.94($\pm$0.26)$\times10^{-3}$\\
O~{\sc ii} adopted & & 4.23($\pm$0.05)$\times10^{-3}$ & 5.37($\pm$0.04)$\times10^{-3}$ & 4.15($\pm$0.03)$\times10^{-3}$\\
\hline
\end{tabular}
\begin{description}
NOTE. -- The adopted O$^{2+}$ abundances are the weighted averages of the results from lines free of blending and bad-pixel contamination.\\
\tablenotemark{\rm{\scriptsize a}} Blended with other nebular emission line(s) that cannot be neglected. (If the blend consist of only O~{\sc ii} lines with effective recombination coefficients, the calculation was carried out using the coefficients of both lines without additional labeling.)\\
\tablenotemark{\rm{\scriptsize b}} Contaminated by bad pixels. \\
\tablenotemark{\rm{\scriptsize c}} Blended with telluric line(s). \\
\end{description}
\end{center}
\end{table*}

\begin{table*}
\begin{center}
\caption{Ne$^{2+}$ ORL abundances}
\label{tab:neii_ionic}
\begin{tabular}{llccc}
\hline\hline
Line & Mult. & \multicolumn{3}{c}{Abundance (X$^{i+}$/H$^+$)} \\
\cline{3-5}
(\AA) & & Hf\,2-2 & M\,1-42 & NGC\,6153\\
\hline
Ne~{\sc ii} $\lambda$3694.21 & M1 & 1.44($\pm$0.23)$\times10^{-3}$ & 1.79($\pm$0.08)$\times10^{-3}$ & 1.31($\pm$0.06)$\times10^{-3}$\\
Ne~{\sc ii} $\lambda$3709.62 & M1 & 1.30($\pm$0.47)$\times10^{-3}$ & 9.29($\pm$3.58)$\times10^{-4}$ & 1.02($\pm$0.19)$\times10^{-3}$\\
Ne~{\sc ii} $\lambda$3766.26 & M1 & 5.66($\pm$0.15)$\times10^{-3}$\tablenotemark{\rm{\scriptsize a}} & 1.30($\pm$0.13)$\times10^{-3}$ & 8.54($\pm$0.73)$\times10^{-4}$\\
Ne~{\sc ii} $\lambda$3777.14\tablenotemark{\rm{\scriptsize b}} & M1 & 1.97($\pm$0.47)$\times10^{-3}$ & 1.20($\pm$0.13)$\times10^{-3}$ & 1.08($\pm$0.13)$\times10^{-3}$\\
Ne~{\sc ii} Mult. 1 &  & 1.39($\pm$0.22)$\times10^{-3}$ & 1.20($\pm$0.10)$\times10^{-3}$ & 9.52($\pm$0.65)$\times10^{-4}$\\
Ne~{\sc ii} $\lambda$3327.15\tablenotemark{\rm{\scriptsize b}} & M2 & $\cdots$ & $\cdots$ & 3.28($\pm$2.01)$\times10^{-4}$\\
Ne~{\sc ii} $\lambda$3334.84 & M2 & 8.86($\pm$1.60)$\times10^{-4}$ & 1.51($\pm$0.06)$\times10^{-3}$ & 1.07($\pm$0.07)$\times10^{-3}$\\
Ne~{\sc ii} $\lambda$3344.40 & M2 & $\cdots$ & 9.74($\pm$4.27)$\times10^{-4}$ & 6.38($\pm$2.61)$\times10^{-4}$\\
Ne~{\sc ii} $\lambda$3355.02 & M2 & 1.13($\pm$0.22)$\times10^{-3}$ & 1.79($\pm$0.19)$\times10^{-3}$ & 1.17($\pm$0.09)$\times10^{-3}$\\
Ne~{\sc ii} $\lambda$3360.60\tablenotemark{\rm{\scriptsize b}} & M2 & $\cdots$ & 5.70($\pm$1.48)$\times10^{-4}$\tablenotemark{\rm{\scriptsize a}} & 4.75($\pm$2.00)$\times10^{-4}$\tablenotemark{\rm{\scriptsize a}}\\
Ne~{\sc ii} Mult. 2 &  & 9.89($\pm$1.31)$\times10^{-4}$ & 1.44($\pm$0.07)$\times10^{-3}$ & 9.79($\pm$0.78)$\times10^{-4}$\\
Ne~{\sc ii} $\lambda$3713.08\tablenotemark{\rm{\scriptsize b}} & M5 & 1.50($\pm$0.18)$\times10^{-3}$ & 1.46($\pm$0.09)$\times10^{-3}$ & 1.03($\pm$0.04)$\times10^{-3}$\\
Ne~{\sc ii} $\lambda$3568.50\tablenotemark{\rm{\scriptsize c}} & M9 & 1.72($\pm$0.95)$\times10^{-1}$ & 8.54($\pm$1.13)$\times10^{-2}$ & 2.45($\pm$0.26)$\times10^{-2}$\\
Ne~{\sc ii} $\lambda$3218.19\tablenotemark{\rm{\scriptsize b}} & M13 & $\cdots$ & $\cdots$ & 1.30($\pm$0.15)$\times10^{-3}$\\
Ne~{\sc ii} $\lambda$3244.10 & M13 & 6.03($\pm$3.36)$\times10^{-4}$\tablenotemark{\rm{\scriptsize a}} & 9.52($\pm$3.09)$\times10^{-4}$ & 6.73($\pm$1.54)$\times10^{-4}$\\
Ne~{\sc ii} $\lambda$3388.42 & M20 & $\cdots$ & 5.11($\pm$1.78)$\times10^{-4}$ & 5.40($\pm$3.14)$\times10^{-4}$\\
Ne~{\sc ii} $\lambda$3417.69 & M20 & 1.30($\pm$0.48)$\times10^{-3}$ & 1.05($\pm$0.18)$\times10^{-3}$ & 8.62($\pm$1.37)$\times10^{-4}$\\
Ne~{\sc ii} $\lambda$4391.99 & 3d$-$4f & 2.61($\pm$0.37)$\times10^{-3}$ & 2.89($\pm$0.12)$\times10^{-3}$ & 2.20($\pm$0.09)$\times10^{-3}$\\
Ne~{\sc ii} $\lambda$4409.30 & 3d$-$4f & 1.96($\pm$0.97)$\times10^{-3}$ & 3.31($\pm$0.28)$\times10^{-3}$ & 2.42($\pm$0.18)$\times10^{-3}$\\
Ne~{\sc ii} $\lambda$4428.61 & 3d$-$4f & 3.03($\pm$1.31)$\times10^{-3}$ & 3.85($\pm$0.26)$\times10^{-3}$ & 2.58($\pm$0.17)$\times10^{-3}$\\
Ne~{\sc ii} $\lambda$4219.75 & 3d$-$4f & 3.40($\pm$0.70)$\times10^{-3}$ & 3.34($\pm$0.34)$\times10^{-3}$ & 2.84($\pm$0.21)$\times10^{-3}$\\
Ne~{\sc ii} $\lambda$4397.99 & 3d$-$4f & $\cdots$ & 2.71($\pm$0.27)$\times10^{-3}$ & 1.90($\pm$0.28)$\times10^{-3}$\\
Ne~{\sc ii} $\lambda$4430.94 & 3d$-$4f & 3.65($\pm$1.96)$\times10^{-3}$ & 3.28($\pm$0.33)$\times10^{-3}$ & 2.80($\pm$0.26)$\times10^{-3}$\\
Ne~{\sc ii} $\lambda$4413.22\tablenotemark{\rm{\scriptsize b}} & 3d$-$4f & $\cdots$ & 5.37($\pm$0.74)$\times10^{-3}$ & 2.88($\pm$0.40)$\times10^{-3}$\\
Ne~{\sc ii} $\lambda$4233.85 & 3d$-$4f & 6.33($\pm$2.41)$\times10^{-3}$ & 3.84($\pm$0.90)$\times10^{-3}$ & 3.69($\pm$0.46)$\times10^{-3}$\\
Ne~{\sc ii} $\lambda$4457.05\tablenotemark{\rm{\scriptsize b}} & 3d$-$4f & $\cdots$ & 5.61($\pm$0.95)$\times10^{-3}$ & 4.44($\pm$0.86)$\times10^{-3}$\\
Ne~{\sc ii} $\lambda$4231.64\tablenotemark{\rm{\scriptsize b}} & 3d$-$4f & 5.74($\pm$2.06)$\times10^{-3}$ & 7.22($\pm$0.79)$\times10^{-3}$ & 5.00($\pm$0.56)$\times10^{-3}$\\
Ne~{\sc ii} $\lambda$4250.64\tablenotemark{\rm{\scriptsize b}} & 3d$-$4f & $\cdots$ & 5.34($\pm$1.54)$\times10^{-3}$ & 4.15($\pm$0.84)$\times10^{-3}$\\
Ne~{\sc ii} 3d$-$4f &  & 2.81($\pm$0.35)$\times10^{-3}$ & 3.19($\pm$0.10)$\times10^{-3}$ & 2.50($\pm$0.07)$\times10^{-3}$\\
Ne~{\sc ii} adopted &  & 1.16($\pm$0.12)$\times10^{-3}$ & 1.03($\pm$0.08)$\times10^{-3}$ & 8.99($\pm$0.46)$\times10^{-4}$\\
\hline
\end{tabular}
\begin{description}
NOTE. -- The adopted Ne$^{2+}$ abundances are the weighted averages of the results from lines free of blending, and also excluding the abundances derived from 3d$-$4f transitions and M9 multiplet, which are systematically higher. The available atomic data for 3d$-$4f transitions are from private communication, so we rather prefer not using them. For the M9 transition, the inner-shell electron is in an excited state, and numbers of Ne$^{2+}$ in such state may be affected by the nebular environment.\\
\tablenotemark{\rm{\scriptsize a}} Low SNR line. \\
\tablenotemark{\rm{\scriptsize b}} Blended with other nebular emission line(s) that cannot be neglected. (If the blend consist of only Ne~{\sc ii} lines with effective recombination coefficients, the calculation was carried out using the coefficients of both lines without additional labeling.)\\
\tablenotemark{\rm{\scriptsize c}} Inner-shell electron is in an excited state.\\
\end{description}
\end{center}
\end{table*}

\begin{figure*}[ht!]
\begin{center}
\includegraphics[width=18 cm,angle=0]{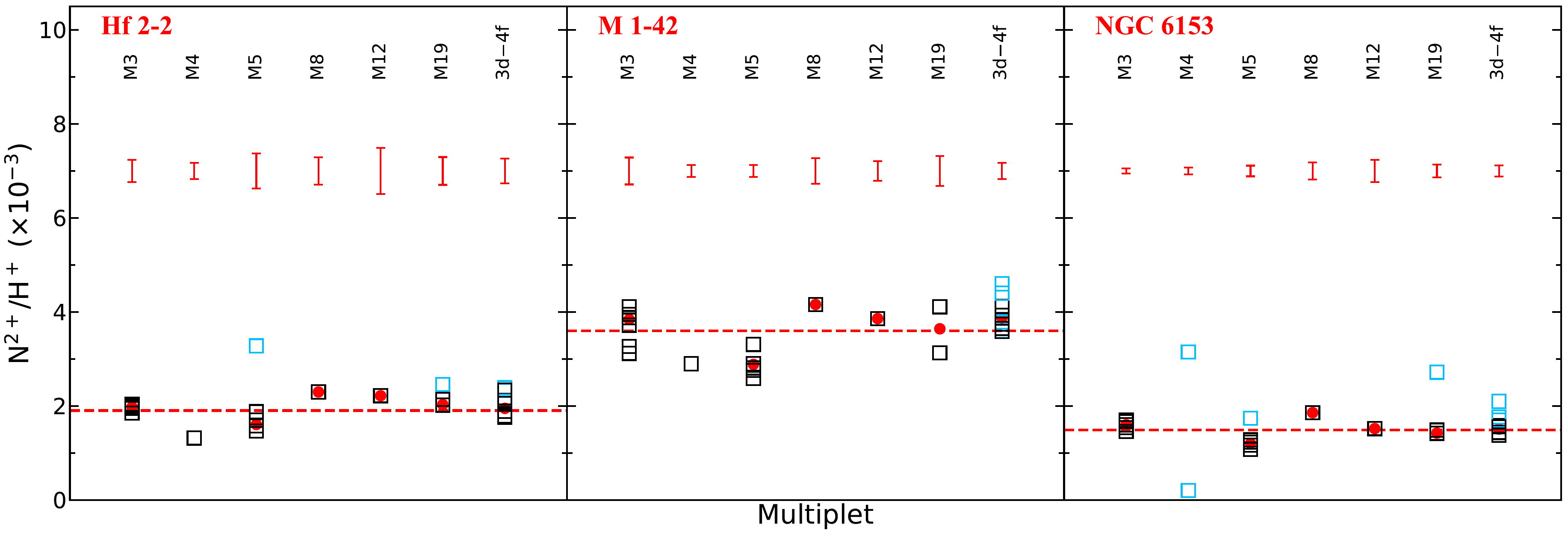}
\caption{Comparison of the results of N~{\sc ii} abundances of Hf\,2-2 (left), M\,1-42 (middle) and NGC\,6153 (right) derived from lines from different multiplets. The black open squares are the results from unblended lines, while the skyblue open squares represent the data from lines that are either blended with other lines or have a low signal-to-noise ratio. The red circles are the adopted abundances of each multiplet and the red dashed lines correspond to the final adopted N$^{2+}$ abundances. The error bars above are the typical errors of each multiplet.} 
\label{fig:NII_abund} 
\end{center}
\end{figure*}

\begin{figure*}[ht!]
\begin{center}
\includegraphics[width=18 cm,angle=0]{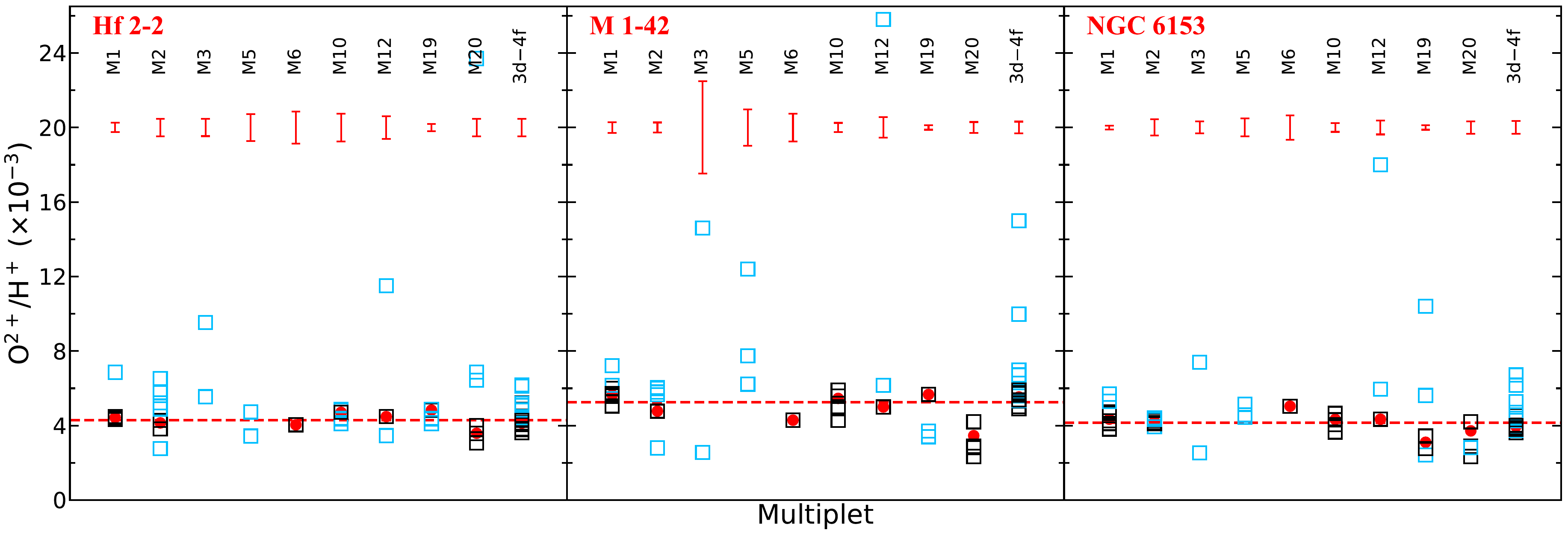}
\caption{Same as Figure\,\ref{fig:NII_abund} but for O~{\sc ii} abundances.} 
\label{fig:OII_abund} 
\end{center}
\end{figure*}

\begin{figure*}[ht!]
\begin{center}
\includegraphics[width=18 cm,angle=0]{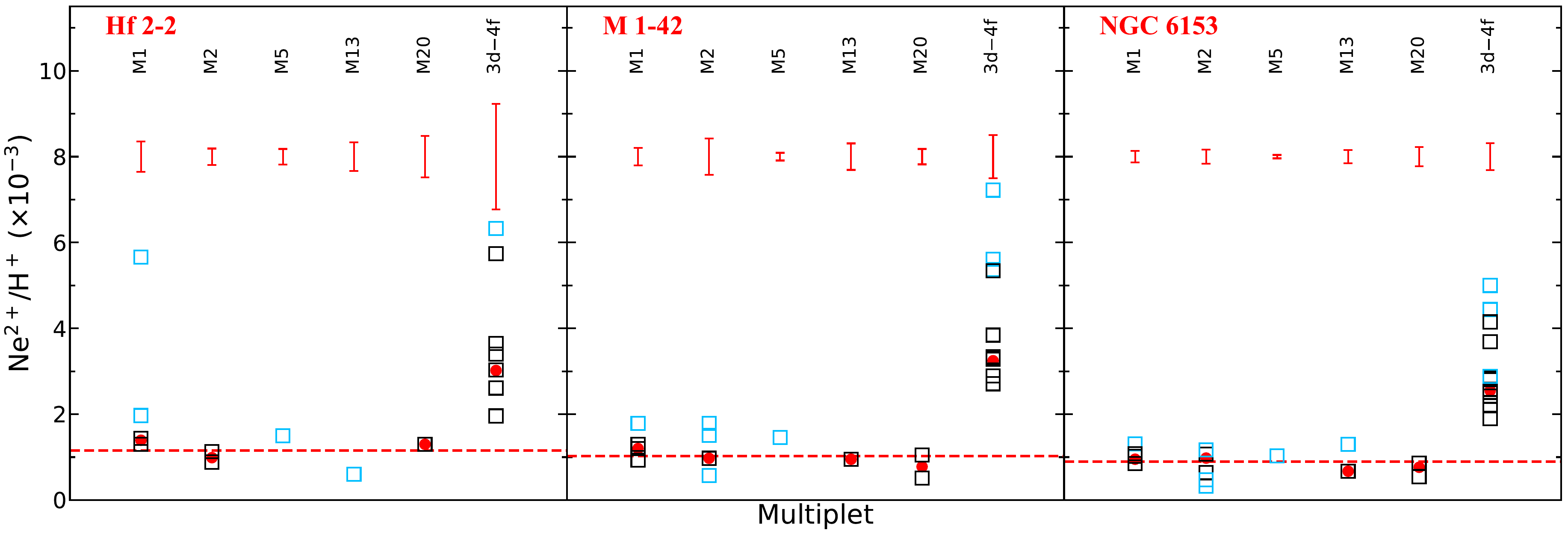}
\caption{Same as Figure\,\ref{fig:NII_abund} but for Ne~{\sc ii} abundances. The results of M9 multiplet were not presented (for reason, see text for details).} 
\label{fig:NeII_abund} 
\end{center}
\end{figure*}

\begin{table*}
\begin{center}
\caption{ORL Abundances of O$^{+}$, C$^{3+}$, N$^{3+}$, O$^{3+}$ and C$^{4+}$}
\label{tab:orl_ionic}
\begin{tabular}{lllll}
\hline\hline
Line & Transition & \multicolumn{3}{c}{Abundance (X$^{i+}$/H$^+$)} \\
\cline{3-5}
(\AA) & & Hf\,2-2 & M\,1-42 & NGC\,6153\\
\hline
O~{\sc i} $\lambda$7771.94 & $3s\ ^5S^o_2 - 3p\ ^5P_3$ & 6.16($\pm$0.15)$\times10^{-3}$\,\tablenotemark{\rm{\scriptsize a}} & 1.59($\pm$0.05)$\times10^{-3}$ & 6.63($\pm$0.47)$\times10^{-4}$\,\tablenotemark{\rm{\scriptsize a}} \\
O~{\sc i} $\lambda$7774.17 & $3s\ ^5S^o_2 - 3p\ ^5P_2$ & 5.85($\pm$0.23)$\times10^{-3}$ & 2.03($\pm$0.07)$\times10^{-3}$\,\tablenotemark{\rm{\scriptsize b}} & 8.48($\pm$0.80)$\times10^{-4}$ \\
O~{\sc i} $\lambda$7775.39 & $3s\ ^5S^o_2 - 3p\ ^5P_1$ & 5.99($\pm$0.41)$\times10^{-3}$ & $\cdots$ & 7.51($\pm$1.55)$\times10^{-4}$ \\
O~{\sc i} $\lambda$9260.84 & $3p\ ^5P_1-3d\ ^5D^o_{0,1,2}$ & 2.60($\pm$0.50)$\times10^{-3}$\,\tablenotemark{\rm{\scriptsize a}}  & 1.55($\pm$0.17)$\times10^{-3}$\,\tablenotemark{\rm{\scriptsize c}} & 2.27($\pm$0.35)$\times10^{-4}$\,\tablenotemark{\rm{\scriptsize a}} \\
O~{\sc i} $\lambda$9262.67 & $3p\ ^5P_2-3d\ ^5D^o_{1,2,3}$ & 4.72($\pm$0.48)$\times10^{-3}$ & 1.14($\pm$0.08)$\times10^{-3}$ & 6.71($\pm$0.84)$\times10^{-4}$ \\
O~{\sc i} $\lambda$9265.94 & $3p\ ^5P_3-3d\ ^5D^o_{2,3,4}$ & 4.73($\pm$0.35)$\times10^{-3}$ & 1.79($\pm$0.10)$\times10^{-3}$\,\tablenotemark{\rm{\scriptsize a}} & 6.59($\pm$0.60)$\times10^{-4}$ \\
O~{\sc i} adopted & & 5.41($\pm$0.18)$\times10^{-3}$ & 1.50($\pm$0.05)$\times10^{-3}$ & 7.24($\pm$0.42)$\times10^{-4}$ \\
C~{\sc iii} $\lambda$4186.90\tablenotemark{\rm{\scriptsize d}} & $4f\ ^1F^o - 5g\ ^1G$ & $\cdots$ & 1.72($\pm$0.20)$\times10^{-4}$ & 1.56($\pm$0.19)$\times10^{-4}$ \\
C~{\sc iii} $\lambda$4647.42 & $3p\ ^3P^o_1– 3s\ ^3S_2$ & $\cdots$ & 1.36($\pm$0.10)$\times10^{-4}$ & 1.85($\pm$0.06)$\times10^{-4}$ \\
C~{\sc iii} $\lambda$8196.50 & $5g\ ^{1,3}G -6h\ ^{1,3}H^o$ & $\cdots$ & 1.30($\pm$0.03)$\times10^{-4}$ & 1.46($\pm$0.05)$\times10^{-4}$ \\
C~{\sc iii} adopted &  & $\cdots$ & 1.36($\pm$0.10)$\times10^{-4}$ & 1.85($\pm$0.06)$\times10^{-4}$ \\
N~{\sc iii} $\lambda$4379.11 & $4f\ ^2F^o-5g\ ^2G$ & 2.01($\pm$0.23)$\times10^{-4}$ & 9.09($\pm$0.08)$\times10^{-4}$ & 3.46($\pm$0.05)$\times10^{-4}$ \\
O~{\sc iii} $\lambda$3260.86 & $3p\ ^3D_2-3d\ ^3F^o_3$ & $\cdots$ & 1.91($\pm$0.45)$\times10^{-4}$ & 8.71($\pm$2.21)$\times10^{-5}$ \\
O~{\sc iii} $\lambda$3265.33 & $3p\ ^3D_3-3d\ ^3F^o_4$ & $\cdots$ & 1.46($\pm$0.25)$\times10^{-4}$ & 8.15($\pm$2.47)$\times10^{-5}$ \\
O~{\sc iii} adopted & & $\cdots$ & 1.62($\pm$0.23)$\times10^{-4}$ & 8.45($\pm$1.65)$\times10^{-5}$ \\
C~{\sc iv} $\lambda$4658.30 & $5g\ ^2G-6h\ ^2H^o$ & $\cdots$ & $\cdots$ & 5.65($\pm$0.47)$\times10^{-5}$ \\
\hline
\end{tabular}
\begin{description}
NOTE. -- A number of telluric emission lines are present near these O~{\sc i} features, and the suitability of a given O~{\sc i} line for abundance averaging depends on the location and systemic velocity of the nebula relative to the telluric features. The degree of blending varies among the different PNe. Only O~{\sc i} lines without warning notes were included in the averaged abundance.\\
\tablenotemark{\rm{\scriptsize a}} Blend with telluric lines (see Figure\,\ref{fig:telluric}).\\
\tablenotemark{\rm{\scriptsize b}} The O~{\sc i} $\lambda\lambda$7774.17, 7775.39 lines in M\,1-42 are partially blended. The abundance derived from the former may be unreliable due to this blending, while the later is severely contaminated by sky emission and was therefore excluded from the abundance analysis. \\
\tablenotemark{\rm{\scriptsize c}} The O~{\sc i} $\lambda\lambda$9260.84, 9262.67 lines in M\,1-42 are partially blended. The sum of the two line fluxes yielded an abundance of 1.29($\pm$0.12)$\times10^{-3}$, which was used in the computation of the adopted O$^+$ abundance.\\
\tablenotemark{\rm{\scriptsize d}} Blend with O~{\sc iii} $\lambda$4187.
\end{description}
\end{center}
\end{table*}

\begin{table}
\begin{center}
\caption{The $b(J_i,J_f)$ factors and transitions of O~{\sc i} lines and C~{\sc iii} $\lambda$4647.42}
\label{tab:bfactor}
\begin{tabular}{lcc}
\hline\hline
Line & Transition & $b(J_i,J_f)$ \\
\hline
O~{\sc i} $\lambda$7771.94 & $3s\ ^5S^o_2 - 3p\ ^5P_3$ & 7/15 \\
O~{\sc i} $\lambda$7774.17 & $3s\ ^5S^o_2 - 3p\ ^5P_2$ &1/3 \\
O~{\sc i} $\lambda$7775.39 & $3s\ ^5S^o_2 - 3p\ ^5P_1$ &1/5 \\
O~{\sc i} $\lambda$9260.84 & $3p\ ^5P_1-3d\ ^5D^o_{0,1,2}$ & 1/5 \\
O~{\sc i} $\lambda$9262.67 & $3p\ ^5P_2-3d\ ^5D^o_{1,2,3}$ &1/3 \\
O~{\sc i} $\lambda$9265.94 & $3p\ ^5P_3-3d\ ^5D^o_{2,3,4}$ &7/15 \\
C~{\sc iii} $\lambda$4647.42 & $3p\ ^3P^o_1– 3s\ ^3S_2$ & 5/9 \\
\hline
\end{tabular}
\end{center}
\end{table}

\begin{figure*}[ht!]
\begin{center}
\includegraphics[width=18 cm,angle=0]{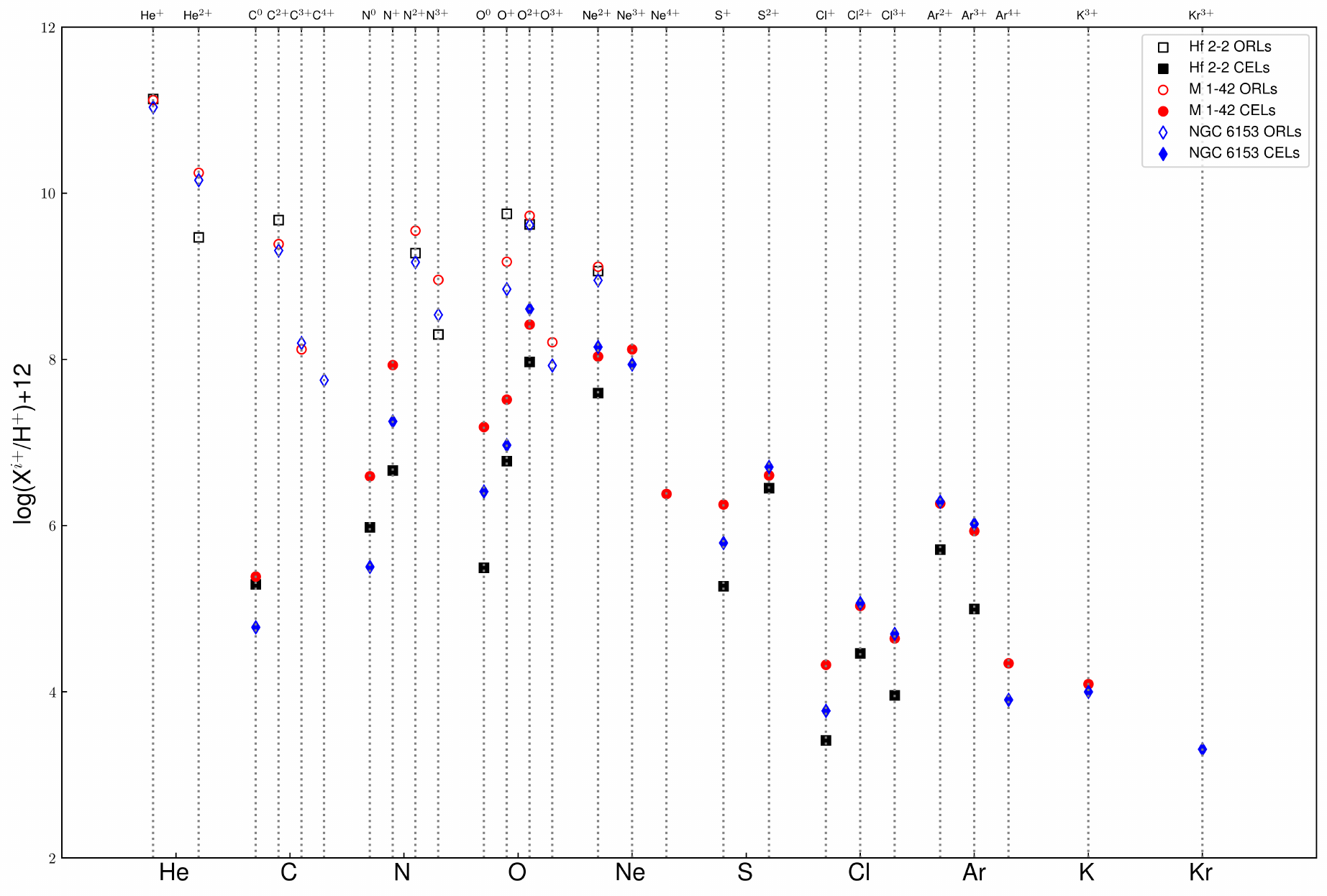}
\caption{Comparison of the results of ionic abundances of Hf\,2-2 (black), M\,1-42 (red) and NGC\,6153 (blue). The solid markers represent the CEL abundances, while the open markers represent the ORL abundances.} 
\label{fig:abund_comparison}
\end{center}
\end{figure*}

CEL ionic abundances were computed with the \texttt{Atom.getIonAbundance} method of \texttt{PyNeb}. We used recombination-corrected [N~{\sc ii}] temperatures and [S~{\sc ii}] densities for singly-ionized and neutral species, while [O~{\sc iii}] temperatures were applied to higher ionization states. For intermediate-ionized ions, we employed [Cl~{\sc iii}] densities, whereas [Ar~{\sc iv}] densities were used for higher ionization levels. The hydrogen temperature in each calculation was assumed the same as the ion. Calculations prioritized the higher-flux emission lines for each ion, with all results listed in Table \ref{tab:cel_ionic}.

The high ADFs in our objects indicate significant recombination contributions in [N~{\sc ii}] and [O~{\sc ii}] lines, and even in [O~{\sc iii}] lines. For N$^+$ abundances, calculations used the recombination-corrected [N~{\sc ii}] $\lambda$5755 fluxes from Section \ref{sec:celdiagnostics}, while the two [N~{\sc ii}] nebular lines were taken directly from the measured fluxes owing to their strength and negligible recombination contributions. The similar approach was applied to the O$^{2+}$ CEL abundance calculations. For O$^+$ CEL abundances, the [O~{\sc ii}] auroral lines within our extraction region are dominated by recombination excitation, limiting the analysis to the  [O~{\sc ii}] $\lambda\lambda$3726,3729 nebular lines. These lines, however, are subject to both recombination contributions and collisional de-excitation. As a result, applying recombination corrections can lead to underestimated [O~{\sc ii}] abundances. We finally adopted O$^+$ CEL abundances derived from the recombination-corrected [O~{\sc ii}] $\lambda$3726 fluxes, which is less affected by collisionally de-excitation and yields smaller abundance underestimations. 

For ions with only one detected line (like Ar$^{4+}$) or a single unblended transition (like Cl$^{3+}$), the abundance derived from that specific spectral line is exclusively utilized in subsequent elemental abundance calculations. For ions with strong CELs, like N$^+$, O$^{2+}$ and S$^{2+}$, the results from the strongest lines were adopted. Neutral atoms are less abundant and their lines are weak, requiring weighted averaging across all available transitions listed in Table \ref{tab:cel_ionic}, with weights proportional to the inverses of the abundance uncertainties. This approach extends to some weakly emitting ionized species, like Ne$^{2+}$, K$^{3+}$ and Kr$^{3+}$, in which case the weighted averages were derived based on the results from their relatively strong transitions. In addition, due to the different brightnesses of our sources, some ions, such as S$^+$, Cl$^{2+}$ and Ar$^{3+}$, have fewer lines detected in the spectra of Hf\,2-2 than in the other two PNe. Therefore, results or averages of the lines detected simultaneously in all three objects have been adopted for the sake of consistency. See the table-notes of Table \ref{tab:cel_ionic} for more details of the adoptions of ionic abundances.

\subsection{Ionic Abundances from ORLs} 
\label{sec:orls_ionic}

To determine the He$^+$ abundances, we employed the He~{\sc i} temperatures computed in Section \ref{sec:orl_diagnostics} and the updated atomic data from \citet{2022MNRAS.513.1198D}. For He$^{2+}$, which originates in regions close to the central star (see Section \ref{sec:structure}), [O~{\sc iii}] temperatures were adopted. The abundances of these two ions are also shown in Table\,\ref{tab:cel_ionic}.

Atomic data from \citet{2000A&AS..142...85D} were adopted to calculate C$^{2+}$/H$^+$, using the O~{\sc ii} temperature.  However, the lowest temperature available from this atomic data set is 500\,K, which is still higher than the O\,{\sc ii} temperature of Hf\,2-2.  The C$^{2+}$ abundance for Hf\,2-2 was thus computed assuming $T_{\rm e}$ = 500\,K.  In all three PNe, we adopted $N_{\rm e}$ = 10$^4$\,cm$^{-3}$, which was the only density considered in the calculations by \citet{2000A&AS..142...85D}.  This choice of $T_{\rm e}$ and $N_{\rm e}$ introduces only negligible bias in the ionic abundance of C$^{2+}$, because the emissivities of recombination lines have similar dependences on electron temperature\footnote{The emissivity of a recombination line, whose excitation is radiative recombination dominated, has a weak power-law dependence on electron temperature, $\epsilon(\lambda)\propto\,T_{\rm e}^{-\alpha}$, where $\alpha\sim$1.}.  The C$^{2+}$ abundances are presented in Table\,\ref{tab:cii_ionic}. 

The results obtained from the two strongest C~{\sc ii} lines, $\lambda$4267 and $\lambda$7325+, exhibit significant discrepancies. The abundances derived from the latter are affected by optical depth effects. C~{\sc ii} $\lambda$7325+ yields higher abundance results under Case A but lower under Case B, suggesting that actual nebular conditions may be intermediate. Additionally these lines have significant continuum pumping fluorescent contributions in addition to recombination, as discussed in \citet[Sect. 4.2]{2024A&A...687A..97R}. In contrast, the emissivities of C~{\sc ii} $\lambda$4267, $\lambda$6151 and $\lambda$5342 arise exclusively from recombinations with no fluorescent contributions, making their derived abundances more consistent. Therefore, ionic abundances obtained from the strongest C~{\sc ii} $\lambda$4267 line were adopted in the following elemental abundance calculations.

ORL abundances of low- and intermediate-ionized species (O$^{+}$, O$^{2+}$, N$^{2+}$, Ne$^{2+}$) were derived using O~{\sc ii} temperatures and densities. N~{\sc ii}, O~{\sc ii}, and Ne~{\sc ii} have multiple multiplets with numerous lines that yield individual abundance determinations. We performed weighted averaging across multiplets using the unblended lines. The final adopted ORL abundances, representing multiplet-weighted averages along with abundances from individual lines and multiplets, are tabulated in Tables \ref{tab:nii_ionic}, \ref{tab:oii_ionic}, and \ref{tab:neii_ionic} for N$^{2+}$, O$^{2+}$ and Ne$^{2+}$, respectively. Comparisons of N$^{2+}$, O$^{2+}$ and Ne$^{2+}$ ORL abundances across different lines, multiplets, and PNe are presented in Figures \ref{fig:NII_abund}, \ref{fig:OII_abund} and \ref{fig:NeII_abund}, respectively.

\begin{deluxetable*}{lccccccccccc}
\tablecaption{The Logarithmic ADFs of O$^+$, O$^{2+}$ and Ne$^{2+}$ of this Work in Comparison with Those Obtained from MUSE and Long-slit Spectroscopy.}
\label{tab:ion_adf}
\tablehead{
Ions & & Hf\,2-2 & & & & M\,1-42 & & & & NGC\,6153 \\
\cline{2-4}
\cline{6-8}
\cline{10-12}
 & This work & MUSE\tablenotemark{\rm{\scriptsize a}} & Long-slit\tablenotemark{\rm{\scriptsize b}} & & This work & MUSE\tablenotemark{\rm{\scriptsize a}} & Long-slit\tablenotemark{\rm{\scriptsize c}} & & This work & MUSE\tablenotemark{\rm{\scriptsize d}} & Long-slit\tablenotemark{\rm{\scriptsize e}}
 }
\startdata
O$^{+}$ & 2.96$\pm$0.03 & 1.87$\pm$0.57 & $\cdots$ & & 1.66$\pm$0.03 & 1.02$\pm$0.12 & $\cdots$ & & 1.89$\pm$0.03 & 1.24$\pm$0.14 & $\cdots$\\
O$^{2+}$ & 1.65$\pm$0.01 & 1.26$\pm$0.08 & 1.85 & & 1.31$\pm$0.01 & 0.90$\pm$0.06 & 1.35 & & 1.01$\pm$0.03 & 0.76$\pm$0.06 & 0.97\\
Ne$^{2+}$ & 1.48$\pm$0.05 & $\cdots$ & 1.80 & & 0.98$\pm$0.04 & $\cdots$ & 1.33 & & 0.80$\pm$0.03 & $\cdots$ & 1.04\\
\enddata
\tablecomments{\tablenotemark{\rm{\scriptsize a}}\citet{2022MNRAS.510.5444G}; \tablenotemark{\rm{\scriptsize b}}\citet{2006MNRAS.368.1959L} (results of 2 arcsec slitwidth spectroscopy); \tablenotemark{\rm{\scriptsize c}}\citet{2001MNRAS.327..141L}; \tablenotemark{\rm{\scriptsize d}}\citet{2024AA...689A.228G} (results of recipe 2); \tablenotemark{\rm{\scriptsize e}}\citet{2000MNRAS.312..585L} (results of minor axis).  ADF(O$^{+}$) of MUSE spectroscopy were calculated based on O~{\sc i} $\lambda$7773+ and [O~{\sc ii}] $\lambda$7319+, which is different from ours (O~{\sc i} $\lambda$7773+ and [O~{\sc ii}] $\lambda$3726). The ionic abundances of long-slit spectroscopy often have no uncertainty.}
\end{deluxetable*}

The O$^{2+}$ and N$^{2+}$ abundances from different multiplets show a relatively good consistency between them.  On the other hand, Ne$^{2+}$ 3d$-$4f abundances systematically exceed 3$-$3 transitions values for all targets. \citet{2000MNRAS.312..585L} mentioned that these differences may be caused by uncertainties in the effective recombination coefficients of the 3d$-$4f transitions, so that only the weighted average results of 3$-$3 transitions were adopted. The results for multiplet M9 (a single line, Ne~{\sc ii} $\lambda$3568.50 3s$^\prime$ $^2$D – 3p$^\prime$ $^2$F$\rm^o$) are also significantly higher than the abundances derived from other 3$-$3 transitions. The inner-shell electrons of both the upper and lower energy levels involved in this transition are in excited states, so the intensity of this line may be affected by the NLTE conditions in PNe, resulting in an overestimated abundance. Consequently, the M9 data were excluded from final Ne~{\sc ii} abundance determinations and not presented in Figure\,\ref{fig:NeII_abund}.

Our spectra reveal two O\,{\sc i} multiplets (about 7773 \AA\ and 9264 \AA), both involving $s=2$ transitions that differ from the spin quantum number ($s=1$) of O$^+$ ground state, precluding fluorescence contributions. The effective recombination coefficients of these multiplets are available in \citet{1991A&A...251..680P}, but not j-resolved. In order to derive the effective recombination coefficients of transitions between two j-resolved levels, $SL_iJ_i$ and $SL_fJ_f$, we use the relation:
\begin{equation}
\alpha_{eff}(SL_iJ_i,SL_fJ_f) = \alpha_{eff}(SL_i,SL_f)b(J_i,J_f),
\end{equation}
where the $b(J_i,J_f)$ factor can be calculated using:
\begin{equation}
b(J_i,J_f)={(2J_i+1)(2J_f+1)\over (2S+1)}
\left\{
\begin{array}{ccc}
J_i & J_f & 1 \\
L_f & L_i & S 
\end{array}
\right\}^2
\end{equation}
where \{\} is the 6-j symbol \citep[refer to][for details in the methodology]{10.1093/oso/9780198517597.001.0001}. The $b(J_i,J_f)$ for the O~{\sc i} lines and for the C~{\sc iii} $\lambda$4647.42 line are presented in Table \ref{tab:bfactor}.

The O~{\sc i} lines fall within a wavelength range densely populated by telluric absorption lines (see Figure\,\ref{fig:telluric}).  Radial velocity disparities between our target PNe ($\sim-$90 km s$^{-1}$ for M\,1-42, and $\sim+$40 km s$^{-1}$ for Hf\,2-2 and NGC\,6153) induce differential blending between telluric and nebular emissions across targets. In addition, the expansion velocities of different nebulae are not the same. In the case of M\,1-42, the blending of O~{\sc i} $\lambda\lambda$7774.17,7775.39 lines is more severe in the spectrum of this PN. Consequently, differential blending with telluric or nearby O~{\sc i} lines necessitates source-specific O~{\sc i} line selections for the adopted O$^+$ abundance derivation for the different PNe, as illustrated in Table \ref{tab:orl_ionic}.

High-ionization species like O$^{3+}$ exhibit spatial distributions inconsistent with recombination emission patterns of lower-ionized ions, showing proximity to central stars rather than existing in cold plasma regions characterized by O~{\sc ii} temperatures and densities (see the location of O~{\sc iii} recombination contribution in the PV diagrams described in Section \ref{sec:2d_O3_temp}). Therefore, we need to adopt [O~{\sc iii}] temperatures and [Ar~{\sc iv}] densities for this species' abundances calculations. For C~{\sc iii} $\lambda$4186.90 and $\lambda$4647.42, we incorporated the branching ratios from \citet{1991A&A...251..680P} to account for non-100\% downward transitions from upper energy levels when calculating recombination coefficients. ORL abundances for C$^{3+}$, C$^{4+}$, N$^{3+}$ and O$^{3+}$ are also presented in Table \ref{tab:orl_ionic}. Despite the low SNRs of these lines and the reliance on atomic data not j-resolved, the unblended transitions for each ion show consistency within uncertainties. Detected S~{\sc ii} and possibly Ar~{\sc ii} lines remain unquantified due to the lack of available atomic data.

Similar to CEL abundance calculations, the temperatures applied to hydrogen are identical to those used for the ions in the ORL abundance calculations, except for the calculations of the O$^{+}$, N$^{2+}$, O$^{2+}$, and Ne$^{2+}$ abundances of Hf\,2-2, where we adopted $T_{\rm e}=500$\,K for H$\beta$ \citep[the lowest elecgtron temperature considered in the calculations by][]{1995MNRAS.272...41S}, rather than the 400\,K adopted for the O~{\sc ii} ORLs.  However, nebular hydrogen emission should inherently maintain a uniform thermal conditions throughout a given PN, independent of the adopted calculation method. As shown in Section \ref{sec:HI_diagnostics}, the temperatures derived from the Balmer and Paschen discontinuities lie between the CEL and ORL temperature regimes. If these results accurately reflect the actual physical conditions of H~{\sc i} line-emitting region, then the $T_{\rm e}$(H~{\sc i}) used in CEL abundance determinations may be overestimated. This would lead to an underestimation of H$\beta$ emissivities and, consequently, CEL abundances. Conversely, this overestimated temperature would lead to overestimated ORL abundances, and ultimately resulting in an overestimation of ADF values. For consistency with the literature, we retained the methodology described in this section.

Ionic abundance uncertainties (given in parentheses) encompass only emission-line flux errors, excluding the contribution of uncertainties of temperature and density adopted in the calculations, and are therefore likely underestimated. Figure \ref{fig:abund_comparison} reveals systematic discrepancies between ORL and CEL abundances. The solid squares (CELs) are systematically lower than their open counterparts (ORLs) for the same ionic species. In addition, the abundance trends show a progressive decline with increasing nuclear charge.

\subsection{Ionic Abundance Discrepancies}

Both CEL and ORL abundances were obtained for O$^+$, O$^{2+}$ and Ne$^{2+}$, enabling ADF determinations for these ions, which are presented in Table \ref{tab:ion_adf}. The ADFs derived from MUSE spectroscopy and long-slit spectra from the literature are also shown in Table \ref{tab:ion_adf} for comparison. Where the references do not provide direct ADFs for the ions, we calculated them using the ionic abundances reported in those papers. It is worth noticing that our ADF(O$^+$) may not be reliable, owing to inaccurate O$^+$ CEL abundances and/or low SNRs of O~{\sc i} lines. 

ADF(O$^{+}$) values systematically exceed ADF(O$^{2+}$) across all three targets, both in our study and in the literature. Under the dual-phase hypothesis, O$^{+}$/O$^{2+}$ ratios are expected to diverge between phases. Metal-rich cold clumps (ORL-dominated) exhibit elevated O$^{+}$/O$^{2+}$ ratios, consistent with low-temperature environments. However, in high-ADF PNe where [O~{\sc ii}] auroral lines are recombination-dominated and CEL emission is weak (e.g., \citet{2022MNRAS.510.5444G}), ADFs calculated from MUSE spectroscopy have larger uncertainties. O$^+$ CEL abundances derived from [O~{\sc ii}] nebular lines are affected by collisional de-excitation, making the corresponding ADFs unreliable. In addition, ADFs do not clearly reflect the chemical composition of different gas phases \citep{2020MNRAS.497.3363G}. Taking the MUSE results for Hf\,2-2 as an example, the proportions of O$^{+}$ and O$^{2+}$ in the cold and warm gases are $M_{c}/M_{w}$(O$^{+}$) = 1.0 and $M_{c}/M_{w}$(O$^{2+}$) = 0.9, respectively \citep{2022MNRAS.510.5444G}. This suggests that the differences in O$^{+}$/O$^{2+}$ ratios between the two plasmas are not significant, and thus it is not possible to conclude that the ratio is systematically higher in the cold plasma than in the warm phase.

Our ADFs are not fully consistent with those reported in the literature. Both ADF(O$^{+}$) and ADF(O$^{2+}$) derived from our UVES data exceed the values obtained from MUSE, likely due to spatial extraction bias (Section \ref{sec:extraction}). Our UVES apertures preferentially sample ORL-enhanced regions (Section \ref{sec:structure} or the ADF maps of \citet{2022MNRAS.510.5444G}), resulting in amplified ADF values. While our ADF(O$^{2+}$) are more consistent with those reported from long-slit spectroscopy, ADF(Ne$^{2+}$) shows a significant discrepancy compared to the long-slit results. Long-slit spectra are typically weighted toward the inner nebular regions (unlike MUSE’s spatially comprehensive IFU coverage), and they yield higher ADF(O$^{2+}$) values that match our measurements. For Ne$^{2+}$, literature long-slit studies often employed 3d$-$4f transition lines, which produce ORL abundances and ADF(Ne$^{2+}$) values that are higher compared to our results based on 3$-$3 transitions.

\subsection{Elemental Abundances} 
\label{sec:abundances}

The elemental abundances of CELs and ORLs are presented in Table \ref{tab:elem_abund}, with their uncertainties that account only for line flux errors. The He abundances were determined by summing the abundances of He$^{+}$ and He$^{2+}$. For elements with higher atomic numbers, it is essential to apply ionization correction factors (ICFs) in the calculations.  The ICFs of \citet{2014MNRAS.440..536D} were prioritized.  The applicabilities of these ICFs are determined by the ionic fractions $\upsilon$ = He$^{2+}$/(He$^{2+}$+He$^{+}$) and $\Omega$\footnote{Here we use $\Omega$, instead of the Greek letter $\omega$ originally used in \citet[][Equation\,5 therein]{2014MNRAS.440..536D}, to avoid confusion with the parameter $\omega$ that we defined, at the end of this section, as the fractional contribution of H$\beta$ emission from the cold plasma component in PNe (see also Section\,\ref{sec:H_weight}).} = O$^{2+}$/(O$^{2+}$+O$^{+}$).  Hf\,2-2, M\,1-42 and NGC\,6153 have $\upsilon=0.021,\ 0.117,$ and $0.117$, and $\Omega=0.940,\ 0.889,$ and $0.978$, respectively.  The $\Omega$ value for NGC\,6153 lies slightly outside the applicability range.  The ICFs from \citet{2001ApJ...562..804K} or \citet{1994MNRAS.271..257K} were considered if there was no suitable ICF in \citet{2014MNRAS.440..536D} for a particular element. 

The CEL abundances of oxygen were derived from the adopted O$^{+}$/H$^{+}$ and O$^{2+}$/H$^{+}$ listed in Table \ref{tab:cel_ionic}, using,
\begin{eqnarray}
\rm log\ ICF(O^{+} + O^{2+}) = {0.08\upsilon + 0.006\upsilon^2 \over 0.34-0.27\upsilon} \nonumber.
\end{eqnarray}
The ORL O$^{+}$/H$^{+}$ was also derived, we utilized the above equation to determine the oxygen ORL abundances, although the physical condition of the region emitting CELs and ORLs differ significantly. However, considering that the He$^{+}$ abundances are much higher than that of He$^{2+}$ in these three PNe, the corresponding ICF($O^{+} + O^{2+}$) values are very close to 1. Therefore even if the amount of He$^{2+}$ in the recombination-emitting region is small, the errors introduced by using the same ICF are negligible.

\begin{deluxetable*}{lccccccccccc}
\tablecaption{Elemental Abundances and Comparison with Those Obtained from MUSE and Long-slit Spectroscopy.}
\label{tab:elem_abund}
\tablehead{
Elem. & & Hf\,2-2 & & & & M\,1-42 & & & & NGC\,6153 \\
\cline{2-4}
\cline{6-8}
\cline{10-12}
 & This work & MUSE\tablenotemark{\rm{\scriptsize a}} & Long-slit\tablenotemark{\rm{\scriptsize b}} & & This work & MUSE\tablenotemark{\rm{\scriptsize a}} & Long-slit\tablenotemark{\rm{\scriptsize c}} & & This work & MUSE\tablenotemark{\rm{\scriptsize d}} & Long-slit\tablenotemark{\rm{\scriptsize e}}
 }
\startdata
He (ORLs) & 11.15$\pm$0.01 & 11.08 & 11.02 & & 11.18$\pm$0.01 & 11.16 & 11.17 & & 11.09$\pm$0.01 & 10.95$-$11.00 & 11.13 \\
C (ORLs) & 9.76$\pm$0.01 & 9.64 & 9.63 & & 9.49$\pm$0.01 & 9.22 & 9.35 & & 9.43$\pm$0.01 & 10.22 & 9.42 \\
N & 7.99$\pm$0.03 & 8.14 & 7.77 & & 8.91$\pm$0.02 & 8.76 & 8.68 & & 8.93$\pm$0.01 & 8.35 & 8.36 \\
N (ORLs) & 9.35$\pm$0.02 & $\cdots$ & 9.52 & & 9.70$\pm$0.03 & $\cdots$ & 9.59 & & 9.27$\pm$0.01 & $\cdots$ & 9.32 \\
O & 8.00$\pm$0.01 & 8.52 & 8.11 & & 8.50$\pm$0.01 & 8.79 & 8.63 & & 8.65$\pm$0.01 & 8.70 & 8.69\\
O (ORLs) & 9.99$\pm$0.01 & 10.01 & 9.94 & & 9.87$\pm$0.01 & 9.74 & 9.79 & & 9.72$\pm$0.01 & 10.63 & 9.66 \\
Ne & 7.63$\pm$0.01 & $\cdots$ & 7.62 & & 8.39$\pm$0.03 & $\cdots$ & 8.12 & & 8.36$\pm$0.02 & $\cdots$ & 8.25\\
Ne (ORLs) & 9.09$\pm$0.05 & $\cdots$ & 9.52 & & 9.04$\pm$0.04 & $\cdots$ & 9.40 & & 8.98$\pm$0.03 & $\cdots$ & 9.29 \\
S & 6.73$\pm$0.01 & 6.74 & 6.36 & & 6.96$\pm$0.01 & 7.20 & 7.08 & & 7.14$\pm$0.01 & 7.18 & 7.23 \\
Cl & 4.61$\pm$0.03 & 5.10 & $\cdots$ & & 5.24$\pm$0.01 & 5.47 & 5.26 & & 5.24$\pm$0.01 & 5.49 & 5.62\\
Ar & 5.82$\pm$0.01 & 6.21 & 6.13 & & 6.48$\pm$0.01 & 6.67 & 6.56 & & 6.49$\pm$0.01 & 6.68 & 6.40\\
Kr & $\cdots$ & $\cdots$ & $\cdots$ & & $\cdots$ & $\cdots$ & $\cdots$ & & 3.63$\pm$0.05 & 4.03 & $\cdots$ \\
\enddata
\tablecomments{Elemental abundances are in logarithm, 12+log(X/H).  The tablenote markers \tablenotemark{\rm{\scriptsize a}} \tablenotemark{\rm{\scriptsize b}} \tablenotemark{\rm{\scriptsize c}} \tablenotemark{\rm{\scriptsize d}} \tablenotemark{\rm{\scriptsize e}} have the same meanings as in Table \ref{tab:ion_adf}.}
\end{deluxetable*}

In the case of carbon ORL abundance, in Hf\,2-2, only C~{\sc ii} emission lines were observed, necessitating corrections for both C$^{+}$ and C$^{3+}$ abundances, employing
\begin{eqnarray}
\rm ICF(C^{2+}/O^{2+})= 0.05 + 2.21\Omega - 2.77\Omega^2 + 1.74\Omega^3  \nonumber.
\end{eqnarray}
In the case of M\,1-42, the C$^{3+}$ abundance was also derived, so the total C abundance was estimated using,
\begin{eqnarray}
\rm ICF(C) = \rm {O^{+}+O^{2+}\over O^{2+}} \nonumber,
\end{eqnarray}
the Eq. A11 in \citet{1994MNRAS.271..257K}. In NGC\,6153, the C$^{4+}$ abundance was also derived from the C~{\sc iv} $\lambda$4658 line. However, since no suitable ICF is available for this case in neither \citet{2014MNRAS.440..536D} nor \citet{1994MNRAS.271..257K}, we used the same equation as for M\,1-42 to derive the total carbon abundance in NGC\,6153.

We estimate the elemental nitrogen abundance with only the N$^+$ abundance, using,
\begin{eqnarray}
\rm log\ ICF(N) = -0.16\Omega(1 + log\ \upsilon)  \nonumber.
\end{eqnarray}
For the ORL abundances of nitrogen, both N$^{2+}$ and N$^{3+}$ abundances were derived. However, the commonly used ICFs for nitrogen are designed for situations where only N$^{+}$ is available.  Assuming that the contributions from N$^{4+}$ and higher ionization stages to the total N abundance is negligible and that the relative ionic abundances of nitrogen under CEL and ORL conditions remain similar, we computed the nitrogen recombination abundances by utilizing,
\begin{eqnarray}
\rm{N_{ORL}^{2+}+N_{ORL}^{3+} \over N_{ORL}} = {N_{CEL}^{2+}+N_{CEL}^{3+} \over N_{CEL}} \approx {N_{CEL}-N_{CEL}^+\over N_{CEL}}\nonumber.
\end{eqnarray}
where N$\rm^{+}_{CEL}$ and N$\rm_{CEL}$ represent the N$^{+}$ and N abundance derived from CELs, respectively.

In Hf\,2-2, only Ne$^{2+}$ CEL abundance was derived, and we used
\begin{eqnarray}
\rm ICF (Ne^{2+}/O^{2+}) = \Omega + ({0.014\over \upsilon^{\prime}} + 2\upsilon^{\prime2.7})^3 \nonumber \\
 \times (0.7 + \rm 0.2\Omega - 0.8\Omega^2) \nonumber.
\end{eqnarray}
to calculate the total neon abundance, where $\upsilon^{\prime}=\upsilon$ when $\upsilon>0.015$. Neon abundances from ORLs for all three PNe were also derived based on the equation above. Weak [Ne~{\sc iv}] lines were detected in both M\,1-42 and NGC\,6153. These lines are very faint, and the Ne$^{3+}$ abundances derived from the different lines are highly inconsistent, yet they are essentially of the same order of magnitude, comparable to the Ne$^{2+}$ abundances in the same PNe. There is no suitable ICF available for the case of simultaneous detection of Ne$^{2+}$ and Ne$^{3+}$. If Ne$^{3+}$ abundances were not used, it would be impossible to obtain a total neon abundance larger than the sum of Ne$^{2+}$/H$^{+}$ and Ne$^{3+}$/H$^{+}$, regardless of which ICF is used. Therefore, we calculated the Ne CEL abundance of NGC\,6153 by summing the abundance of all neon ions. [Ne~{\sc v}] $\lambda$3426 is detected in the spectrum of M\,1-42, and total neon abundance was estimated by summing Ne$^{2+}$, Ne$^{3+}$ and Ne$^{4+}$ abundances.

For the third-period elements, we calculated their abundances only using CELs. For sulfur, both S$^{+}$/H$^{+}$ and S$^{2+}$/H$^{+}$ were derived in our targets, so we calculate the abundances of this element using,
\begin{eqnarray}
\rm log\ ICF((S^{+} + S^{2+})/O^+) \nonumber \\
=  {-0.02 - 0.03\Omega - 2.31\Omega^2 + 2.19\Omega^3 \over 0.69 + 2.09\Omega - 2.69\Omega^2}\nonumber.
\end{eqnarray}

For chlorine, Cl$^{+}$,Cl$^{2+}$ and Cl$^{3+}$ abundances were all obtained, and the ICF of this case was developed by \citet{2014MNRAS.440..536D}. According to their Figure 12, the ICF is slightly greater than but close to unity when the He$^{2+}$/He$^+$ ratio is near zero, a condition similar to all our targets. Therefore, we neglected the ICF and determined the chlorine elemental abundance by summing all the ionic chlorine abundances,
\begin{eqnarray}
\rm {Cl\over H} &=& \rm {Cl^{+}\over H^{+}}+{Cl^{2+}\over H^{+}}+{Cl^{3+}\over H^{+}} \nonumber .
\end{eqnarray}

In the spectrum of Hf\,2-2, we detected only [Ar~{\sc iii}] and [Ar~{\sc iv}] lines, so we calculated the argon elemental abundance of this PN based on the ICF method provided by equation (4g) of \citet{2001ApJ...562..804K}, 
\begin{eqnarray}
\rm ICF(Ar) = \rm {1\over 1-(N^{+}/N)}\times{He^{+}+He^{2+}\over He^{+}} \nonumber.
\end{eqnarray}
The presence of highly ionized Ar$^{4+}$ ion was confirmed in M\,1-42 and NGC\,6153. Therefore, we calculated argon abundances for these two PNe using the Eq. A30 of \citet{1994MNRAS.271..257K}, expressing as,
\begin{eqnarray}
\rm ICF(Ar) = \rm {1\over 1-(N^{+}/N)} \nonumber.
\end{eqnarray}

We also determined the Kr$^{3+}$ abundance in NGC\,6153 based on [Kr~{\sc iv}] $\lambda$5346 and [Kr~{\sc iv}] $\lambda$5868. \citet{2015ApJS..218...25S} provided ICF expressions for some $s$-process elements. In the case where only Kr$^{3+}$ is detected, the total Kr abundance is calculated using,
\begin{eqnarray}
\rm {Kr\over H} &=& \rm ICF(Kr)\times {Kr^{3+}\over H^{+}} \nonumber \\
\rm &=& (0.06681+1.05x+0.7112x^{2}-0.907x^{3})^{-1} \times {\rm Kr^{3+}\over H^{+}} \nonumber.
\end{eqnarray}
where $x$ is Ar$^{3+}$/Ar abundance ratio. 

Elemental abundances obtained from MUSE and long-slit spectroscopy are also presented in Table \ref{tab:elem_abund} for comparison. The different wavelength coverage of different spectra leads to the use of distinct diagnostics for determining $T_{\rm e}$ and $n_{\rm e}$, which are subsequently employed to calculate ionic abundances. Additionally, different ICFs and ionic abundances must be used to estimate elemental abundances due to the detection of ions with different ionization degrees in the various spectra. Consequently, these factors can result in unneglectable discrepancies in the reported abundances across different studies. 

For Hf\,2-2, the CEL abundances from our results are generally lower than those reported by \citet{2022MNRAS.510.5444G}, but similar to those of \citet{2006MNRAS.368.1959L}.  Differences in the adopted electron temperatures are one of the main culprits of these abundance differences.  Much more similar abundances will be obtained if the same temperature is adopted.  The [N~{\sc ii}] and [S~{\sc iii}] temperatures used by \citet{2022MNRAS.510.5444G} in abundance calculations are 8200\,K and 6820\,K, respectively, which are lower than the [N~{\sc ii}] and [O~{\sc iii}] temperatures we adopted.  This leads to higher abundances in their work compared to ours and those reported by \citet{2006MNRAS.368.1959L}.  The ORL abundances of He, C, N and O obtained in our work generally agree with those of \citet{2022MNRAS.510.5444G} and \citet{2006MNRAS.368.1959L}, but the Ne abundance we derived is significantly lower than that of \citet{2006MNRAS.368.1959L}. A similar discrepancy is found for the other two sources. This is because we did not use the abundances derived from Ne~{\sc ii} 3d$-$4f transitions (which significantly deviate from the 3$-$3 transition results, see Section \ref{sec:orls_ionic} for more details) when calculating the average ORL abundance of Ne$^{2+}$.

For M\,1-42, most of the CEL abundances in this work are about 0.2 dex lower than those from the MUSE spectroscopy \citep{2022MNRAS.510.5444G} but close to the results of \citet{2001MNRAS.327..141L}. The reason is the same as in Hf\,2-2; the temperatures adopted in the abundance calculations by \citet{2022MNRAS.510.5444G} are lower. However, our nitrogen abundance is higher, even though our N$^{+}$ abundance is lower. We speculate that the higher total N abundance results from uncertainties in the ICF. The ionic abundance of O$^{+}$ is inaccurately affected by both recombination excitation and collisionally de-excitation, which in turn increases the uncertainty of the ICF(N). Our ORL-derived abundances closely align with those reported in the literature, albeit slightly higher. This discrepancy may due to the fact that the extraction region of our spectra is close to the central star (Section \ref{sec:extraction}), where ORL emission is more enhanced.

Finally for NGC\,6153, most of our CEL abundances are close to the results from \citet{2024AA...689A.228G} and \citet{2000MNRAS.312..585L}.  The Ne abundance we derived for this PN is also higher, mainly due to uncertainties in ICF.  For the C and O abundances from ORLs, our results are consistent with those of \citet{2000MNRAS.312..585L} but systematically lower than those of \citet{2024AA...689A.228G}.  The elemental abundances calculated and adopted by \citet{2024AA...689A.228G} have been multiplied by 1/$\omega$.  Here $\omega$ is the fractional contribution of H$\beta$ emission from the cold component in a PN; applying $1/\omega$ yields a \emph{local} abundance for the cold gas and increases the value by $\sim$0.9 dex compared to the integrated (nebula-wide) abundance.  Using the same offset on our ORL results of C and O brings them close to the values of \citet{2024AA...689A.228G}.  We also estimated the $\omega$ values based on the UVES spectra to test their results in Section\,\ref{sec:H_weight}.

\section{Analysis of 2D spectrum} 
\label{sec:2d}

Most PNe exhibit main shell expansion velocities of around 30 km s$^{-1}$, requiring high-resolution spectroscopy to study their kinematics. The spatially resolved UVES spectra (R$\sim$20000) enable both kinematical and spatial analysis of PNe not only through strong CELs but also through weak ORLs, providing insights into the centrally concentrated ORL-emitting gas in high-ADF PNe (e.g., \citet{2015ApJ...803...99C, 2016MNRAS.455.3263J, 2022MNRAS.510.5444G}). Additionally, plasma diagnostics and abundance calculations based on these 2D spectra allow for a detailed characterization of the physical conditions and chemical compositions across different nebular structures. In this section, we present 2D spectral analysis for Hf\,2-2 and M\,1-42. For NGC\,6153, we performed the same 2D spectral analysis and obtained results that are broadly consistent with \citet{2022AJ....164..243R}; consequently, we do not present the analysis in our paper.

\subsection{PV Diagrams Construction}
\label{sec:PV_construction}

Spectral analysis of emission lines of gaseous nebulae requires prior subtraction of the continuum. Before constructing PV diagrams, we estimated the continuum emission in the 2D spectra by fitting regions unaffected by emission lines. The continuum at each spatial coordinate was estimated independently and then subtracted from the 2D spectra. This spatially resolved subtraction simultaneously eliminates continuum contributions from both the nebula and the central star. However, spectra from the field stars containing numerous absorption lines result in incomplete emission removal, producing intermittent structures at the spatial positions around +3.5$^{\prime\prime}$ and $-$6$^{\prime\prime}$ along the wavelength (velocity) direction, which are clearly visible in the PV diagrams of weak lines in M\,1-42. Residual emission from the central star is also apparent in some cases, for example in the O~{\sc ii} $\lambda$4649 of Hf\,2-2 (see Section \ref{sec:AD}), where companion-star emission \citep{2016AJ....152...34H} creates a bright feature coinciding with the central source position.

\begin{figure*}[ht!]
\begin{center}
\includegraphics[width=18 cm,angle=0]{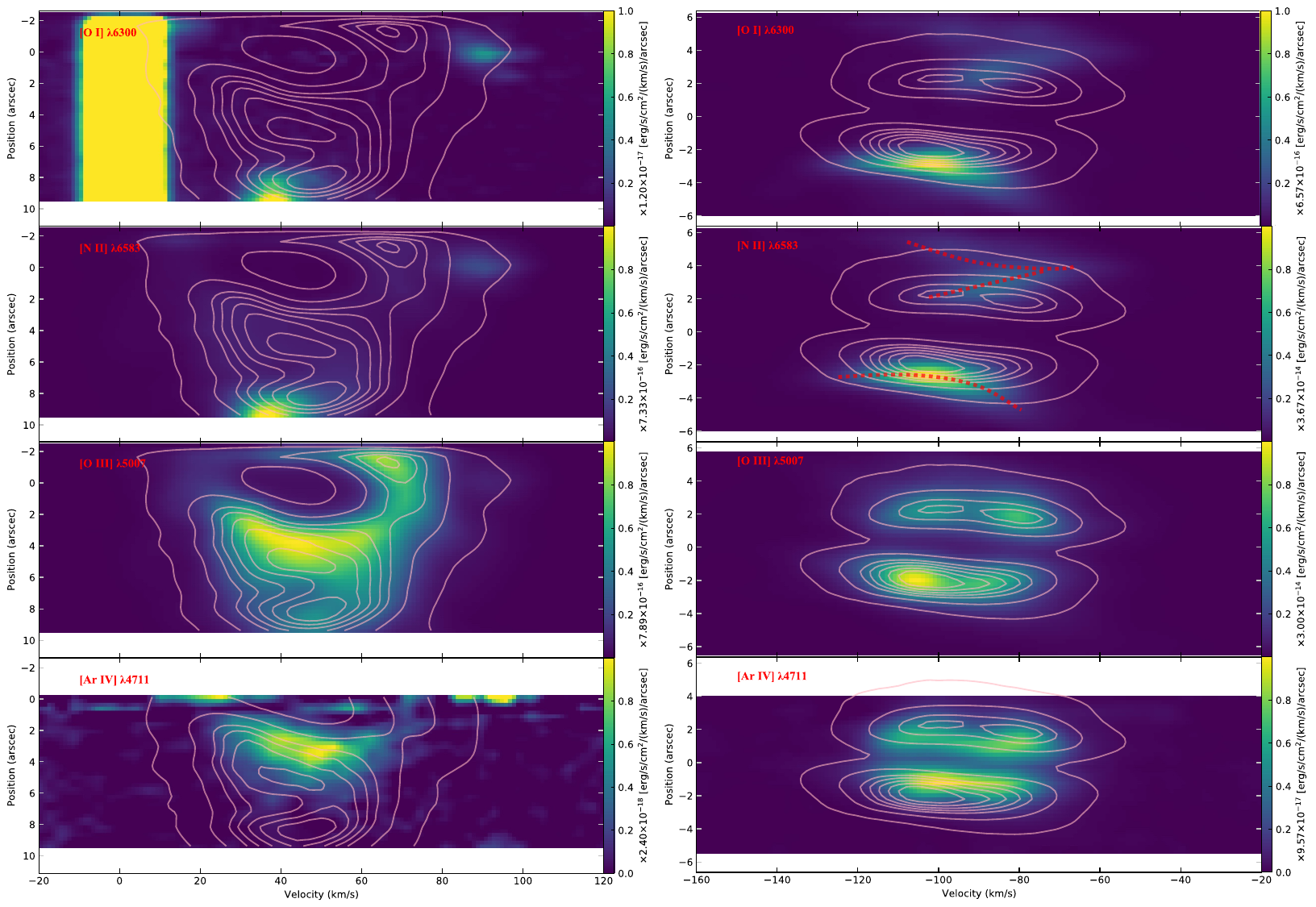}
\caption{PV diagrams of CELs in Hf\,2-2 (left) and M\,1-42 (right). From top to bottom are [O~{\sc i}] $\lambda$6300, [N~{\sc ii}] $\lambda6583$, [O~{\sc iii}] $\lambda$5007, and [Ar~{\sc iv}] $\lambda$4711, representing species from neutral to highly ionized ions. [Ar~{\sc v}] $\lambda$7006 was also observed in M\,1-42, but it is not shown due to its poor quality. The light pink contours correspond to the H$\alpha$ PV diagrams, except for the [Ar~{\sc iv}] PV map of Hf\,2-2, where H$\beta$ contours are shown instead (see text). The values displayed in the colorbar label indicate the peak intensities in each PV diagram. The vertical bright band in the PV diagrams of [O~{\sc i}] is a telluric line, visible only in Hf\,2-2 due to its systemic velocities.  In the PV diagram of the [N\,\textsc{ii}] $\lambda6583$ nebular line of M\,1-42, red-dotted curves are overplotted to visually delineate the hyperbolic-shaped (or X-shaped) arc structures.  Extinction correction has not been applied to these PV diagrams.} 
\label{fig:pv_cel} 
\end{center}
\end{figure*}

\begin{figure*}[ht!]
\begin{center}
\includegraphics[width=18 cm,angle=0]{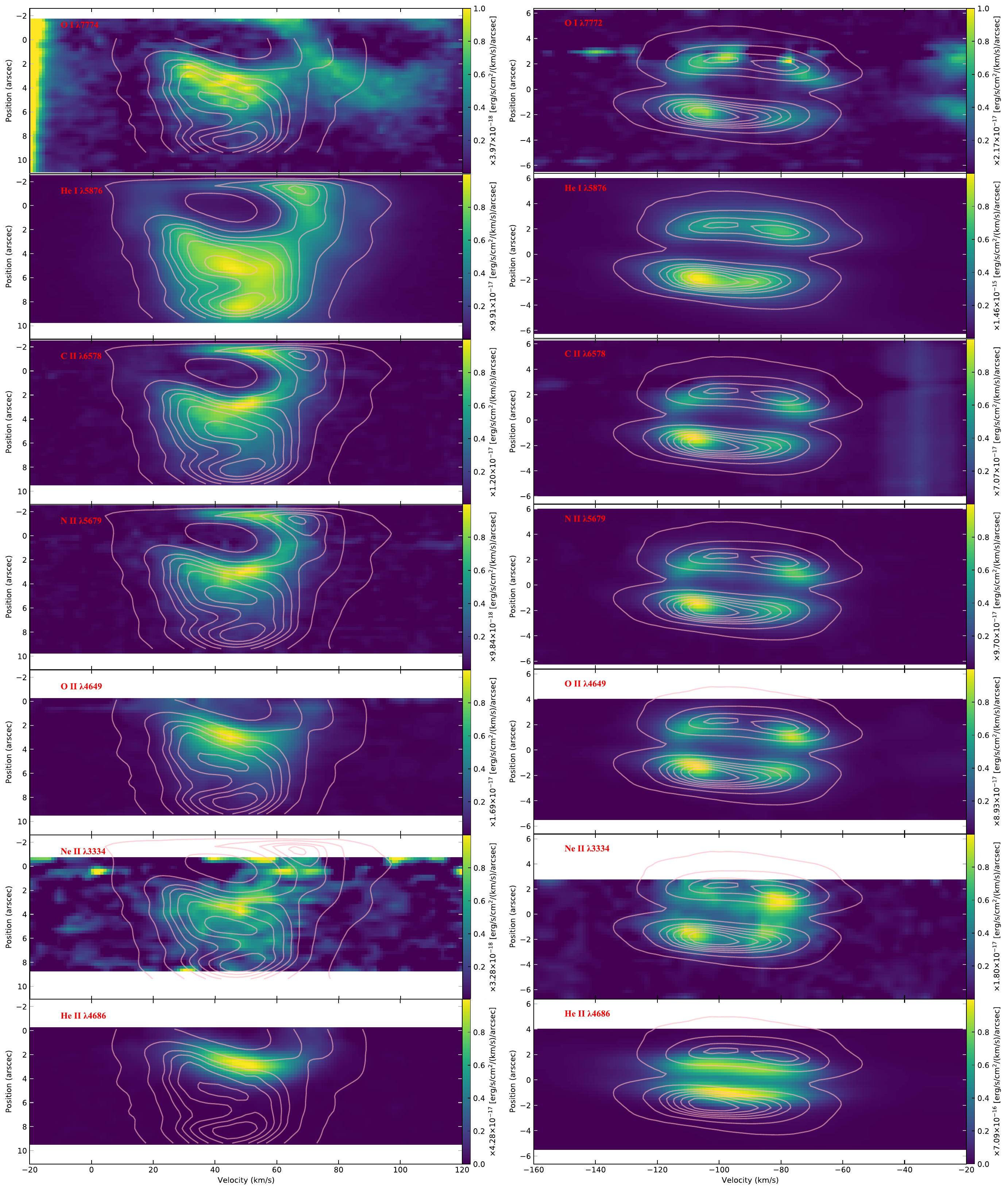}
\caption{Same as Figure\,\ref{fig:pv_cel}, but for PV diagrams of selected ORLs. From top to bottom: O~{\sc i}, He~{\sc i} $\lambda$5876, C~{\sc ii} $\lambda$6578, N~{\sc ii} $\lambda$5679, O~{\sc iii} $\lambda$4649, Ne~{\sc ii} $\lambda$3334, and He~{\sc ii} $\lambda$4686. We show different O~{\sc i} lines for each object in order to minimize potential contamination from telluric features. Although the ionization potential of He$^{+}$ is higher than C$^{2+}$, we place C$^{2+}$ under He$^{+}$ for an easier comparison with other heavy-element ORLs. Intermittent structures along the velocity direction at +3.5$^{\prime\prime}$ and $-6^{\prime\prime}$ in PV diagrams of M\,1-42 are originate from imperfect subtraction of field stars spectra (see text).} 
\label{fig:pv_orl} 
\end{center}
\end{figure*}

Because the spatial coverage and resolution differ between blue and red spectra, we standardized the spatial resolution and velocity range of the PV maps for each PN. Central stars (whose positions were determined before continuum subtraction) were aligned to the zero-point of each diagram. We adopted a spatial resolution of 0$\farcs$25 arcsec, which is slightly finer than the original UVES sampling, in order to preserve spatial detail. The spatial range of the PV maps for each source was carefully selected (from $\sim -3^{\prime\prime}$ to $+$11$^{\prime\prime}$ for Hf\,2-2, and from $\sim -6^{\prime\prime}$ to $+$6$^{\prime\prime}$ for M\,1-42) to accommodate stellar position variations within the slits and encompass the full extents of the nebulae. For PV diagrams from the spectra with smaller spatial coverage (blue arm), we left blank regions at space coordinates where no data. Subsequent multi-line analyses (e.g., plasma diagnostics) were restricted to spatially overlapping regions, with non-overlapping areas left blank.

We set the velocity resolution to 1 km s$^{-1}$, which is smaller than the instrumental spectral resolution to avoid losing details along the velocity direction.  The differing systemic velocities of the sample PNe (about $+$45 km s$^{-1}$ for Hf\,2-2 and $-$95 km s$^{-1}$ for M\,1-42) requires source-specific velocity ranges in the PV maps (from $-20$ to $+$120 km s$^{-1}$ for Hf\,2-2, and from $-160$ to $-20$ km s$^{-1}$ for M\,1-42). Interpolation was applied to the 2D spectra around the lines to construct PV diagrams with the aforementioned spatial and velocity ranges and resolutions. Each pixel unit is converted to erg s$^{-1}$ cm$^{-2}$ (km s$^{-1}$)$^{-1}$ arcsec$^{-1}$.

\begin{figure}[ht!]
\begin{center}
\includegraphics[width=8.5 cm,angle=0]{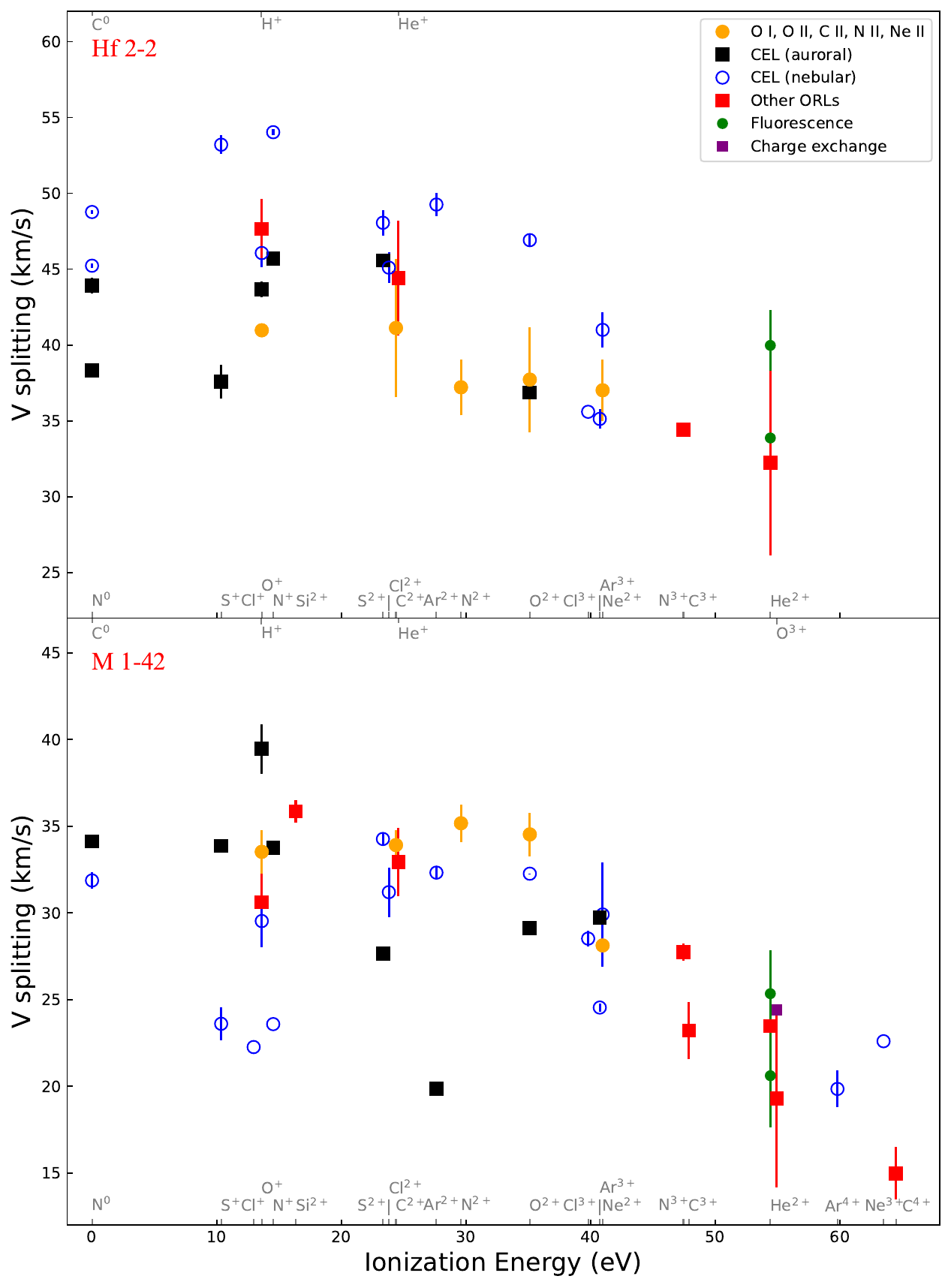}
\caption{Wilson diagrams of Hf\,2-2 (upper panel) and M\,1-42 (bottom panel) showing the relationship between velocity splitting of emission lines and ionization potentials of corresponding species. Different types of emission lines are indicated by different markers, as shown in the legend in the upper panel. The short gray vertical lines mark the positions of ionization potentials of ions labeled nearby. Heavy element ORLs lines, as well as auroral lines with recombination excitation contribution, have velocity splittings slowly varied with ionization potentials, while nebular lines and other ORLs decrease with increasing ionization potentials.} 
\label{fig:wilson}
\end{center}
\end{figure}

Uncertainties in wavelength calibration and in the adopted laboratory wavelengths can introduce artificial structure in PV diagrams generated through mathematical operations involving multiple PV maps. Additionally, the faintness of the central star spectrum in Hf\,2-2 at longer wavelengths, and the extreme faintness of the central star in M\,1-42, may lead to artificial spatial structures due to uncertainties in stellar position measurements. For the contruction of emission-line PV diagrams, all lines were initially aligned to H$\alpha$. However, this method is not fully applicable to Hf\,2-2, due to a slight silt position offset during the second exposure with Dichroic 2 (see Section \ref{sec:correction}). For the Dichroic 2 data of Hf\,2-2, alignment was instead performed using H$\beta$. The velocity-direction alignment was based on the profiles of lines extracted from regions near the central star (for M\,1-42, the alignment relied on the overall 2D spectrum profiles due to weaker nebular emission near the central star), while spatial alignment followed line profiles along the slit. Although approximate (given kinematic variations between lines), this method corrected major misallignements and minimized errors. For operations involving multiple images requiring higher precision, further alignment was performed using emission lines from the same ion with near-identical structures, ensuring robustness.

\begin{figure}[ht!]
\begin{center}
\includegraphics[width=8.5 cm,angle=0]{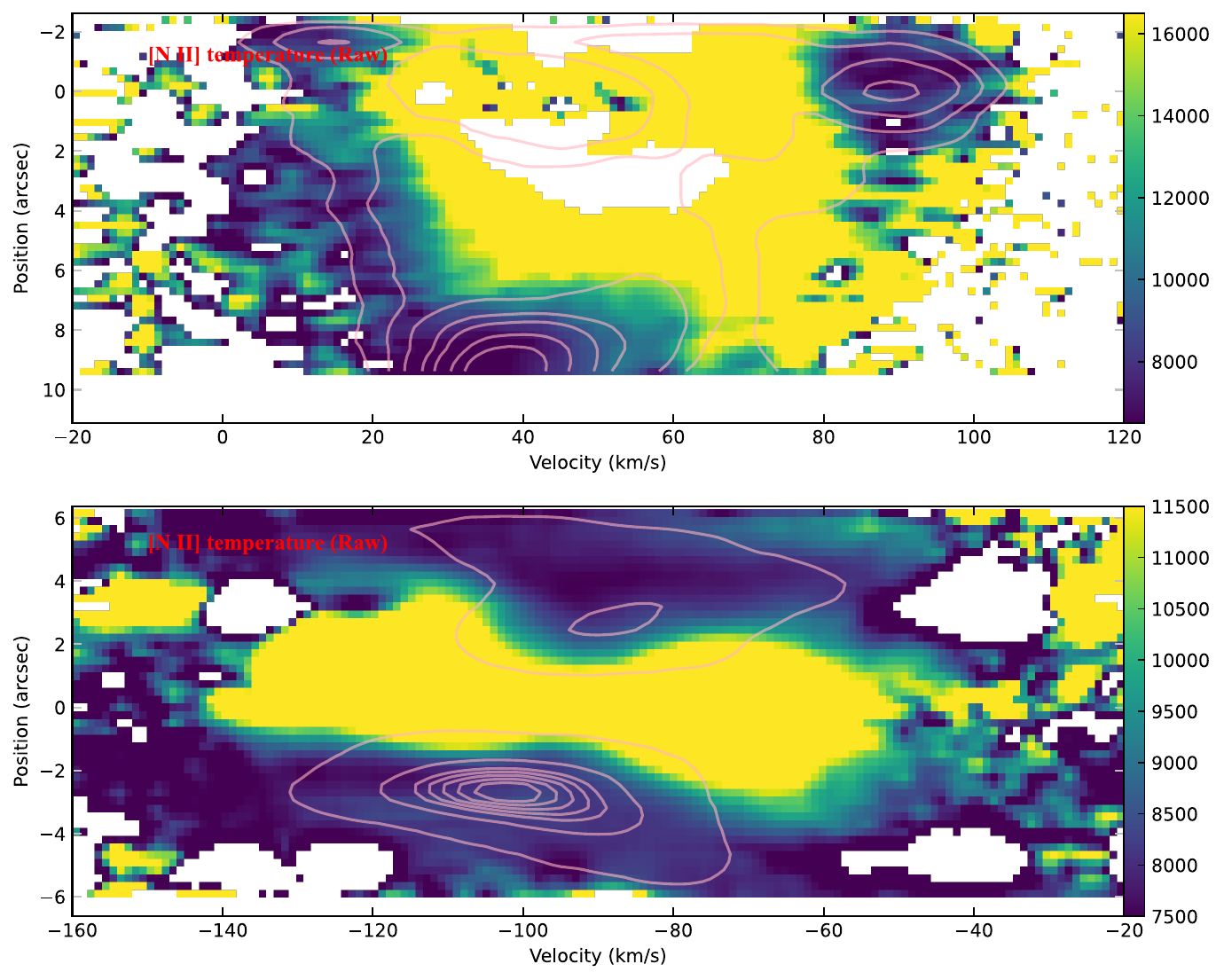}
\caption{[N~{\sc ii}] temperatures for Hf\,2-2 (upper panel) and M\,1-42 (bottom panel) calculated using the raw PV diagrams of [N~{\sc ii}] $\lambda$5755 and [N~{\sc ii}] $\lambda\lambda$6548,6583. The temperatures in the central part of the nebulae (where [N~{\sc ii}] nebular lines are weak) are extremely high, which is caused by the un-subtracted contribution from recombination excitation.} 
\label{fig:NII_temp_raw} 
\end{center}
\end{figure}

PV diagrams are affected by atmospheric dispersion. While shifts along the slit direction were corrected by aligning the central star position to the zero point, displacements perpendicular to the slit remain uncorrected. For Hf\,2-2 (observed at low airmass, $\sim$1.01$-$1.08), atmospheric refraction at Paranal\footnote{https://www.eso.org/gen-fac/pubs/astclim/lasilla/diffrefr.html} indicates minimal wavelength-dependent image shifts. In contrast, M\,1-42 was observed over a wider airmass range ($\sim$1.09$-$1.98), with a slit position angle of 120$^{\circ}$, aligning the direction of the atmospheric dispersion nearly parallel to the slit at large airmass (about 17$^{\circ}$ at maximum airmass). This limited the perpendicular shift to 0$\farcs$56 between 3500 and 5000 \AA. Although larger shifts are expected at bluer wavelengths, only a few lines in that range are used in the 2D analysis --especifically O~{\sc iii} $\lambda\lambda$3260, 3265 -- which are very weak and therefore have a negligible impact on the results. Overall, the effect of atmospheric dispersion in the both nebulae is small at most wavelengths. However, for precision, subsequent analyses prioritized spectral lines with similar wavelengths and simultaneous observations at identical airmass.

\subsection{Ionization Structure and Kinematics} 
\label{sec:structure}

PV maps of nebular emission lines trace the spatial and kinematic distribution of their emission regions. Comparing lines from species with different ionization potentials reveals the nebular ionization structure and kinematics. In high-ADF PNe, ORL-emitting gas is more centrally concentrated than the CEL-dominated main shell \citep[e.g.][]{2015ApJ...803...99C, 2016MNRAS.455.3263J, 2022MNRAS.510.5444G}, indicating distinct structural and kinematical properties between components. Figures \ref{fig:pv_cel} and \ref{fig:pv_orl} show high SNR PV maps of selected CELs and ORLs, respectively. Both figures arrange from low- to high-ionization species vertically (except He$^{+}$ and C$^{2+}$ in Figure \ref{fig:pv_orl}, where the former has a 0.2 eV higher ionization potential). 

Figure\,\ref{fig:pv_cel} presents PV maps of [O~{\sc i}] $\lambda$6300, [N~{\sc ii}] $\lambda$6583, [O~{\sc iii}] $\lambda$5007 and [Ar~{\sc iv}] $\lambda$4711 (from top to bottom), revealing distinct kinematics and ionization structures among CELs with different ionization potentials. In Hf\,2-2, neutral and singly ionized species are concentrated in two low-ionization structures (LISs). One LIS is spatially offset from the central star and exhibits a velocity differing by 10$-$20 km s$^{-1}$ from the systemic velocity. The other is located close to the central star along the line-of-sight, with a radial velocity offset of about 40 km s$^{-1}$ from the systemic velocity, about 20 km s$^{-1}$ beyond the main shell expansion velocity, exceeding the speed of sound in the ionized gas of PNe (about 17 km/s at $T_{\rm e}$ = 10000 K, \citet{2000oepn.book.....K}). This suggests that the enhanced [O~{\sc i}] and [N~{\sc ii}] emission in this region may be shock-excited. For M\,1-42, neutral and singly ionized species exhibit hyperbolic (or X-shaped) arcs, in contrast to the flattened, elliptical arcs seen in the highly ionized CELs. Similar hyperbolic kinematics in [N~{\sc ii}] $\lambda$6583 have been observed in other PNe, like NGC\,6537 (the ``Red" Spider Nebula), and can be reproduced by shock outflow model \citep{1995A&A...304..475C}.

The PV maps of [O~{\sc iii}] broadly follow the H~{\sc i} recombination line structures (see contours), although the emission peaks do not fully coincide, particularly in Hf\,2-2. The 2D spectral diagnostics presented in Section \ref{sec:2d_O3_temp} for Hf\,2-2 reveal elevated temperatures in regions where [O~{\sc iii}] emission dominates over H~{\sc i}, potentially due to the higher temperature at the center or shock excitation \citep{2000oepn.book.....K}. Higher-ionization species ([Ar~{\sc iv}]) are concentrated near the systemic velocity and the central star. This matches Hubble-type flow kinematics, where slower-moving, highly ionized gas resides closer to the nebular center.

Figure\,\ref{fig:pv_orl} displays PV maps of O~{\sc i}, He~{\sc i} $\lambda$5876, C~{\sc ii} $\lambda$6578, N~{\sc ii} $\lambda$5679, O~{\sc iii} $\lambda$4649, Ne~{\sc ii} $\lambda$3334, C~{\sc iii} $\lambda$4647 and He~{\sc ii} $\lambda$4686 (from top to bottom). He~{\sc i} $\lambda$5876 spatially coincides with the H~{\sc i} recombination emission in both PNe, while He~{\sc ii} $\lambda$4686 originates closer to the center, resembling CEL morphologies. In contrast, the behavior of heavy element ORLs is completely different. Although their parent ions span a wide range of ionization potentials from 13.6 eV for O$^{+}$ to 41.0 eV for Ne$^{2+}$ -- their PV diagrams exhibit remarkably similar spatiokinematical distributions. This stands in stark contrast to CELs, which cover a comparable ionization potential range (e.g., from [N~{\sc ii}], 14.5 eV for N$^{+}$, to [Ar~{\sc iv}], 40.7 eV for Ar$^{3+}$), but display remarkably different structures in their PV maps. These findings highlight the significantly different ionization structures traced by heavy-element ORLs and CELs.

\begin{figure*}[ht!]
\begin{center}
\includegraphics[width=18 cm,angle=0]{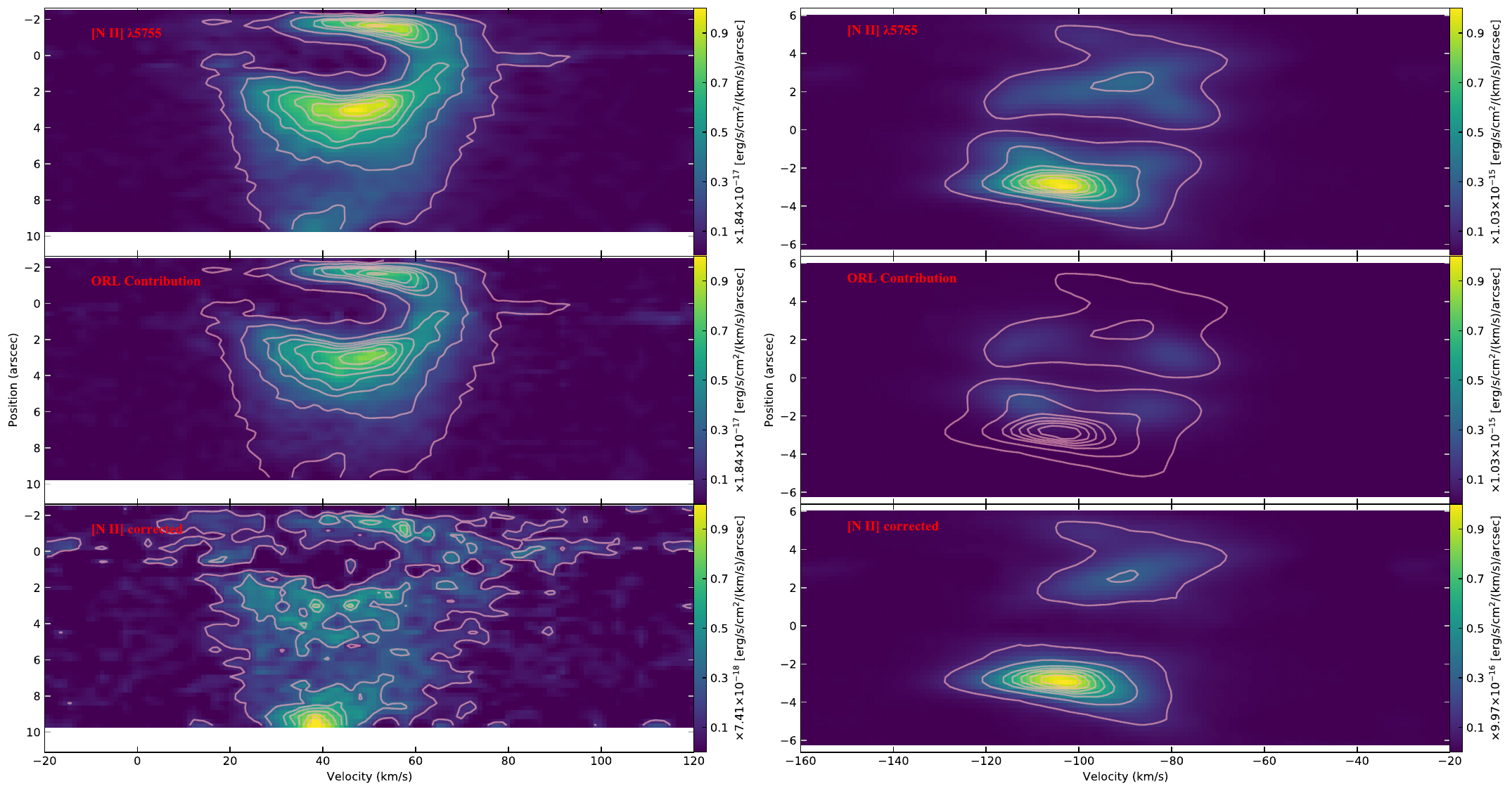}
\caption{Recombination excitation correction of [N~{\sc ii}] $\lambda$5755 for Hf\,2-2 (left) and M\,1-42 (right). The subplots from top to bottom are PV diagrams of [N~{\sc ii}] $\lambda$5755 lines, recombination excitation contributions and the PV diagrams of [N~{\sc ii}] $\lambda$5755 after corrections. The pink contours presents in the top and bottom subplots are the PV maps of [N~{\sc ii}] $\lambda$5755 lines, while in the bottom panels, the contours represent the PV maps of corrected [N~{\sc ii}] $\lambda$5755 intensities.} 
\label{fig:NII_corr} 
\end{center}
\end{figure*}

PV diagrams allow for a qualitative analysis of nebular structure and kinematics, though quantitative analysis remains challenging. To enable a more systematic comparison, we measured velocity splittings due to nebular expansion in the main shell and constructed Wilson diagrams (\citealt{1950ApJ...111..279W}; Figure \ref{fig:wilson}) by plotting these splittings against ionization potentials \citep[adopted from][]{kramida_nist_2024}. In Hf\,2-2, clear nebular emission is presented near the central star (Figure \ref{fig:pv_cel} and \ref{fig:pv_orl}), allowing reliable determination of velocity differences between the approaching and receding sides of the main nebular shell by measuring the double-peaks profiles extracted from regions near central star position. For M\,1-42, faint central emissions make such features barely discernible. In this case, we used the 1D spectra described in Section \ref{sec:extraction} and applied double-Gaussian fitting to estimate expansion velocities. When multiple lines were available for a single ion, those with higher-SNR lines were prioritized. The resulting mean velocity splittings and their standard deviations are shown in Figure \ref{fig:wilson}. However, some points lack uncertainties because they could not be quantified reliably. 

Wilson diagrams clearly reveal kinematic differences between CELs and ORLs. For most nebular CELs and ORLs (excluding O~{\sc i}, C~{\sc ii}, N~{\sc ii}, O~{\sc ii}, and Ne~{\sc ii}), velocity splittings decrease with increasing ionization potential -- consistent with standard PN ionization structure, in which faster-expending outer layers are less ionized. N~{\sc iii} and O~{\sc iii} Bowen fluorescence lines, energized by He~{\sc ii} Ly$\alpha$ photons, exhibit velocity splittings (plotted at He$^{2+}$’s ionization potential) that closely match those of He~{\sc ii}, indicating similar kinematics. Similarly, O~{\sc iii} charge exchange lines follow the sameahe velocity-ionization potential trend, reinforcing this correlation.

In contrast, O~{\sc i}, C~{\sc ii}, N~{\sc ii}, O~{\sc ii}, and Ne~{\sc ii} ORLs exhibit nearly flat velocity splitting trends with ionization potential, except for Ne$^{2+}$ in M\,1-42, which shows a smaller splitting than other ORLs. The splittings of these ORLs are smaller than those of CELs in Hf\,2-2, but larger in M\,1-42. This aligns with the PV diagrams of M\,1-42, which show smoother velocity-direction continuity in CELs compared to ORLs. Notably, Both nebulae converge at Ne$^{2+}$ ionization potential, where ORL and CEL velocity splittings match. 

Auroral lines with significant recombination contributions, such as [O~{\sc ii}] $\lambda\lambda$7320,7330, [N~{\sc ii}] $\lambda$5755, and [O~{\sc iii}] $\lambda$4363 in M\,1-42, show velocity splittings that differ from those of their nebular counterparts. In Hf\,2-2, the [O~{\sc iii}] auroral line also shows smaller velocity splitting than the nebular [O~{\sc iii}] lines, despite the absence of detected O~{\sc iii} ORLs. This can be explained by the presence of higher temperatures in central regions, which favor excitation to higher energy levels (producing auroral lines), while the slower expansion in these regions leads to smaller velocity splittings. Moreover, auroral lines of third-period elements are intrinsically weaker, leading to less accurate velocity splitting measurements. Both PV maps and velocity splitting quantifications consistently indicate distinct ionization structures and kinematics between ORL- and CEL-emitting regions.

\subsection{Plasma Diagnostics} 
\label{sec:PD}

\begin{figure*}[ht!]
\begin{center}
\includegraphics[width=18 cm,angle=0]{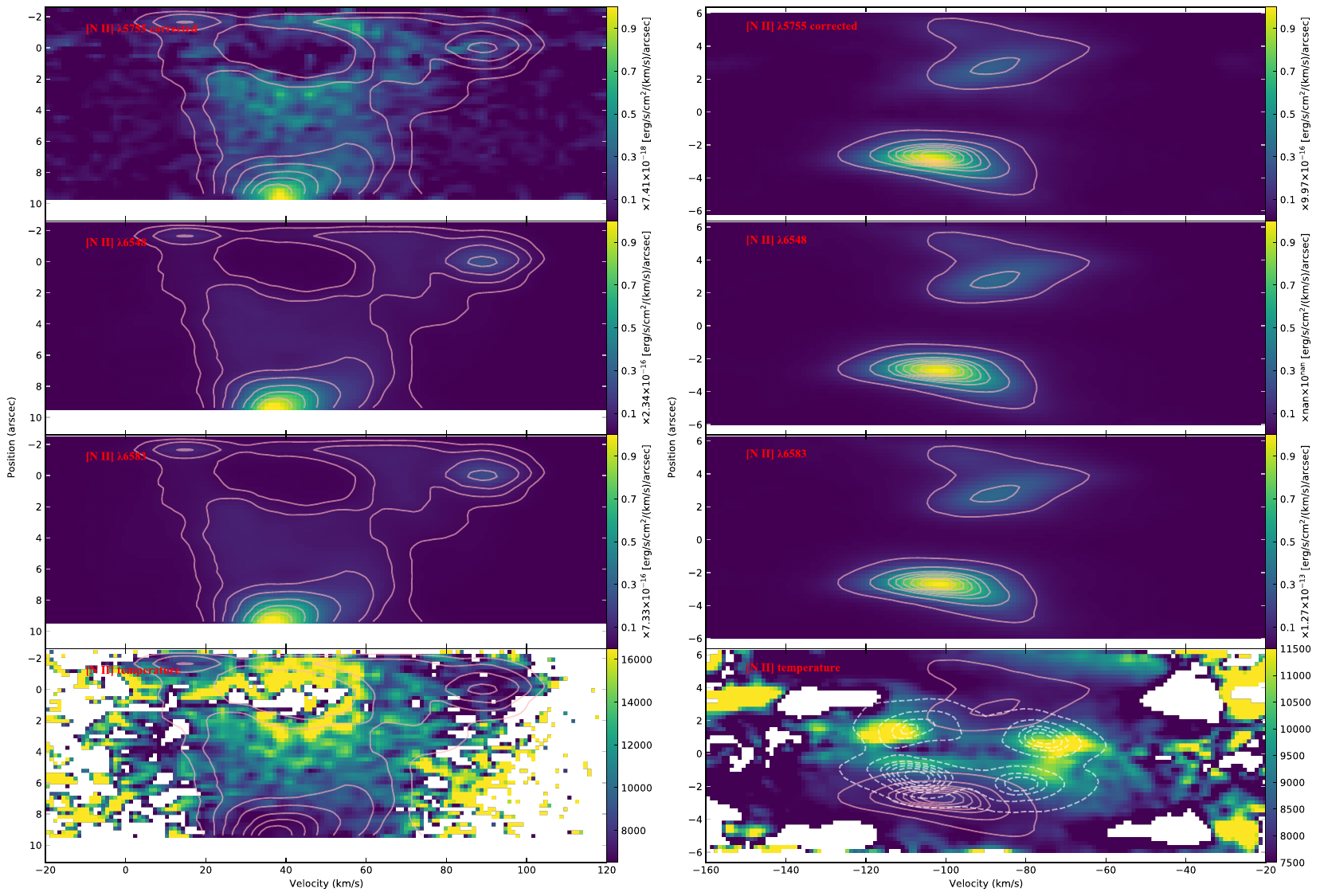}
\caption{PV maps of [N~{\sc ii}] temperatures for Hf\,2-2 (left) and M\,1-42 (right). The subplots from top to bottom are PV diagrams of corrected [N~{\sc ii}] $\lambda$5755, observed [N~{\sc ii}] $\lambda\lambda$6548,6583 lines and [N~{\sc ii}] temperatures. The pink contours represent the PV maps of [N~{\sc ii}] $\lambda$6548. In the bottom panel of M\,1-42, we also present the PV map of N~{\sc ii} $\lambda$5679 with white dashed contours to illustrate that the region with apparent high temperature result from to the the imperfect subtraction of the recombination contribution (see text).} 
\label{fig:NII_temp} 
\end{center}
\end{figure*}

Plasma diagnostics serve as an effective tool for probing nebular physical conditions. Section \ref{sec:diagnostics} demonstrated distinct physical conditions in ORL- and CEL-emitting regions through 1D spectral analysis, with ORL-emitting regions being colder and denser. However, 1D spectra only provide spatially averaged results, masking spatial and kinematic variations. To resolve these details, we implemented 2D spectral diagnostics, generating PV maps of electron temperature and density distributions.

\subsubsection{The [N~{\sc ii}] Temperature}

[N~{\sc ii}] temperature diagrams were derived using [N~{\sc ii}] $\lambda$5755 and $\lambda\lambda$6548,6583 lines. Plasma diagnostics require [N~{\sc ii}] line with pure collisionally excitation, and recombination contributions (\citealt{1986ApJ...309..334R}) needs to be accounted for. Figure \ref{fig:NII_temp_raw} shows [N~{\sc ii}] temperatures from raw CEL PV diagrams, revealing abnormally high central values; a similar effect is clearly seen in the temperature maps of \citet{2022MNRAS.510.5444G} (their Figure 8). While recombination effects are negligible for strong nebular lines, the faint $\lambda$5755 auroral line required recombination correction to ensure accuracy.

Warm plasmas also produce ORLs, though these contribute minimally in high-ADF PNe. Following \citet{2022AJ....164..243R}, we estimated N~{\sc ii} $\lambda$5679 intensities from warm plasmas using [O~{\sc iii}] $\lambda$5007 fluxes, emissivities at [O~{\sc iii}] temperatures, and N$^{2+}$/O$^{2+}$ ratios. For Hf\,2-2 and M\,1-42, the intensity ratios of N~{\sc ii} $\lambda$5679 from warm and cold plasmas are about 0.01 and 0.05, respectively, which means that warm plasma contributions in ORL emission are negligible. Thus, the [N~{\sc ii}] $\lambda$5755 recombination corrections in Figure \ref{fig:NII_corr} were applied directly from the N~{\sc ii} $\lambda$5679 PV diagrams, not split into component-wise corrections as in \citet{2022AJ....164..243R}. For consistency with previous ionic abundance calculations, 1D-derived N~{\sc ii} temperatures were applied for corrections. Final [N~{\sc ii}] temperature PV maps are shown in Figure \ref{fig:NII_temp}.

\begin{figure}[ht!]
\begin{center}
\includegraphics[width=8.5 cm,angle=0]{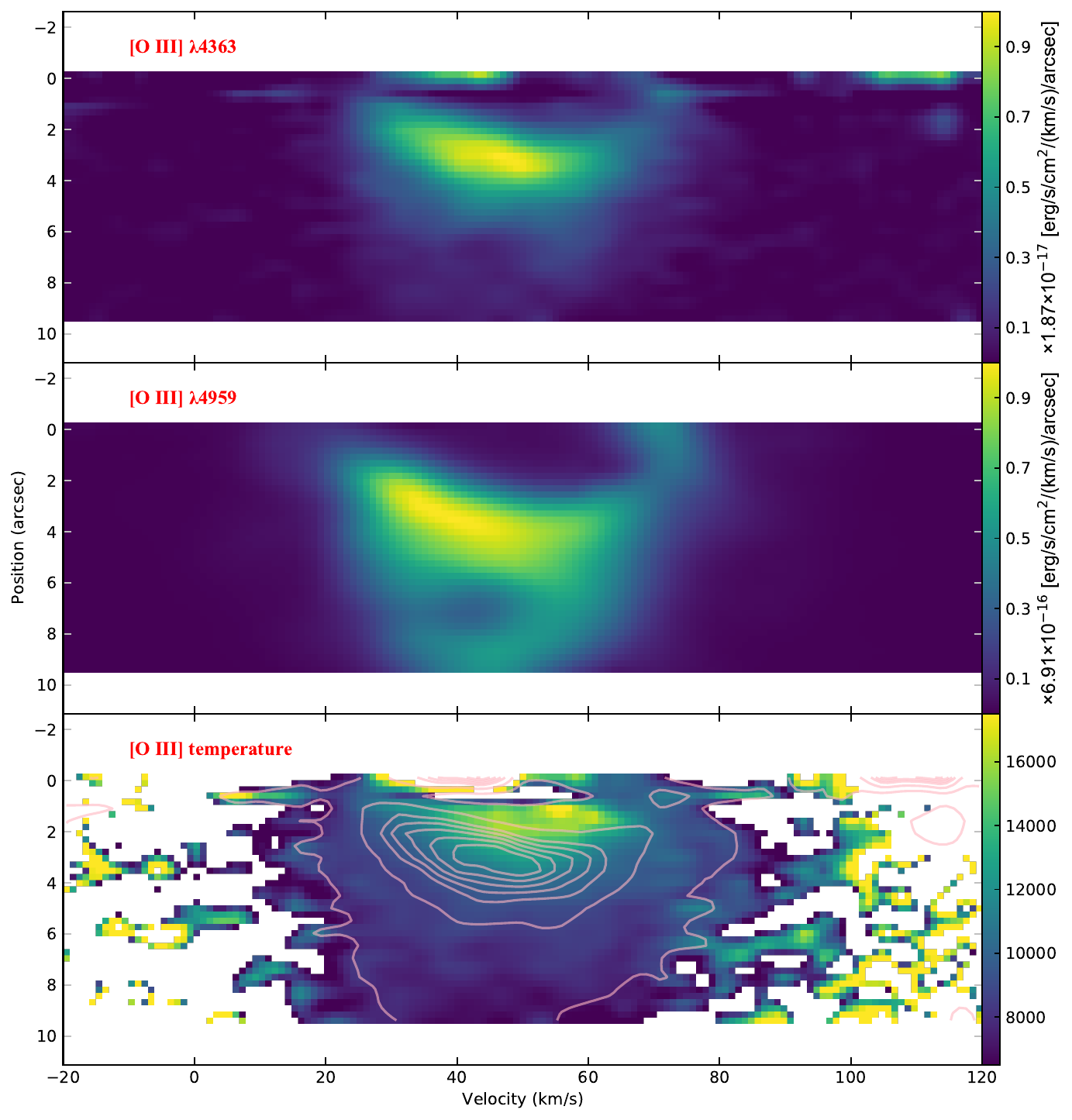}
\caption{Determination of the [O~{\sc iii}] temperature for Hf\,2-2. From top to bottom: PV diagrams of [O~{\sc iii}] $\lambda$4363, [O~{\sc iii}] $\lambda$4959, and the [O~{\sc iii}] temperature. The contours in the bottom panel correspond to  the PV diagram of the [O~{\sc iii}] auroral line. The electron temperatures are higher near the central star, and also appear to be higher at radial velocities close to the systemic velocity. which is similar, though not identical, to the [N~{\sc ii}] temperature distribution for the same target (Figure\,\ref{fig:NII_temp}).} 
\label{fig:Hf22_OIII} 
\end{center}
\end{figure}

For Hf\,2-2, we applied a $T_{\rm e}$(N~{\sc ii}) = 2500 K obtained from the 1D analysis (see Section \ref{sec:orl_diagnostics} and Table \ref{temden}) to correct recombination contributions. As shown in the left panels of Figure\,\ref{fig:NII_corr}, [N~{\sc ii}] $\lambda$5755 is dominated by recombination processes. After correction, the primary emission region of this line shifts from nebular center to the periphery -- a smiliar behavior was found in the MUSE data analysis in \citet{2022MNRAS.510.5444G}. The PV map of the auroral line also becomes morphologically consistent with the [N~{\sc ii}] nebular lines (Figure\,\ref{fig:NII_temp}). Using the corrected [N~{\sc ii}] auroral line, we constructed the [N~{\sc ii}] temperature PV map for Hf\,2-2. Higher temperatures are found in structures spatially close to the central star and near the systemic velocity, although weak auroral line intensities at the center reduce reliability in this region. Temperatures in the 2$^{\prime\prime}-$8$^{\prime\prime}$ spatial range (corresponding to the extraction zone of 1D spectra) are $\sim$10,000 K, consistent with 1D results. The LISs show [N~{\sc ii}] temperatures $<$8000 K, lower than the main nebular shell. While statistical studies of LISs by \citet{2023MNRAS.525.1998M} indicate no systematic [N~{\sc ii}] temperature difference between LISs and Rims/Shells, Hf\,2-2 diverges from this trend but aligns with lower-temperature in LISs reported in several objects in \citet{2023MNRAS.525.1998M}, such as NGC\,7009.

For M\,1-42, the raw [N~{\sc ii}] $\lambda$5755 intensity distribution shown in the upper right panel of Figure \ref{fig:NII_corr}, clearly reflects contributions from both collisionally excitation and recombination. Recombination corrections were applied using the N~{\sc ii} temperature (2150 K) derived from the 1D analysis for this source (see Section \ref{sec:orl_diagnostics} and Table \ref{temden}), which effectively removed the elliptical-shaped recombination component, leaving behind only hyperbolic structures similar to those seen in the [N~{\sc ii}] nebular lines. Due to the weakness of [N~{\sc ii}] CEL emission near the center, small residual recombination contributions can result in spuriously high temperatures in this region. The bottom-right panel of Figure \ref{fig:NII_temp} shows that the morphology of high-temperature zones resembles the N~{\sc ii} $\lambda$5679 intensity distribution, suggesting incomplete recombination subtraction. Within the pink contours, the average temperature is $\sim$8500 K, slightly lower than the 1D spectral results. This discrepancy arises because the centrally concentrated 1D extraction region retains residual recombination emission, raising the derived temperatures.

Recombination corrections significantly improve [N~{\sc ii}] temperature determinations, particularly by effectively removing spurious-high temperatures caused by recombination processes near the nebular center. This underscores the necessity of subtracting recombination contributions from faint auroral lines to obtain reliable [N~{\sc ii}] temperatures, whether for determining average values across the nebula or for probing localized temperatures. The importance of a proper recombination contribution correction in high-ADF PNe was also highlighted by other studies \citep[e.g.][]{2022AJ....164..243R, 2022MNRAS.510.5444G, 2024AA...689A.228G}. However, even after applying these corrections, spurious high-temperature regions persist near the nebular centers. While a genuinely warmer central zone remains plausible, incomplete subtraction of the recombination component is the more likely explanation. Large uncertainties in the ORL temperatures applied for recombination corrections likely contribute to this residual effect.

\subsubsection{The [O~{\sc iii}] Temperature}
\label{sec:2d_O3_temp}

The [O~{\sc iii}] temperature PV maps were built using the [O~{\sc iii}] $\lambda$4363 and $\lambda$4959 lines, adopting the electron density given by  [S~{\sc ii}] line ratios in Table \ref{temden}. Temperature determination assumes that [O~{\sc iii}] emission arises purely from collisionally excitation. As no charge-exchange or pure recombination O~{\sc iii} features were detected in Hf\,2-2, no correction war applied for this object. The uncorrected [O~{\sc iii}] auroral and nebular line PV maps (Figure \ref{fig:Hf22_OIII}) reveal higher central temperatures where the auroral line emission is concentrated near the central star. Elevated temperatures are also observed at radial velocities approaching the systemic velocity. In the outer nebular regions (1D spectral extraction area), temperatures range 8000$-$10,000 K, consistent with the result in Table \ref{temden}.

The case of M\,1-42 is different. In this object, the detection of O~{\sc iii} lines dominated by charge-exchange or recombination processes implies that [O~{\sc iii}] lines arise from mixed excitation mechanisms. To isolate collisional excitation component in the PV maps, we applied corrections for non-collisional processes contributions using a standard PN temperature of 10,000 K. The methods for subtracting non-collisional processes are detailed separately below.

\begin{figure*}[ht!]
\begin{center}
\includegraphics[width=18 cm,angle=0]{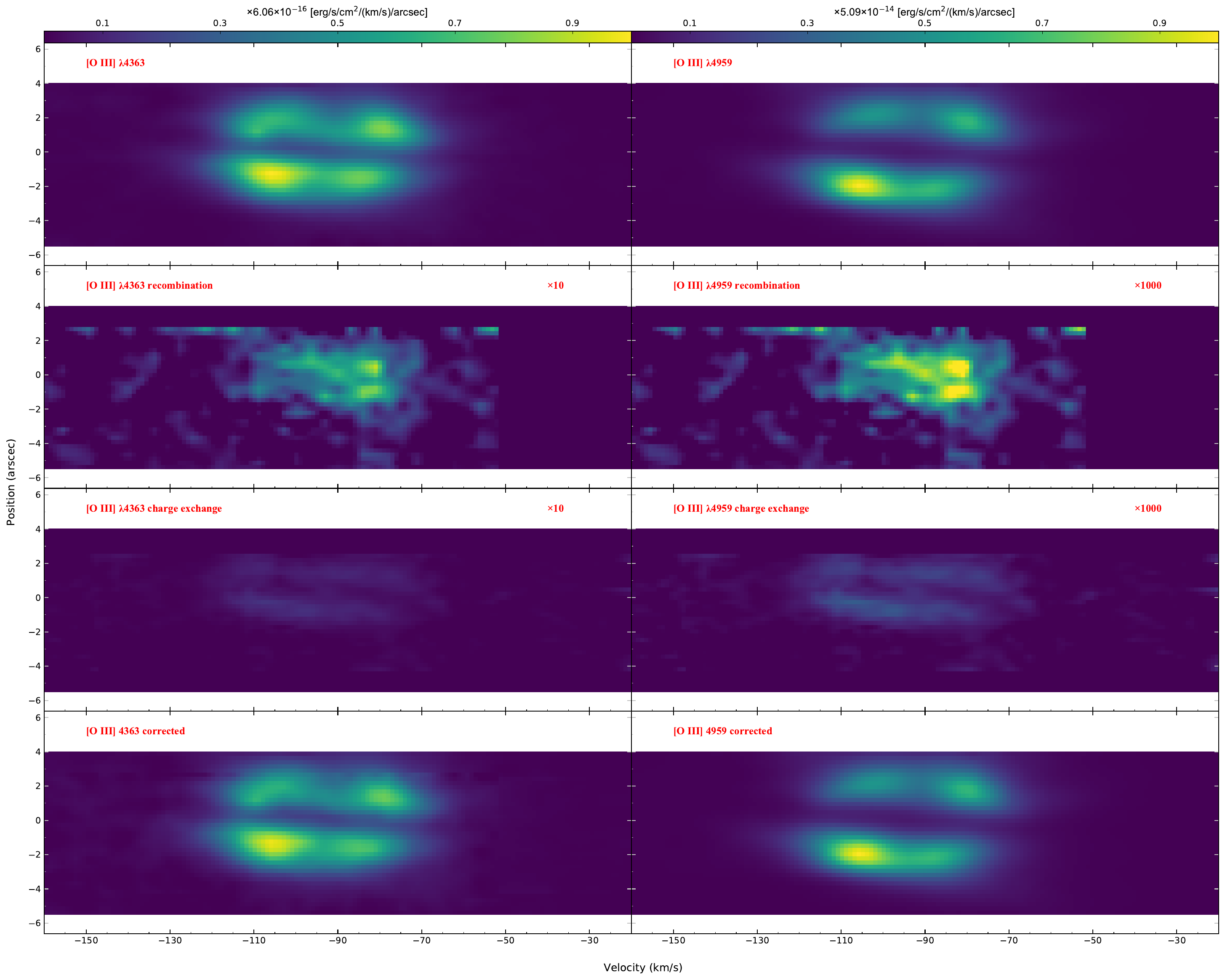}
\caption{Correction of contribution of recombination and charge-exchange processes from [O~{\sc iii}] $\lambda$4363 (left) and [O~{\sc iii}] $\lambda$4959 (right) lines for M\,1-42. From top to bottom are the original PV diagrams, the PV diagrams of recombination and charge-exchange contribution, and the PV diagrams after correction. Contribution of recombination and charge exchange processes in the auroral and nebular lines were magnified 10 and 1,000 times, respectively, to ensure that the same error bar scale could be used. See text for details on the horizontal and vertical truncations in the panels in the second and third rows.} 
\label{fig:M142_OIII_correction} 
\end{center}
\end{figure*}

To estimate recombination excitation contributions, we selected the pure recombination O~{\sc iii} $\lambda$3265 line as a template. Recombination contributions to the [O~{\sc iii}] CELs were calculated using the O~{\sc iii} recombination coefficients from \citet{1991A&A...251..680P}. There are very few O~{\sc iii} lines with available atomic data, and the only lines known to be unaffected by no fluorescence are O~{\sc iii} $\lambda\lambda$3260,3265. However, these lines are at the bluest end of the spectrum, where the SNR is low, thus the stronger line, O~{\sc iii} $\lambda$3265, was selected. Additionally, the spectral dispersion is more significantly distorted perpendicular to the slit at short wavelengths. These factors reduce the reliability of the recombination correction. 

For the charge-exchange correction, only transitions populating the $^1$S and $^1$D upper levels contribute to [O~{\sc iii}] CEL intensities, with O~{\sc iii} $\lambda$5592 being the only line meeting this criterion in UVES spectra. In M\,1-42, recombination contributes $\sim$10\% of the O~{\sc iii} $\lambda$5592 intensity, base on the atomic data from \citet{1991A&A...251..680P}, making charge-exchange dominant, and recombination contributions to this line were neglected. Using O$^{2+}$ transition probabilities from \citet{1982ApJ...257L..87D}, we calculated the intensity ratio between $\lambda$5592 and the [O~{\sc iii}] lines arising from charge exchange process.

\begin{figure}[ht!]
\begin{center}
\includegraphics[width=8.5 cm,angle=0]{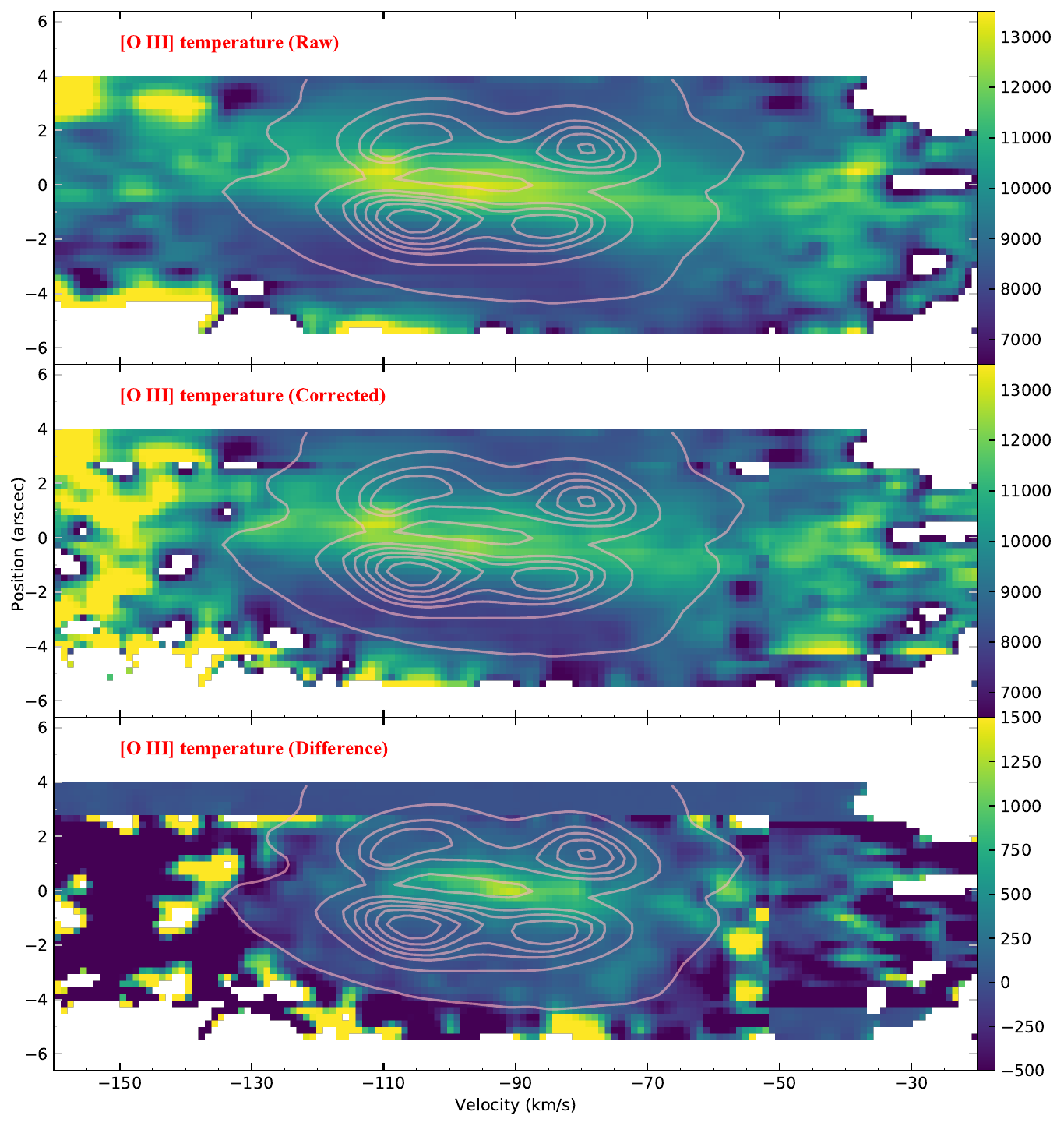}
\caption{PV diagrams of [O~{\sc iii}] temperatures for M\,1-42 derived using the original [O~{\sc iii}] lines (upper panel), recombination and charge-exchange corrected [O~{\sc iii}] lines (middle panel), and their differences (bottom panel). Contours represent the PV diagram of the corrected [O~{\sc iii}] $\lambda$4363 line. The horizontal structures at about 3$^{\prime\prime}$ and -4$^{\prime\prime}$, and the vertical structure at about $-$50 km s$^{-1}$ in the figure, are consistent with those in Figure \ref{fig:M142_OIII_correction}, as detailed in the text. Regardless of correction, the nebular center remains hotter, with temperatures gradually decreasing outward. The temperature is more uniform along the velocity direction. The correction resulted in a temperature reduction of at least 1000 K at the center but only a few hundred Kelvin or less at the outer regions.} 
\label{fig:M142_OIII} 
\end{center}
\end{figure}

Recombination and charge-exchange corrections for the [O~{\sc iii}] lines in M\,1-42 are shown in Figure \ref{fig:M142_OIII_correction}. The position of the central star in the slit is slightly different between CD\#1 and CD\#2 spectra. Thus, the spatial coverage of O~{\sc iii} $\lambda$3265 from CD\#1 spectra differ from the [O~{\sc iii}] lines from CD\#2. Pixels corresponding to spatial regions covered in CD\#2 spectra but not in CD\#1 spectra are set to 0, introducing artificial structures in the PV diagrams of the recombination contribution presented in the second row in Figure \ref{fig:M142_OIII_correction}. The O~{\sc iii} $\lambda$3265 is also affected by skylines and bad pixels, so pixels with velocities $v > -50$ km s$^{-1}$ were set to 0 to avoid contamination, which also introduce artifacts.

\begin{figure}[ht!]
\begin{center}
\includegraphics[width=8.5 cm,angle=0]{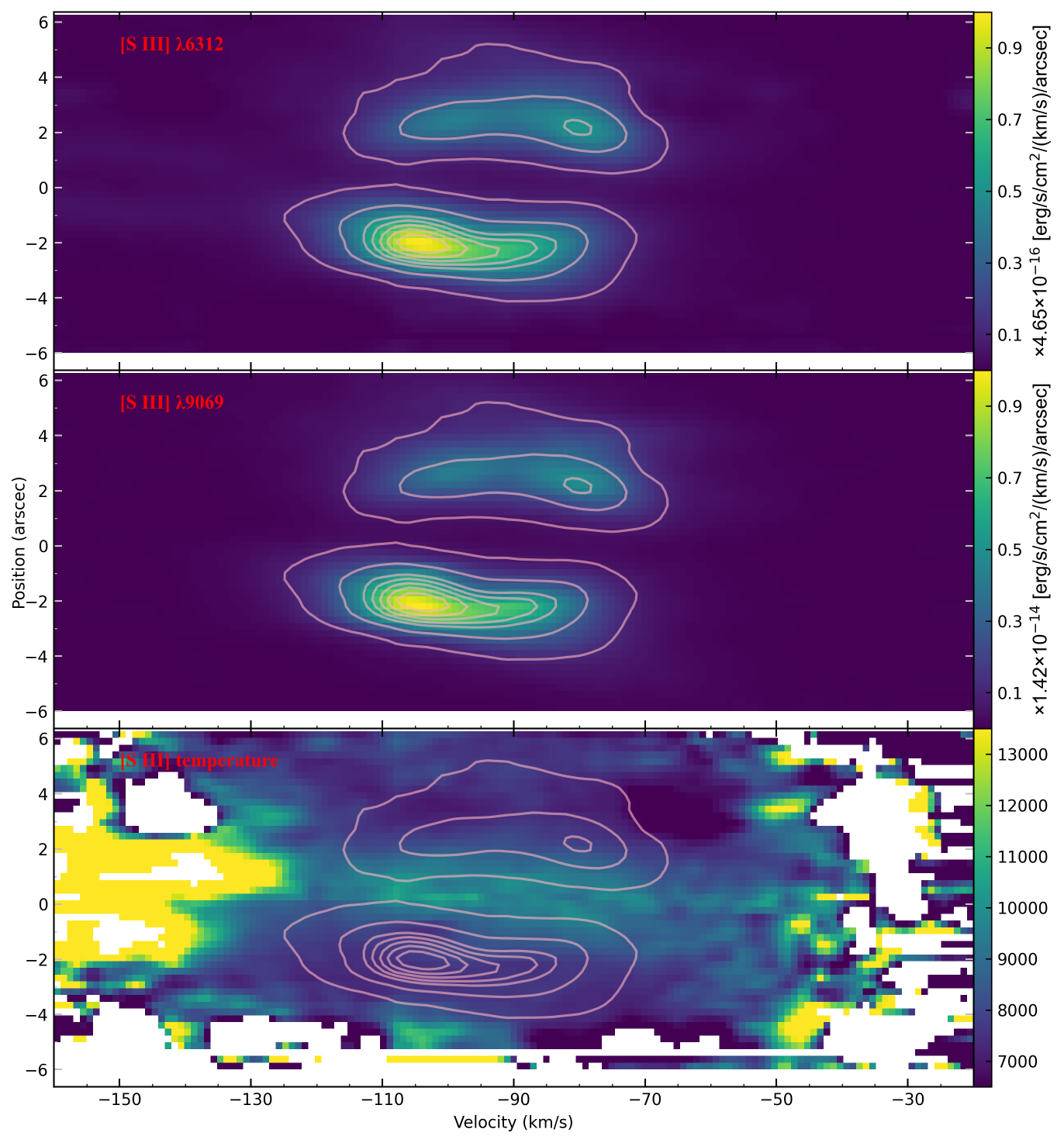}
\caption{PV diagram of the [S~{\sc iii}] $\lambda$6312 (upper panel) and $\lambda$9069 (middle panel) lines in M\,1-42 (bottom panel), and the [S~{\sc iii}] temperature. Contours represent the PV diagram of the [S~{\sc iii}] $\lambda$6312 line. The vertical structures between $-$110 and $-$90 km s$^{-1}$ are residuals from the correction of telluric emission blended with the auroral line. As in the [O~{\sc iii}] temperature of M\,1-42 presented in Figure \ref{fig:M142_OIII}, electron temperatures are higher near the central star.} 
\label{fig:M142_SIII} 
\end{center}
\end{figure}

Field stars are visible in the PV maps of O~{\sc iii} $\lambda$5592. Fortunately, this charge-exchange line mainly appears in the central region of the nebula, and emission from the field stars has negligible impact on this line. The spatial range containing field stars was masked, though two other artificial structures remain in the PV diagrams of the charge-exchange contribution along the velocity axis. These artifacts in the PV diagrams introduce minor artifacts in the temperature diagram. Both processes contribute less than 10\% to the auroral line intensity, and less than one-thousandth of the intensity of the nebular line, which is almost negligible.

The PV diagrams of [O~{\sc iii}] temperature in M\,1-42 are presented in Figure\,\ref{fig:M142_OIII}. Although the contributions of recombination and charge-exchange to the [O~{\sc iii}] lines are relatively small (as shown in Figure \ref{fig:M142_OIII_correction}), they can not be entirely neglected, especially in the region near the central star, where [O~{\sc iii}] emission is weak. The electron temperature remains higher the central star, decreasing outward. In the velocity direction, temperatures remain uniform. The differences between the PV maps before and after applying the corrections are more pronounced near the nebular center, where the temperature decreases by $\sim$1000 K. In contrast, other regions show minimal change. The contributions from recombination and charge-exchange to the [O~{\sc iii}] CEL fluxes, while modest, remain non-negligible, particularly in nebular cavities where CEL emission is the weakest.

\subsubsection{Temperatures from Other CEL Diagnostics}

\begin{figure}[ht!]
\begin{center}
\includegraphics[width=8.68 cm,angle=0]{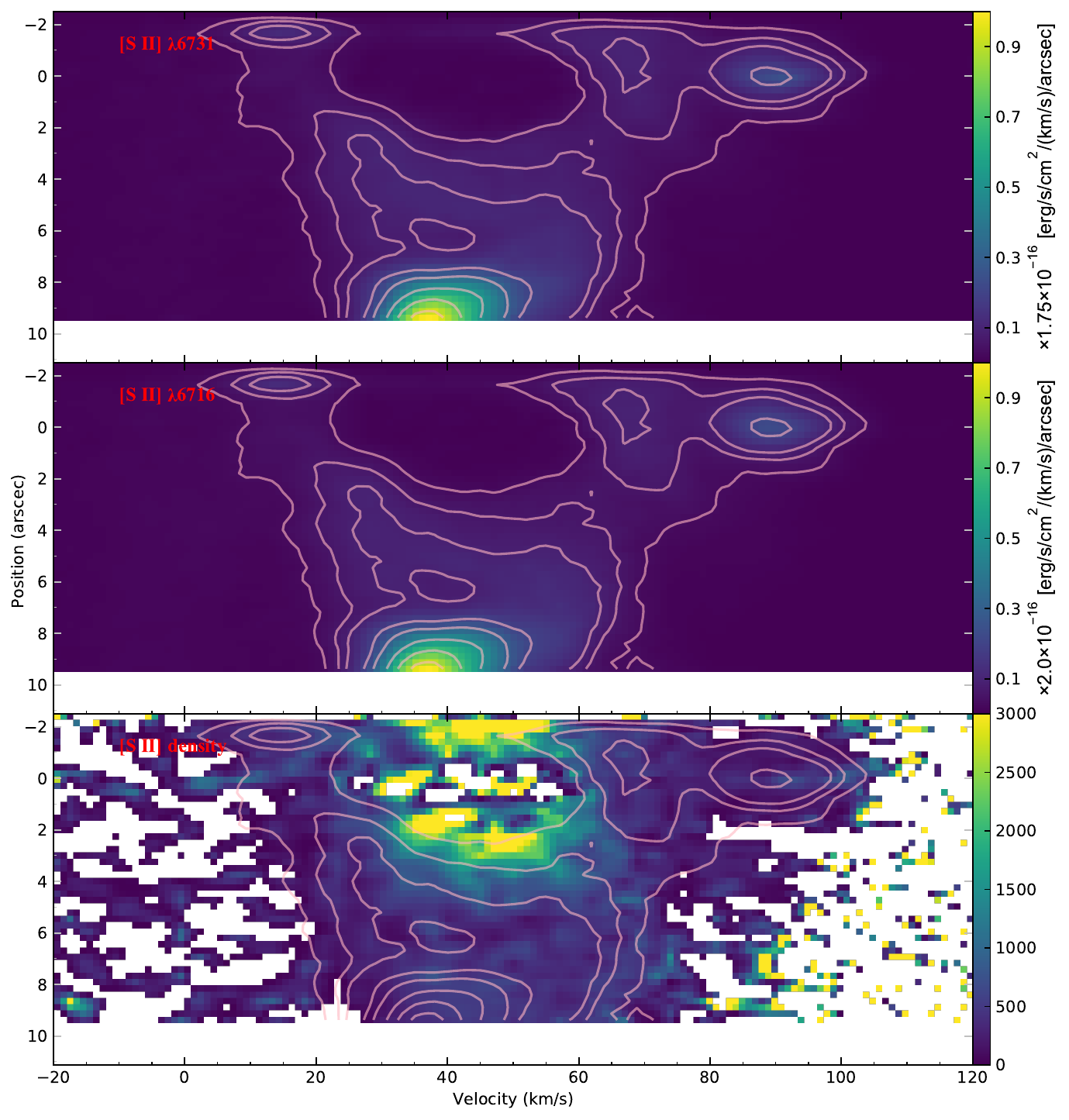}
\caption{Determination of the [S~{\sc ii}] electron density PV diagram for Hf\,2-2 (bottom panel) based on [S~{\sc ii}] $\lambda\lambda$6716,6731 (middle and upper panel respectively) lines. Contours represent the PV diagram of the [S~{\sc ii}] $\lambda$6731 line. Densities are about 500 cm$^{-3}$ in the pixels where [S~{\sc ii}] line emission is strong.} 
\label{fig:Hf_SII} 
\end{center}
\end{figure}

PV diagrams of CELs from third-row elements were also examined for temperature diagnostics. Two diagnostics were considered: [S~{\sc iii}] and [Ar~{\sc iii}] diagnostics. The [Ar~{\sc iii}] auroral line suffers from very low SNR -- \citet{2022AJ....164..243R} report a noisy [Ar~{\sc iii}] temperature PV diagram for NGC\,6153 using the same dataset -- and is undetected in Hf\,2-2 and only marginally detected in M\,1-42 which is noisy. Hence we do not present any [Ar~{\sc iii}] temperature maps in this paper. In contrast, the [S~{\sc iii}] diagnostic is more robust. Recombination corrections could not be applied for [S~{\sc iii}] lines because the relevant recombination lines are too weak and the necessary atomic data are lacking. Moreover, source-to-source variations in systemic velocity further limit the applicability of the [S~{\sc iii}] diagnostics. Therefore, a reliable [S~{\sc iii}] temperature PV map could be obtained only for M\,1-42.

The [S~{\sc iii}] $\lambda\lambda$9069,9531 lines, which theoretically exhibit an intensity ratio $I$(9531)/$I$(9069) $\approx$ 2.47 due to their shared upper-level transitions, consistently show lower observed ratios in the 1D spectra and divergent PV diagram distributions. Since these [S~{\sc iii}] nebular lines fall within a the wavelength range heavily affected by telluric emission and absorption, flux measurement accuracy is compromised. However, the differences in radial velocities of our targets allow us to verify where the telluric lines exist. While [S~{\sc iii}] $\lambda$9531 is blended with telluric absorption in all PNe, [S~{\sc iii}] $\lambda$9069 is affected by telluric features in Hf\,2-2 but remains uncontaminated in M\,1-42.

\begin{figure}[ht!]
\begin{center}
\includegraphics[width=8.5 cm,angle=0]{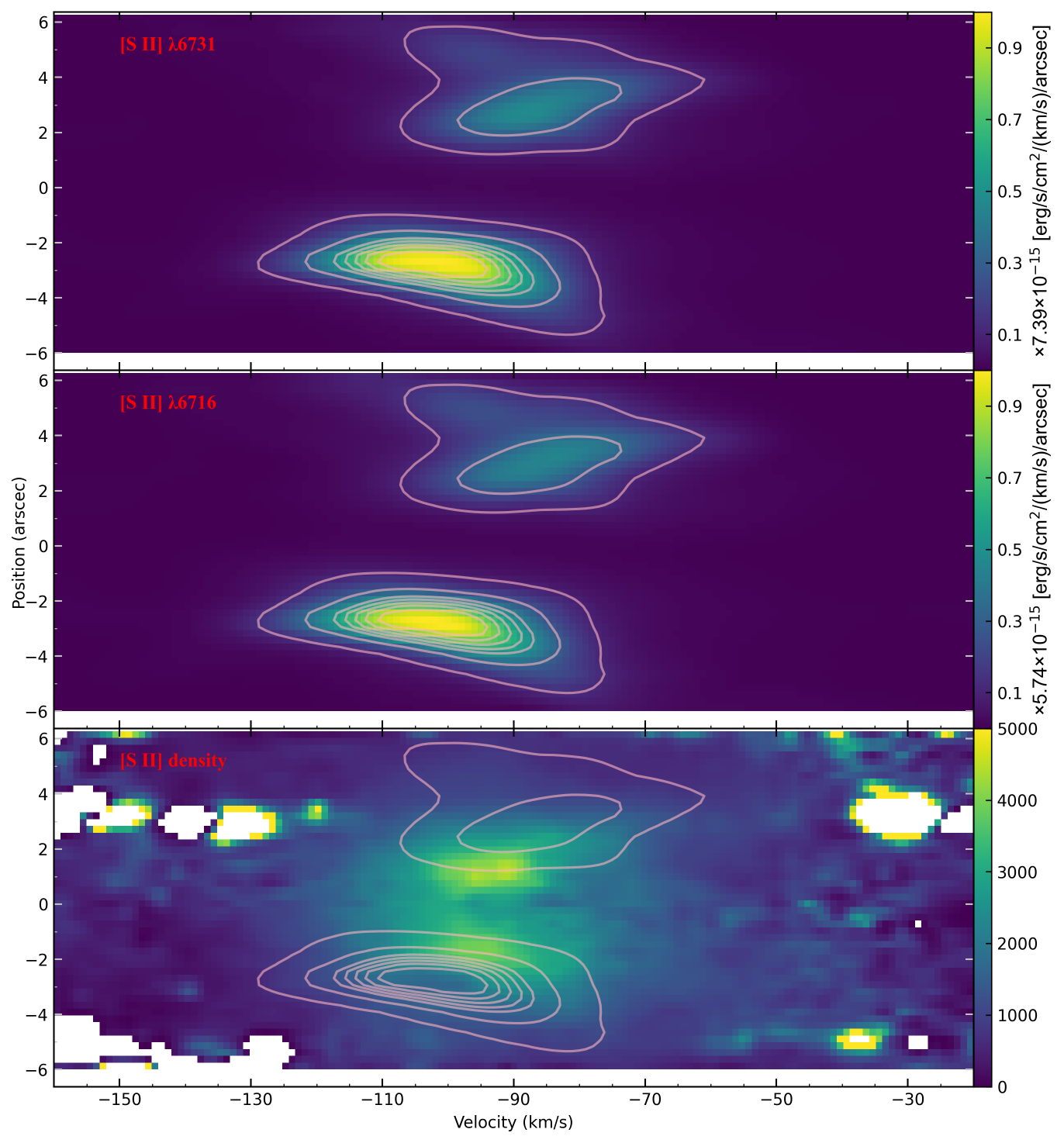}
\caption{Same as Figure \ref{fig:Hf_SII} but for M\,1-42. The main [S~{\sc ii}] emission region exhibits a hyperbolic shape. Densities decrease from the center toward the outer region. This behavior contrasts with the [Cl~{\sc iii}] and [Ar~{\sc iv}] density PV diagrams shown in Figure\,\ref{fig:M142_Cl3} and \ref{fig:M142_Ar4}, which are more uniform in the spatial direction. This difference may also originate from recombination contribution (see text).} 
\label{fig:M142_SII} 
\end{center}
\end{figure}

The [S~{\sc iii}] $\lambda$6312 line in M\,1-42 blends with a telluric emission line, which clearly affects the flux measurements due to the auroral line’s low intensity. We corrected telluric contributions by subtracting median sky values derived from non-nebular emission pixels at each wavelength. However, residual artifacts remain in Figure \ref{fig:M142_SIII} due to skyline intensity variations along the slit. The average [S~{\sc iii}] temperatures in the main shell is $\sim$8000 K, lower than both [O~{\sc iii}] temperatures and the 1D spectral results (the latter affected by telluric absorption in [S~{\sc iii}] $\lambda$9531, which was excluded from the 2D analysis). The temperature structures  resembles that of [O~{\sc iii}], with higher values in the inner cavities and lower values in the main shell, and the distribution along the velocity axis remain uniform.

\subsubsection{CEL Density Diagnositcs}

Electron density PV maps derived from [S~{\sc ii}], [Cl~{\sc iii}], and [Ar~{\sc iv}] diagnostics face observational limitations across the targets. The low ionization state and relatively low surface brightness of Hf\,2-2, result in insufficient SNR for both [Cl~{\sc iii}] and [Ar~{\sc iv}] lines, preventing the construction of reliable PV maps. Therefore, only the [S~{\sc ii}] density map is presented for Hf\,2-2. Additionally, residuals from the bright central star of Hf\,2-2 contaminate the density measurements in the nebular center, where the nebular emission is weak.

\begin{figure}[ht!]
\begin{center}
\includegraphics[width=8.5 cm,angle=0]{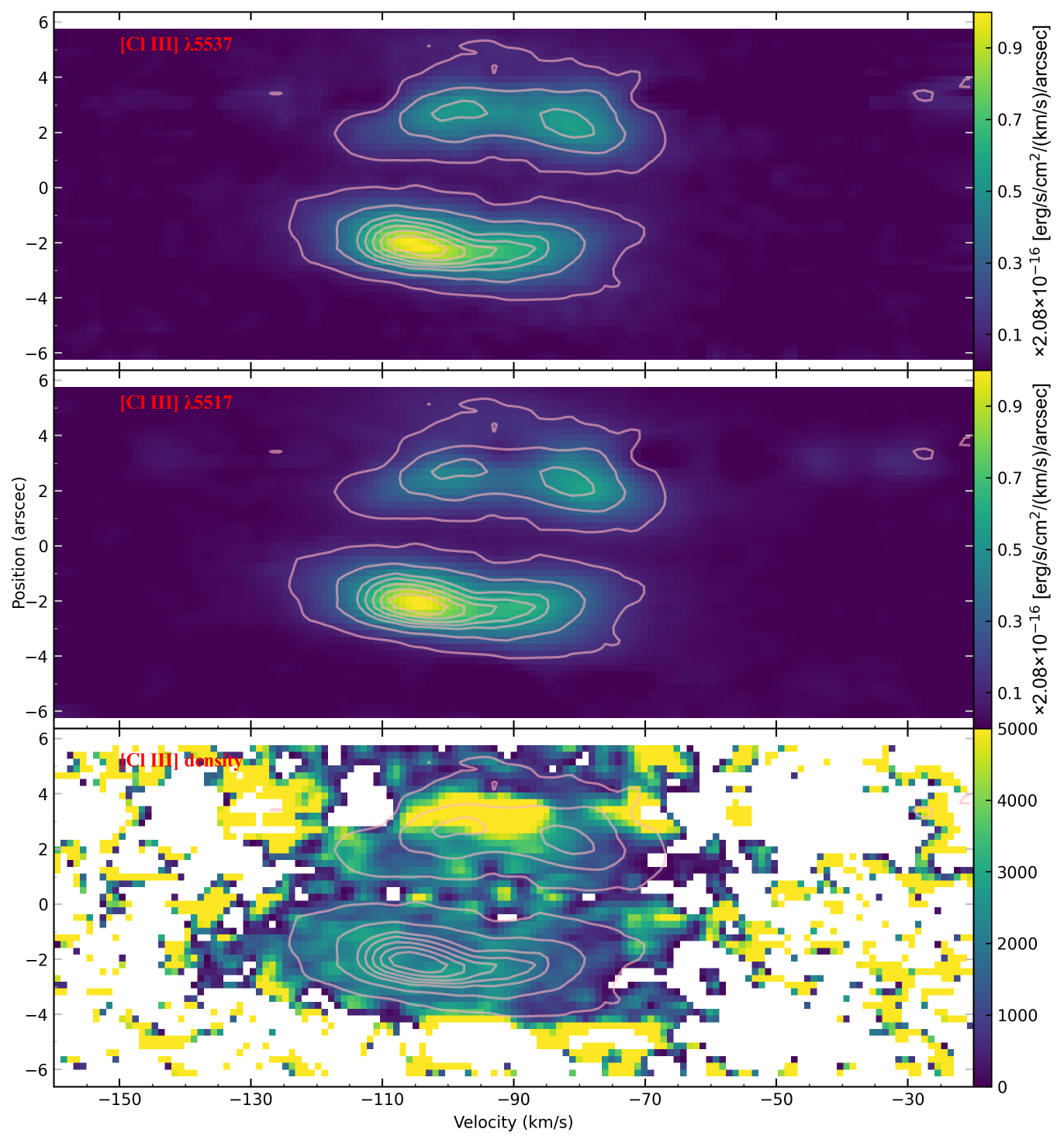}
\caption{PV diagrams of [Cl~{\sc iii}] $\lambda\lambda$5517,5537 (middle and upper panel respectively) and the derived [Cl~{\sc iii}] density of M\,1-42. Contours represent the PV diagrams of [Cl~{\sc iii}] $\lambda$5537 line. The high density structure in the bottom panel, at velocity coordinates of about -100 km s$^{-1}$ and position coordinates of about 3$^{\prime\prime}$, results from residuals caused by imperfect subtraction of the continuum of the field star, which contains numerous absorption lines.} 
\label{fig:M142_Cl3} 
\end{center}
\end{figure}

In M\,1-42, weak central nebular emission and contamination from two field stars compromise the reliability of [Cl~{\sc iii}] and [Ar~{\sc iv}] density results in those regions. In addition, all three diagnostics are based on emission lines from third-row elements, for which the recombination contribution is can not be corrected. Results are presented by target in the following paragraphs.

Figure \ref{fig:Hf_SII} shows the [S~{\sc ii}] $\lambda\lambda$6716,6731 PV maps and the derived electron densities for Hf\,2-2, assuming an electron temperature of  9300 K (the 1D [O~{\sc iii}] temperature). Strong-emission pixels within the outermost contour exhibit densities $\sim$500 cm$^{-3}$, matching 1D [S~{\sc ii}] density results. These regions are spatially or kinematically offset from the central star or systemic velocity, confirming low-density conditions in the outer nebula. The central regions show apparent density enhancements, but these coincide with the low SNR regions, rendering central density enhancements uncertain.

\begin{figure}[ht!]
\begin{center}
\includegraphics[width=8.5 cm,angle=0]{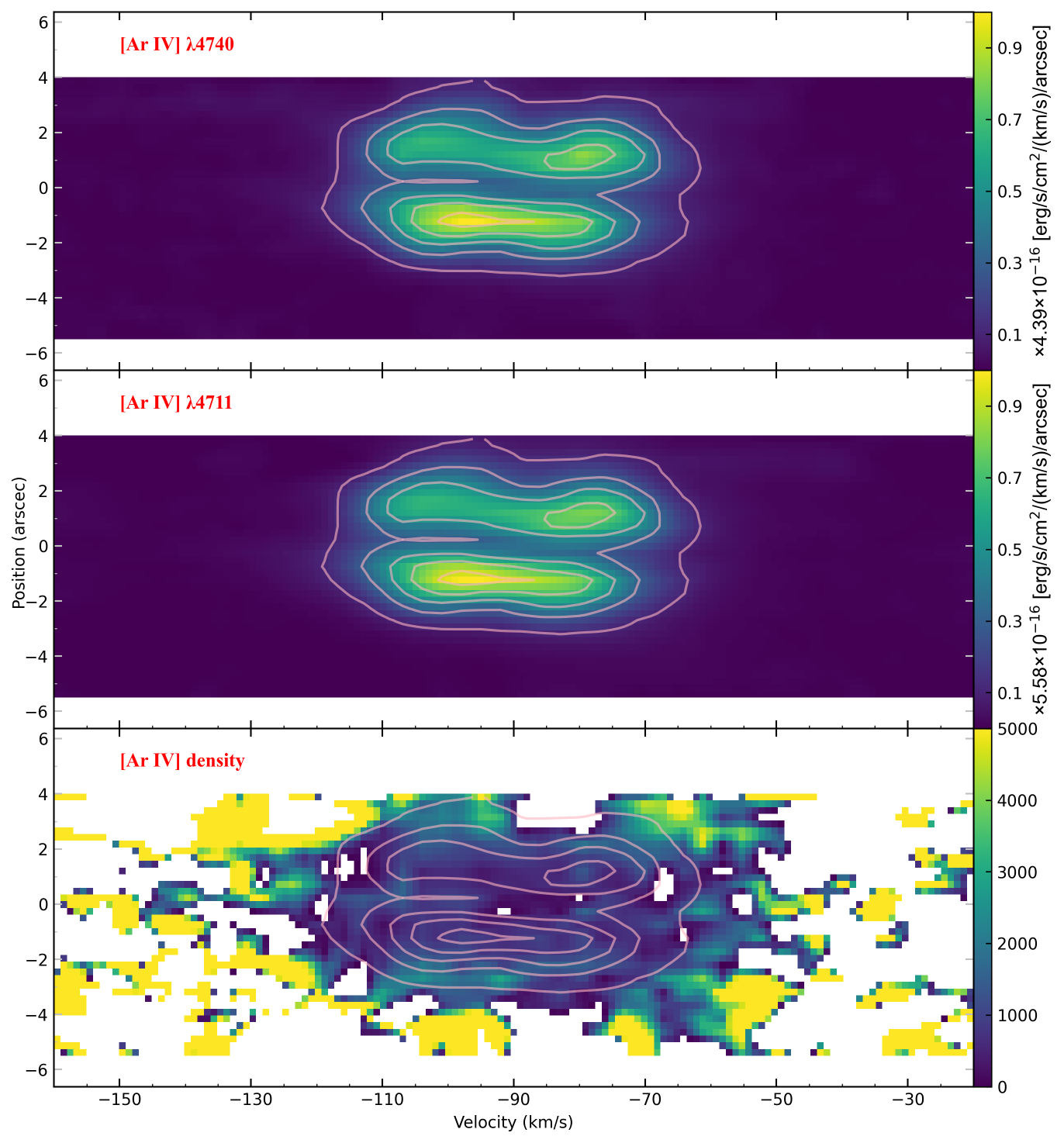}
\caption{PV diagrams of [Ar~{\sc iv}] $\lambda\lambda$4711,4740 (middle and upper panel respectively) and the derived [Ar~{\sc iv}] density of M\,1-42. Contours are the PV diagrams of [Ar~{\sc iv}] $\lambda$4740 line. The white zones within the outer contour correspond to line ratios outside the valid range for density calculations.} 
\label{fig:M142_Ar4} 
\end{center}
\end{figure}

For M\,1-42, density PV maps are organized by the ionization level of corresponding species, calculated using [O~{\sc iii}] temperatures from 1D spectra. Figure \ref{fig:M142_SII} displays [S~{\sc ii}] line PV maps and corresponding densities. The hyperbolic-shaped main emission region shows velocity-uniform densities that increase near the central star ($\sim$2000 cm$^{-3}$ in the lower shell, matching 1D results where this region dominates the flux of 1D spectra) and decreas outward. The central regions exhibit anomalously high densities despite weak [S~{\sc ii}] emissions, in contrast with [Cl~{\sc iii}] and [Ar~{\sc iv}] diagnostics (Figures \ref{fig:M142_Cl3} and \ref{fig:M142_Ar4}) that indicate lower densities in the more highly ionized region near the star. This discrepancy is similar to the [N~{\sc ii}] temperature enhancement near the center prior to applying recombination correction. Given the detection of S~{\sc ii} recombination lines (see the line tables in supplementary material) in the spectrum of M\,1-42, and that these lines originate near the central star, we hypothesize that these apparent density enhancements may originate from uncorrected S~{\sc ii} recombination contributions. However, verification remains impossible due to the lack of accurate S~{\sc ii} recombination atomic data. Given the similar electronic configurations of sulfur and oxygen, the recombination contribution for sulfur is expected to be similar to that for oxygen \citep{2023MNRAS.523.2952M}.

\begin{figure*}[ht!]
\begin{center}
\includegraphics[width=18 cm,angle=0]{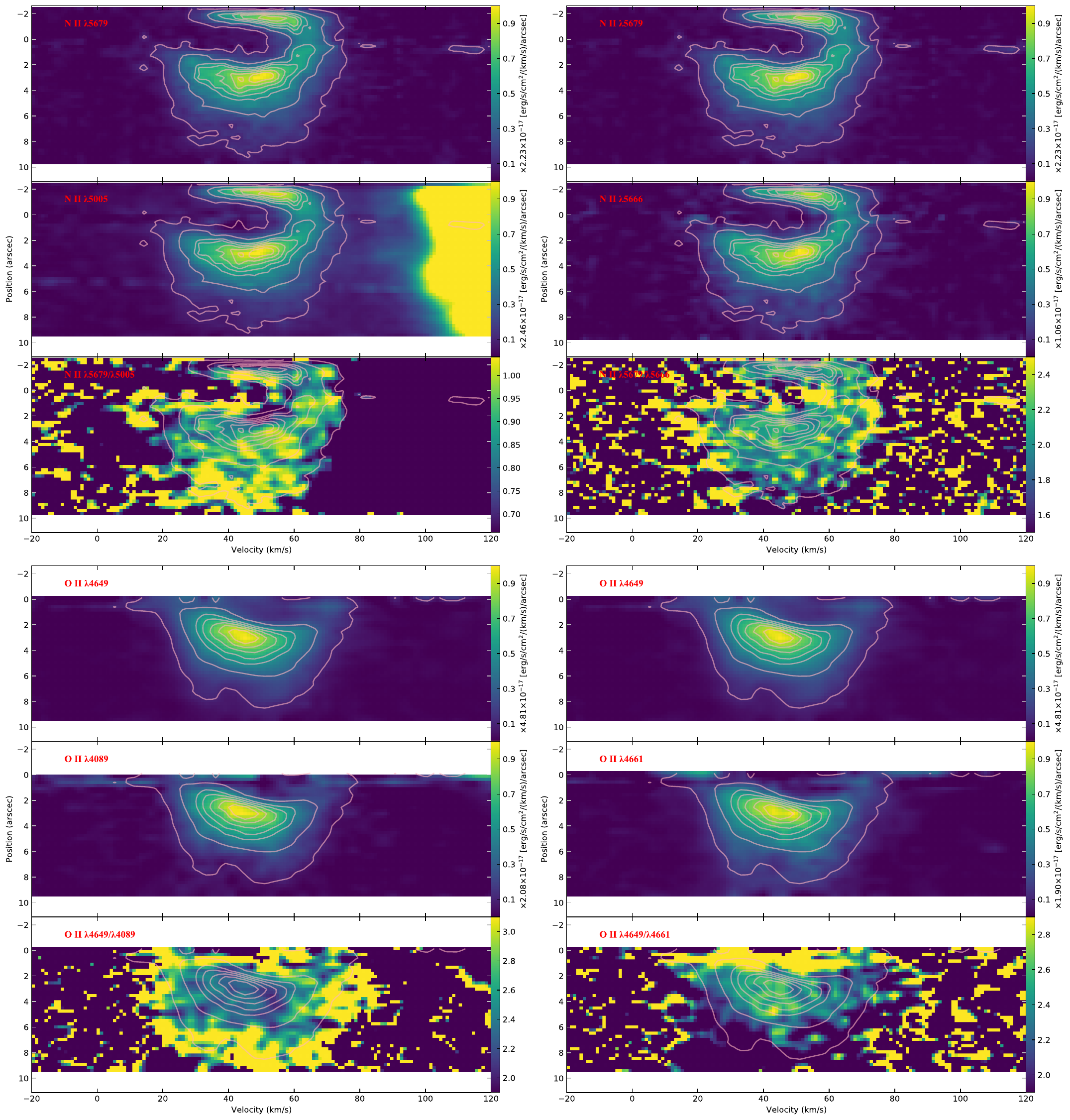}
\caption{PV diagrams of ORL fluxes and ratios for Hf\,2-2. The figure is divided into four blocks. The blocks in the left column present the lines and ratios used for temperature diagnostics, while those in the right are for density. The two upper blocks correspond to N~{\sc ii}, and the two lower blocks to O~{\sc ii}. The contours in each block represent the PV diagram of the line shown in the upper subplot of each block. The label in the upper left corner of each subplot indicates the corresponding line or line ratio. Temperature and density increase monotonically with the respective line ratios.} 
\label{fig:Hf_PVorldiag} 
\end{center}
\end{figure*}

\begin{figure*}[ht!]
\begin{center}
\includegraphics[width=18 cm,angle=0]{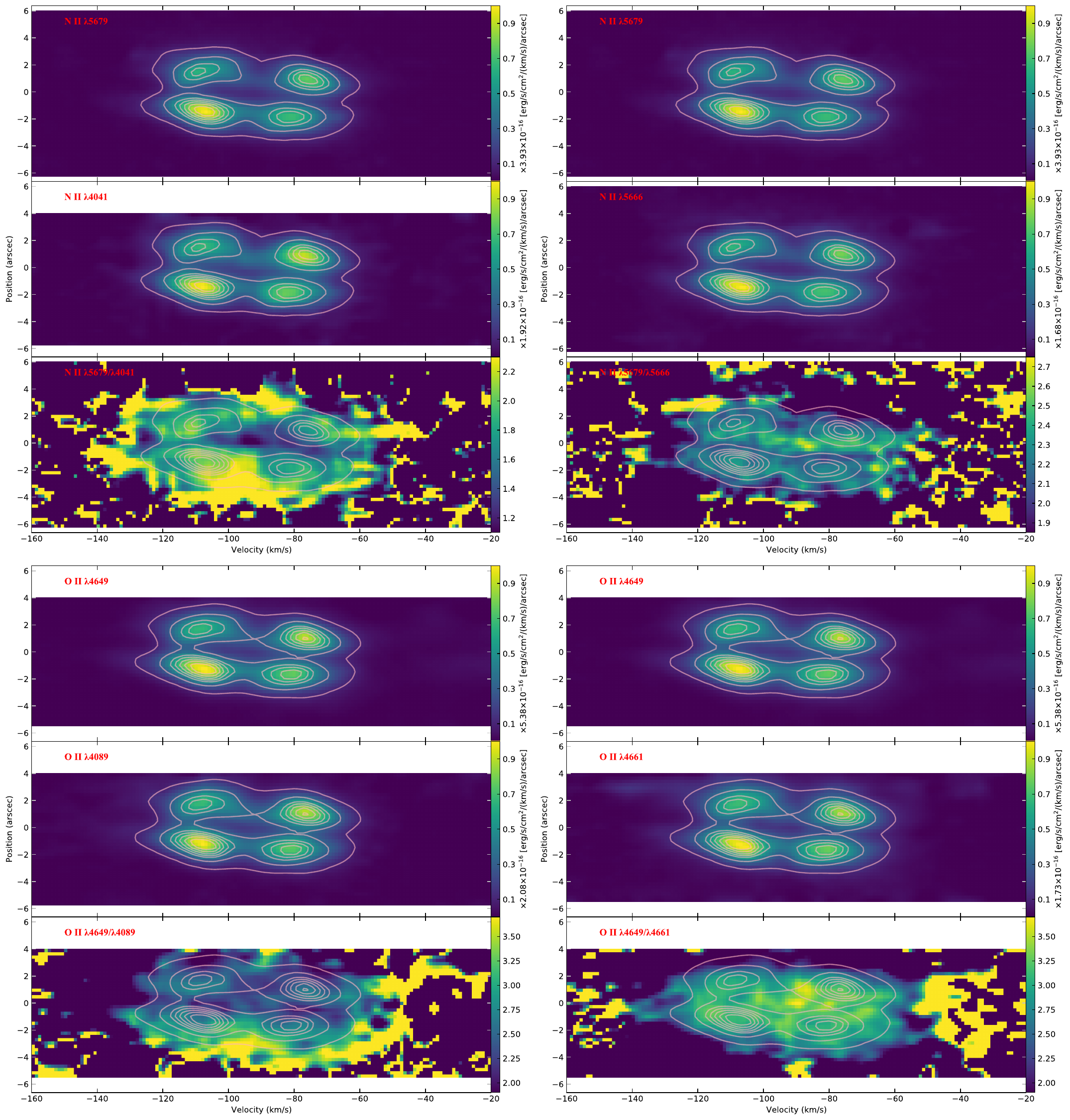}
\caption{Same as Figure\,\ref{fig:Hf_PVorldiag} but for M\,1-42. The structure in blocks is the same as in Figure \ref{fig:Hf_PVorldiag}. Similarly to Hf~2-2, temperature and density also increase monotonically with the line ratios in this object.} 
\label{fig:M_PVorldiag} 
\end{center}
\end{figure*}

Figure\,\ref{fig:M142_Cl3} presents [Cl~{\sc iii}] line maps and derived density map for M\,1-42, showing values $\sim$2000 cm$^{-3}$ values (consistent with 1D results) across most nebular regions, excluding field-star-contaminated upper main shell areas. The observed density fluctuations likely arise from [Cl~{\sc iii}] line SNR limitations, with no clear systematic spatial patterns. Figure \ref{fig:M142_Ar4} reveals lower [Ar~{\sc iv}] densities compared to [S~{\sc ii}] and [Cl~{\sc iii}] diagnostics, including white zones within contours where densities exceed the validity limits of the diagnostic. The [Ar~{\sc iv}] densities appear uniform along the velocity axis and show a tentative increase outward, consistent with the 1D spectra trend, in where high-ionization diagnostics yield lower densities (see Table \ref{temden}). 

Our 2D plasma diagnostics generally agree with the 1D results across different ionization species. However, a notable discrepancy arises in the [S~{\sc ii}] map, where a central density enhancement appears. This contrasts with the more uniform [Cl~{\sc iii}] and [Ar~{\sc iv}] distributions in the same region and with the 1D findings, in which higher ionization lines -- emitted closer to the central star -- consistently yield lower densities. Potential causes include low [S~{\sc ii}] SNR in central regions or uncorrected recombination contributions. Resolving this discrepancy will require recombination coefficients for third-row elements, enabling a more precise analysis of high-resolution spectra to determine the origin of the anomalous central [S~{\sc ii}] densities.

\subsubsection{Temperatures and Densities from ORL Diagnostics}

We used PV maps of N~{\sc ii} and O~{\sc ii} recombination lines to study the physical structures of metal-rich cold regions in PNe. Given the low SNR of the faint ORLs involved in these calculations, the PV diagrams show uncertainties and many pixels provide unphysical values in the temperature and density maps.

We used the same diagnostic line ratios as in the 1D analysis for consistency. Although the N~{\sc ii} $\lambda$5679/$\lambda$4041 ratio was calculated using lines from different CCDs, we still adopted it for M\,1-42 because the N~{\sc ii} $\lambda$5005 line was contaminated by stray light from the saturated [O~{\sc iii}] $\lambda$5007 line in the long exposures. Thanks to the monotonic bahavior of the diagnostic curves, we can infer the temperature and density structure of the nebulae from the variation of the line ratios in the PV diagrams. As shown in Figures \ref{fig:NIITe} to \ref{fig:OIINe}, all diagnostic line ratios increase with temperature or density. Therefore, in the PV diagrams, pixels with higher line ratios indicate higher temperature or density -- provided the ordering of the numerator and denominator of the line ratio remains consistent with the 1D analysis.

The PV diagrams of ORLs and their ratios for Hf\,2-2 are shown in Figure \ref{fig:Hf_PVorldiag}. Due to the weakness of the ORLs, the results show considerable variation. In the central region of the ORL emission, there is no clear structure or trend in the line ratios. However, in the outer parts of ORL-emitting region, the temperature appears higher and the density lower than in the inner region. This is particular evident along the spatial direction, where there is a significant increase in the temperature diagnostic line ratio and a pronouced decrease in the density diagnostic line ratio at a position coordinate of about 6$^{\prime\prime}$ in both the N~{\sc ii} and O~{\sc ii} diagnostic PV diagrams. In pixels with velocities deviating from the systemic velocity but position coordinates between 0$^{\prime\prime}$ and 6$^{\prime\prime}$, the O~{\sc ii} temperature and density follow the same trend as in the outer nebula, but this trend is not clear in the N~{\sc ii} temperature and density maps.
 
Figure \ref{fig:M_PVorldiag} shows the ORL fluxes and ratios of M\,1-42, displaying less severe fluctuations than those observed in Hf\,2-2. Similar to Hf\,2-2, the recombination diagnostics of M\,1-42 also indicate high temperatures and low densities in the outer ORL-emitting regions. Within the outermost contours, the N~{\sc ii} and O~{\sc ii} temperature-sensitive line ratios are lower near the center and peak between $-$2$^{\prime\prime}$ and $-$4$^{\prime\prime}$, being particularly evident in N~{\sc ii} temperature PV map. Regarding density, both the N~{\sc ii} and O~{\sc ii} ratios seem higher near the center and decreases outward along the spatial axis. No significant trends are observed along the velocity axis in the nebular main shell.

\begin{figure*}[ht!]
\begin{center}
\includegraphics[width=18 cm,angle=0]{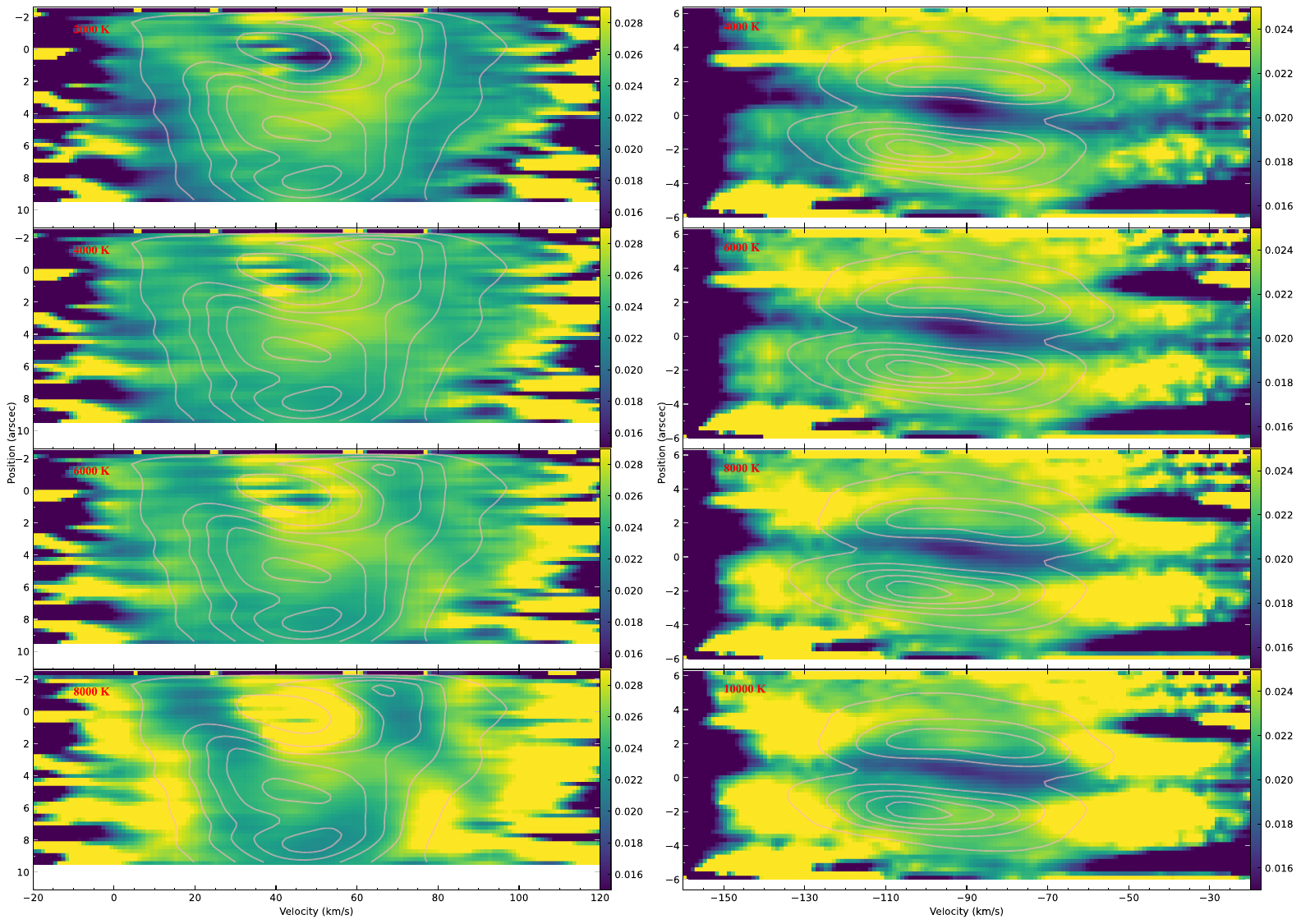}
\caption{PV maps of the ratio between He~{\sc i} $\lambda$6678 and H$\alpha$ for Hf\,2-2 (left) and M\,1-42 (right). To ensure consistency, the PV diagrams of the singlet He~{\sc i} line were used as templates to construct the model PV maps containing the fine structure of H$\alpha$. These model maps were convolved with Gaussian profiles at different assumed temperatures to simulate the thermal broadening of H$\alpha$ (see text for more details). The assumed temperature for each PV diagram is indicated in the upper left corner of each subplot. Contours represent the PV diagrams of H$\alpha$. At the true kinematic temperatures, the ratios should appear relatively uniform, especially in the velocity direction. Thus, the kinematic temperatures are roughly 4000$-$6000 K for Hf\,2-2, and 6000 K for M\,1-42. } 
\label{fig:kine_temp} 
\end{center}
\end{figure*}

Both nebulae exhibit denser and lower-temperature conditions in the outer parts of their ORL-emitting regions.  Due to the intrinsic weakness of ORLs and the fact that significant parameter variations appear at the faint edges of the ORL emission region, it is uncertain whether these trends are real or simply artifacts arising from large observational uncertainties. If genuine, however, this phenomenon is straightforward to interpret. Outside the metal-rich, H-deficient region in the nebula, the contribution of recombination emission from the hotter, lower-density main shell plasmas gradually increases, leading to higher temperature and lower density trends. If the kinematics of the metal-rich plasma follow a Hubble-like flow and the outer region parameter variations are real, then pixels with systemic velocity offsets should exhibit higher temperatures and lower densities. This behavior is partially supported by the O~{\sc ii} temperature ratio, but not by the N~{\sc ii} diagnostics. Therefore, the authenticity of this phenomenon remains unconfirmed.

\subsubsection{Kinematic Temperature} 
\label{k_temp}

We estimate kinematic temperature by exploiting the mass dependence of thermal broadening, $\sigma^2 = kT/m$, following the procedure similar to that of \citet[][their Sect.~3.2.1]{2022AJ....164..243R}. Practically, the PV diagram of the heavier species is Gaussian-convolved over a grid of assumed temperatures until its morphology best matches that of a lighter species; the temperature that minimizes the pixel-wise residuals (the scatter of line ratio PV map) is adopted. Although this method is affected by non-thermal broadening (including nebular kinematics, intrinsic line structure, and instrumental effects) and by the fact that different species used to compute the temperature may originate from distinct ionization zones, its strength is that it relies on simple physics rather than on atomic data with comparatively larger uncertainties.

To apply this method, we require pairs of lines that have enough SNR, share similar spatial distributions, and are recorded under comparable observational conditions (like airmass). H$\alpha$ and He~{\sc i} $\lambda$6678 satisfy these criteria. The emitting-region these line are not entirely identical owing to the presence of He$^{2+}$ region. Moreover, the mass difference of corresponding species of these lies produces a noticeable contrast in thermal broadening.  Although [O\,\textsc{iii}] would, in principle, enhance the thermal contrast owing to the larger mass difference between O$^{2+}$ and H$^+$, we do not use it here because CEL emissivities are strongly temperature dependent; in particular, the pronounced central temperature rise in Hf\,2-2 (Figure~\ref{fig:Hf22_OIII}) would bias the kinematic results.  We account for H$\alpha$ fine structure by constructing model H$\alpha$ maps from the He~{\sc i} template using the normalized component intensities and velocity offsets from \citet{1999AAS..135..359C} (Case B, at 10$^4$ K and appropriate densities, 10$^{2}$ cm$^{-3}$ for Hf\,2-2, 10$^{2}$ cm$^{-3}$ for M\,1-42), then convolving those models across the assumed temperatures for comparison with the observations.

\begin{figure*}[ht!]
\begin{center}
\includegraphics[width=18 cm,angle=0]{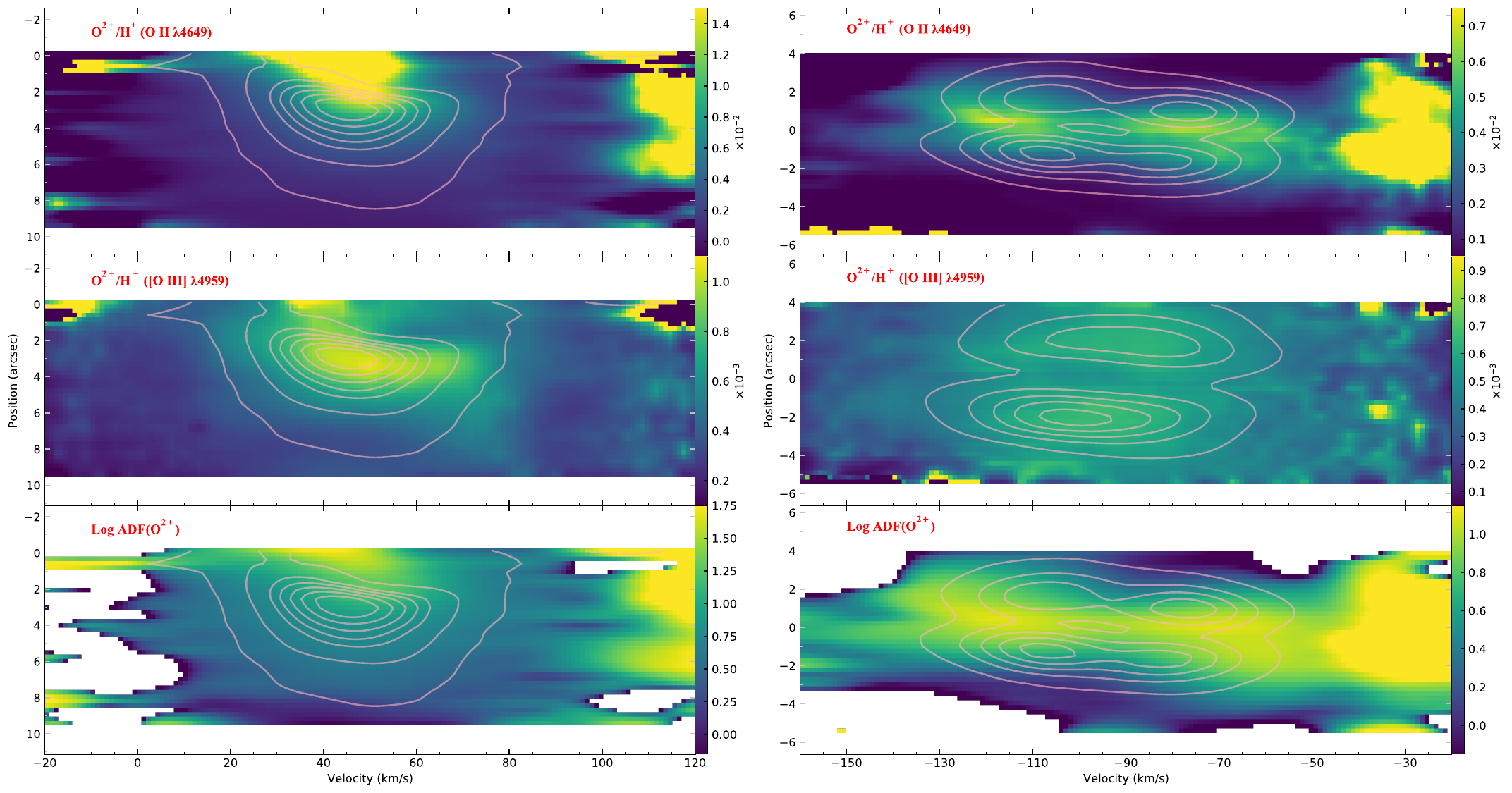}
\caption{PV diagrams of O$^{2+}$ ORL abundances (upper row) and CEL abundances (middle row), derived from O~{\sc ii} $\lambda$4649 and [O~{\sc iii}] $\lambda$4959 lines, respectively, along with the logarithmic ADF diagrams (bottom row). Diagrams for Hf\,2-2 and M\,1-42 are shown in the left and right columns, respectively. Contours represent the PV diagrams of O~{\sc ii} $\lambda$4649 after thermal broadening correction. The high values at the edges of the velocity axis  regions in the ORL abundance and ADF PV maps originate from the O~{\sc ii} $\lambda$4650.84. The labels on the right of the colorbar represent the abundance scale. } 
\label{fig:O2_abund} 
\end{center}
\end{figure*}

Figure \ref{fig:kine_temp} shows the PV diagrams of the ratio between the broadened models (derived from He~{\sc i} $\lambda$6678) and H$\alpha$. If the assumed temperature is close to the true kinematic temperature of the PN, the ratio should appear relatively uniform. Given the different ionization potentials of H$^+$ and He$^{+}$, and the fact that helium in PNe centers is mostly He$^{2+}$, the low ratios near the nebular center are expected. We focus on regions with strong nebular emission, where the high SNRs ensure more reliable results. For Hf\,2-2, pixels with larger velocity deviations from the systemic velocity have lower ratios than those near the systemic velocity at an assumed temperature of 2000 K. The trend reverses at 8000 K, while the ratios are relatively uniform at 4000 K and 6000 K. For M\,1-42, the ratios are more uniform at 6000 K. Thus, we estimate the kinematic temperatures of Hf\,2-2 and M\,1-42 to be 4000–6000 K, and $\sim$6000 K, respectively.

The kinematic temperatures of both sources are lower than their CEL temperatures, but higher than the O~{\sc ii} and N~{\sc ii} recombination line temperatures, as well as the Balmer and Paschen jump temperatures. It appears that the lower the ORL and discontinuity temperatures, the lower the resulting kinematic temperature. This may be because H~{\sc i} and He~{\sc i} lines are affected by both low- and high-temperature plasmas \citep{2015MNRAS.453.1281B, 2020MNRAS.497.3363G, 2023arXiv231114244M}, placing the kinematic temperature between the ORL and CEL values. The fact that the kinematic temperature is higher than the discontinuity temperature may result from the emphasis on maintaining consistent He~{\sc i}/H$\alpha$ ratios inside and outside the nebula, which can artificially amplify the contribution of hot plasma in the outer regions. It should be noted that all of the above interpretations assume reliable kinematic temperatures. However, deriving kinematic temperatures is quite subjective, leading to large uncertainties in the results.

\subsection{Abundance Distribution} 
\label{sec:AD}

Analysis of both 1D and 2D spectra reveals that CEL and ORL emission regions have different physical conditions. CEL-emitting regions have higher temperatures and lower densities, while ORL-emitting regions are colder and denser. Additionally, the outer parts of ORL-emitting regions appear to have higher temperatures than their inner parts. Similar to the 1D spectral analysis, this section explores the ionic abundances distribution in PV maps to highlight differences between plasmas with varying physical conditions, and to calculate the ADF PV diagrams.

When creating abundance PV diagrams, selecting the appropriate temperature is the critical. Temperature serves two roles: as the electron temperature used in abundance calculations, and as a parameter for correcting thermal broadening differences. The electron temperature affects line emissivities and reflects the thermal motion of electrons, while the kinematic temperature relates to ion motions causing emission-line broadening. In gaseous nebulae, elastic scattering cross sections between electrons are usually the largest, allowing electrons to reach the Maxwellian velocity distribution easily \citep{2006agna.book.....O}. Moreover, electron-ion elastic scattering cross sections, driven by Coulomb interactions, are expected to be of similar magnitude, implying that ion thermal motions should also follow a Maxwellian distribution at the same temperature as electrons. Thus, when building ionic abundance PV maps, we should make the temperature used for thermal broadening correction and abundance calculation for the same line as consistent as possible.

We calculated the PV diagrams for O$^{2+}$ ORL and CEL abundances using the O~{\sc ii} $\lambda$4649 and [O~{\sc iii}] $\lambda$4959 lines, respectively. As discussed in the plasma diagnostics in Section \ref{sec:diagnostics}, and kinematic temperatures in Section \ref{k_temp}, we found that different emission lines yield different temperatures, leading to inconsistent thermal broadening effects and distinct electron temperatures. For the ORL abundances, we adopted the O~{\sc ii} temperatures and densities listed in Table \ref{temden} for each PN; these O~{\sc ii} temperatures were also adopted for thermal broadening correction. For the CEL abundances, we used the [O~{\sc iii}] temperatures and [S~{\sc ii}] densities, and the thermal broadening corrections were carried out based on the [O~{\sc iii}] temperatures. 

Ionic abundance calculations also require setting the temperature of hydrogen. This is typically overlooked, and the atomic parameters of hydrogen are assumed to be at the same temperature as the corresponding ions, as was done in our 1D analysis in Section \ref{sec:abundance}. As a result, the hydrogen temperature used in ORL abundance calculations is lower than that used for CELs. However, for each PN, the average temperature of the hydrogen emitting region is an intrinsic property that should remain constant regardless of the calculation method. Various studies \citep[e.g.][]{2022AJ....164..243R, 2022MNRAS.510.5444G, 2024AA...689A.228G}, including this one, indicate that hydrogen continuum temperatures typically lie between the CEL and ORL temperatures. Similarly, the kinematic temperatures estimated from He~{\sc i}/H~{\sc i} ratios also fall within this range. If these temperatures reflect actual temperature of the hydrogen-emitting region, then using an overestimated hydrogen temperature in CEL abundance calculations would lead to an underestimated H$\beta$ emissivity, as thus to underestimated CEL abundances. Conversely, ORL abundances would be overestimated, resulting in an artificially inflated ADF value. Therefore, maintaining a consistent hydrogen temperature in ionic abundance calculations is essential. If the same hydrogen temperature were used for both CEL and ORL abundance calculations, CEL abundances would increase compared to our 1D results, while ORL abundances and ADF values would decrease.

An unavoidable issue is that temperatures derived from the Balmer and Paschen discontinuities are lower than the kinematic temperatures. For our calculations, we adopted kinematic temperatures of 5000 K for Hf\,2-2 and 6000 K for M\,1-42 for hydrogen. The [O~{\sc iii}] and O~{\sc ii} lines were then corrected from their own thermal broadening to that of H$^+$ at the adopted hydrogen temperatures, yielding cleaner PV diagrams without pseudo-structures and ensuring consistency with the hydrogen lines. For abundance calculations, however, we used the Balmer-jump temperatures to derive H$\beta$ emissivities, since hydrogen only appears in the denominator of X$^{i+}$/H$^+$ and the emissivity temperature does not affect the final ADF.

\begin{table*}
\begin{center}
\caption{Mass ratios of O$^+$, O$^{2+}$ and Ne$^{2+}$ between warm and cold plasmas from UVES and MUSE spectroscopy.}
\label{tab:mass_ratio}
\begin{tabular}{llcclcclcc}
\hline\hline
Parameters & & \multicolumn{2}{c}{Hf\,2-2} & & \multicolumn{2}{c}{M\,1-42} & & \multicolumn{2}{c}{NGC\,6153}\\
\cline{3-4}\cline{6-7} \cline{9-10}
 & & UVES & MUSE\tablenotemark{\rm{\scriptsize a}} & & UVES & MUSE\tablenotemark{\rm{\scriptsize a}} & & UVES & MUSE\tablenotemark{\rm{\scriptsize b}}\\
\hline
$T\rm{_e^w}$ (K) & & 9000 & 7000 & & 9000 & 8000 & & 9000 & 8300 \\
$T\rm{_e^c}$ (K) & & 400 & 800 & & 1400 & 800 & & 2200 & 2000 \\
$n\rm{_e^w}$ (cm$^{-3}$) & & 500 & 500 & & 2000 & 1000 & & 4000 & 3400 \\
$n\rm{_e^c}$ (cm$^{-3}$) & & 800 & 2000 & & 6000 & 2000 & & 10000 & 10000 \\
\hline
$M{\rm^c}/M{\rm^w}({\rm O^{2+}})$ & & 2.98 & 0.9 & & 1.52 & 0.7 & & 1.22 & 0.7 \\
$M{\rm^c}/M{\rm^w}({\rm Ne^{2+}})$ & & 1.98 & $\cdots$ & & 1.07 & $\cdots$ & & 1.03 & $\cdots$ \\
\hline
$M{\rm^c}/M{\rm^w}({\rm N^{2+}})$ & & 2.94 & $\cdots$ & & 1.50 & $\cdots$ & & 1.21 & $\cdots$ \\
$M{\rm^c}/M{\rm^w}({\rm O^{2+}})$ & & 2.96 & $\cdots$ & & 1.46 & $\cdots$ & & 1.11 & $\cdots$ \\
$M{\rm^c}/M{\rm^w}({\rm Ne^{2+}})$ & & 1.98 & $\cdots$ & & 1.02 & $\cdots$ & & 0.92 & $\cdots$ \\
\hline
\end{tabular}
\begin{description}
\tablenotemark{\rm{\scriptsize a}}\citet{2022MNRAS.510.5444G}; \tablenotemark{\rm{\scriptsize b}}\citet{2024AA...689A.228G}. The mass ratios in the second block of the table are calculated using Equation (\ref{eq:6}), while those in the last block are calculated using Equation (\ref{eq:7}) and (\ref{eq:8}).\\
\end{description}
\end{center}
\end{table*}

Figure \ref{fig:O2_abund} presents the PV diagrams of O$^{2+}$ CEL and ORL abundances, as well as the logarithmic ADF maps derived from the ratio of the abundance maps. For Hf\,2-2, the ORL abundance increases from the outer nebula toward the central star, rising sharply at the position of the central star. This may be due to O~{\sc ii} $\lambda$4649 emission produced the emission from the companion irradiated by the central star \citep{2016AJ....152...34H}. The substantial rise in [O~{\sc iii}] temperature (see Figure \ref{fig:Hf22_OIII}) at the nebular center causes differences in the PV diagrams of [O~{\sc iii}] $\lambda$4959 and H$\beta$, which results in abundance maps derived from calculations using a single temperature being non-uniform. We attempted to use [O~{\sc iii}] temperature PV maps but encountered two issues. First, the temperature PV map was obtained without thermal broadening correction, making it inconsistent with the present analysis. Second, the temperature maps have poor SNR in the nebular center and outer parts, leading to noisy results. Despite the inhomogeneous CEL abundance, the ADF is still higher in the nebular center and decreases outward, suggesting a metal-rich region at the center of Hf\,2-2 \citep{2022MNRAS.510.5444G}. In the velocity direction, the ADF appears higher in pixels closer to the systemic velocity, particularly in the nebular center.

For M\,1-42, the PV map of O$^{2+}$/H$^+$ derived from [O~{\sc iii}] CEL is quite uniform, but the recombination O$^{2+}$ abundances are higher at  positions near the central star. Interestingly, the location of the strongest O~{\sc ii} emission does not correspond to the highest O$^{2+}$ abundance. The resulting ADF exhibits a saddle-like structure. In the spatial direction, the ADF peaks in the cavity near the central star and gradually decreases outward, similar to the spatial ADF distributions reported by \citet{2022MNRAS.510.5444G}. In the velocity direction, the ADF does not peak at the systemic velocity; instead, it increases away from the systemic velocity, reaching a maximum within the outermost contour before declining again. In pixels located far from the central star and with velocities close to the systemic velocity, the ADF is low and approaches unity. 

In summary, M\,1-42 exhibits a uniform distribution of O$^{2+}$ CEL abundances, whereas CEL abundances of Hf\,2-2 shows inhomogeneous distributions, likely reflecting temperature variations. In both nebulae, ORL abundances peak toward the center, indicating that metal-rich plasma resides in the interior. The ORL plasma expands more slowly in Hf\,2-2 but somewhat faster in M\,1-42, perhaps reflecting their distinct morphologies (circular vs. bipolar). The ADF distributions are relatively smooth, suggesting that hot and cold plasmas may be connected -- through mixing or a common origin -- though limited spectral and spatial resolution could also mimic such continuity.

\section{Discussion} 
\label{sec:discussion}

\subsection{Relative Content of Heavy-element Ions in Two Plasmas}\label{sec:mass_ratio_NONe}

The existence of the metal-rich plasma component proposed by \citet{2000MNRAS.312..585L} has been widely considered responsible for extreme ADFs in PNe \citep[e.g.][]{2003MNRAS.340.1153E,2005MNRAS.364..687T,2006MNRAS.368.1959L,2008MNRAS.383.1639W,2018MNRAS.480.4589W}, as their efficient cooling produces low ORL temperatures and strong ORL emission. Various observations have confirmed their presence \citep[e.g.,][]{2016ApJ...824L..27G, 2022MNRAS.510.5444G}. To assess the degree of metal enrichment, we derived the mass ratios of metals in warm and cold plasmas using both the methods mentioned in \citet{2020MNRAS.497.3363G} and \citet{2022AJ....164..243R}.

Following the methodology of \citet{2020MNRAS.497.3363G}, we assumed that heavy-element ORLs originate exclusively from cold plasmas while CELs only arise from warm plasmas. Following equations (5) and (6) from \citet{2022MNRAS.510.5444G}, the mass ratio of a specific ion in cold to warm plasma can be expressed as:
\begin{equation}\label{eq:6}
{M{\rm^c}\over M{\rm^w}}(X^{i+}) = {j_{\rm CEL}(T{\rm_e^w}, n{\rm_e^w})\over j_{\rm ORL}(T{\rm_e^c}, n{\rm_e^c})}\cdot{I_{\rm ORL}\over I_{\rm CEL}}\cdot{n{\rm_e^w}\over n{\rm_e^c}}
\end{equation}
where $I_{\rm CEL}$ and $I_{\rm ORL}$ represent the CEL and ORL intensities, respectively, and $j$ represents the line emissivity. The superscripts indicate whether warm (w) or cold (c) plasma physical conditions were employed. Table \ref{tab:mass_ratio} lists the adopted physical conditions and derived mass ratios, including comparative data from \citet{2022MNRAS.510.5444G} and \citet{2024AA...689A.228G}. All calculations were based on emission line intensities measured from 1D spectra.

\begin{figure*}[ht!]
\begin{center}
\includegraphics[width=18 cm,angle=0]{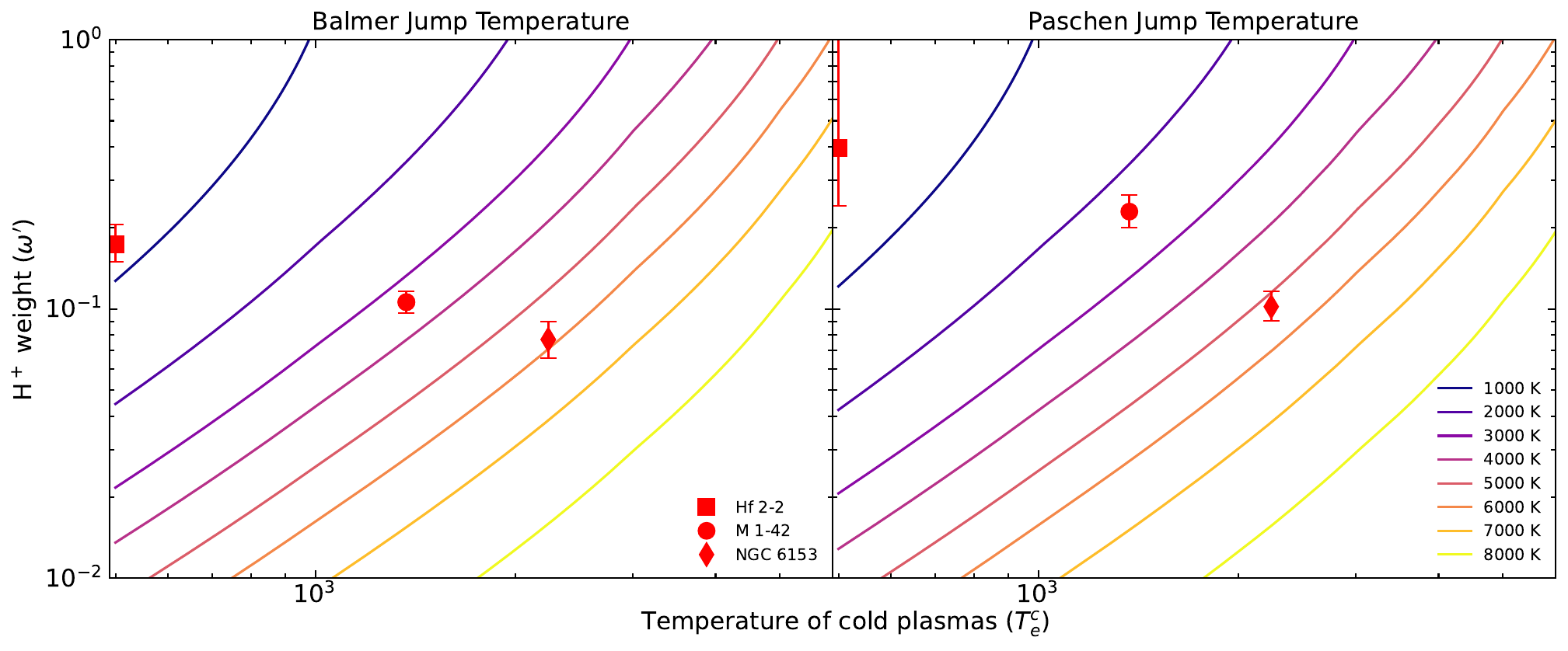}
\caption{Observed constant lines of Balmer jump temperature (left) and Paschen jump temperature (right) in the plane consisting of the cold gas temperature, $T{\rm _e^c}$, and the H$^+$ weight in cold plasmas, $\omega^{\prime}$. The temperature of the warm plasmas, $T{\rm _e^w}$, was assumed as 9000 K. The position of our targets in the $T{\rm _e^c}$-$\omega^{\prime}$ plane calculated based on the adopted $T{\rm _e^c}$ and hydrogen discontinuity temperatures in Table \ref{temden} are also presented. Only the uncertainties of hydrogen discontinuity temperatures are considered. The Paschen temperatures of our targets are systematically lower than the Balmer temperatures, leading to higher H$^+$ weights in the cold plasmas.} 
\label{fig:H_weight} 
\end{center}
\end{figure*}

The derived mass ratios depend on the adopted physical parameters. For the two most uncertain parameters, $T{\rm_e^c}$ and $n{\rm_e^c}$, the ratios scale linearly and inversely with their values, respectively \citep{2022MNRAS.510.5444G}. For the warm plasma conditions, higher temperatures and densities lead to larger mass ratios. Our adopted parameters yield higher $n{\rm_e^w}T{\rm_e^c}/n{\rm_e^c}$ ratios compared to those of MUSE spectroscopy, ultimately resulting in larger derived mass ratios. Adopting identical physical parameters as \citet{2022MNRAS.510.5444G} and \citet{2024AA...689A.228G} would bring our O$^{2+}$ mass ratios agree with theirs, highlighting both the substantial uncertainties inherent in this calculation method and the critical need for precise temperature determinations.

The Ne$^{2+}$ mass ratios remain systematically lower than the corresponding O$^{2+}$ values across all targets, a difference that is not attributable to the temperature or density uncertainties.  This disparity persists even when adopting physical parameters from \citet{2022MNRAS.510.5444G} or \citet{2024AA...689A.228G}.  One plausible explanation involves the use of outdated effective recombination coefficients for Ne\,{\sc ii} lines published by \citet{1998A&AS..133..257K}, whose calculations were conducted in pure LS coupling and considered only the transitions between states with angular momentum quantum number $l\leq$2 (with limited treatment of high-$l$ transitions and adopting the assumption of thermalized populations of ground-term fine-structure levels of the recombining ion Ne$^{2+}$).  Utilizing these Ne\,{\sc ii} data may bias the inferred Ne$^{2+}$ abundances.  However, a genuine physical origin cannot be excluded, particularly given analogous mass-dependent variations such as significant hydrogen depletion observed in cold plasmas.

The comparable amounts of heavy elements in warm and cold plasmas imply that the assumption of ORLs arising exclusively from cold plasma is not accurate. If we can distinguish the recombination contributions from the cold and warm plasmas, we can calculate the mass ratio following the method of \citet{2022AJ....164..243R}, the mass ratio can be expressed as:
\begin{equation}\label{eq:7}
{M{\rm^c}\over M{\rm^w}}(X^{i+}) = {j(\lambda,T{\rm_e^w}, n{\rm_e^w})\over j(\lambda,T{\rm_e^c}, n{\rm_e^c})}\cdot{I_{\rm c}\over I_{\rm w}}\cdot{n{\rm_e^w}\over n{\rm_e^c}}
\end{equation}
where emissivities of the same line at different plasma conditions are compared, and the ORL and CEL fluxes are replaced by the contributions from the cold and warm components of that line compared to Equation (\ref{eq:6}). The warm-plasma flux is given by,
\begin{equation}\label{eq:8}
I_{\rm w} = {j(\lambda,T{\rm_e^w}, n{\rm_e^w})\over j({\rm[O~III]},T{\rm_e^w}, n{\rm_e^w})}\cdot{N(X^{i+}) \over N({\rm O^{2+}})}\cdot I({\rm [O~III]})
\end{equation}
with [O~{\sc iii}] lines observed simultaneously with ORLs on the same CCD were selected to minimize atmospheric dispersion. Both emissivities $j$ are evaluated under warm conditions. Ionic abundance ratios are taken from CEL results, except for N$^{2+}$, where we adopt ORL-based N$^{2+}$/O$^{2+}$ ratios. The resulting mass ratios are listed in Table \ref{tab:mass_ratio}.

Combining Equations (\ref{eq:7}) and (\ref{eq:8}) yields a relation analogous to Equation \ref{eq:6}, but without thermal recombination contributions and with the term $j_{\rm CEL}(T{\rm_e^w}, n{\rm_e^w})/I_{\rm CEL}$ replaced by $N({\rm O^{2+}})/N(X^{i+})$ and $j_{\rm [O~III]}(T{\rm_e^w}, n{\rm_e^w})/I({\rm [O~III]})$. If abundance ratios are accurate, this method should produce slightly lower mass ratios than the results derived from Equation \ref{eq:6}, as seen for Ne$^{2+}$ and O$^{2+}$. The close agreement between N$^{2+}$ and O$^{2+}$ ratios across all sources indicates that N$^{2+}$/O$^{2+}$ remains consistent between plasma phases, supporting our use of ORL-derived ratios. For NGC\,6153, our cold-plasma mass fractions exceed those of \citet{2022AJ....164..243R}, but adopting their parameters reproduces similar results, confirming methodological consistency.

\begin{table}
\begin{center}
\caption{Weights of H$^+$ ($\omega^{\prime}$) and H$\beta$ emission ($\omega$) in cold plasmas, and ACF values.}
\label{tab:H_weight}
\begin{tabular}{llcclcclcc}
\hline\hline
Parameters & Hf\,2-2 & M\,1-42 & NGC\,6153\\
\hline
$\omega^{\prime}$ (BJ) & $0.174^{+0.032}_{-0.024}$ & $0.106^{+0.011}_{-0.009}$ & $0.077^{+0.013}_{-0.011}$ \\
$\omega$ (BJ) & $0.65^{+0.05}_{-0.04}$ & $0.33^{+0.02}_{-0.02}$ & $0.20^{+0.03}_{-0.03}$ \\
$\omega^{\prime}$ (PJ) & $0.396^{+0.604}_{-0.154}$ & $0.230^{+0.035}_{-0.029}$ & $0.103^{+0.014}_{-0.012}$ \\
$\omega$ (PJ) & $0.85^{+0.15}_{-0.11}$ & $0.55^{+0.05}_{-0.04}$ & $0.25^{+0.03}_{-0.02}$ \\
log(ACF(O$^{+}$)) & 2.69 & 1.97 & 2.49 \\
log(ACF(O$^{2+}$)) & 1.38 & 1.62 & 1.61 \\
log(ACF(Ne$^{2+}$)) & 1.21 & 1.29 & 1.40 \\
\hline
\end{tabular}
\end{center}
\end{table}

We did not calculate O$^{+}$ mass ratios from 1D spectra as done by \citet{2022MNRAS.510.5444G,2024AA...689A.228G}, because our extraction regions barely cover the outer nebular layers; consequently, nearly all [O~{\sc ii}] auroral flux is due to recombination and vanishes after correction. \citet{2022MNRAS.510.5444G} showed that the O$^{+}$ and O$^{2+}$ mass ratios are nearly identical in Hf\,2-2, M\,1-42, and NGC\,6778, whereas \citet{2024AA...689A.228G} found that in NGC\,6153 O$^{+}$ is more concentrated in the cold phase. Overall, the amounts of second-period ions in cold and warm plasmas appear comparable, and our results are consistent with previous studies \citep{2020MNRAS.497.3363G,2022AJ....164..243R,2022MNRAS.510.5444G,2024AA...689A.228G}.

\subsection{Relative Mass of H in the Two Plasmas}
\label{sec:H_weight}

In Section \ref{sec:mass_ratio_NONe}, we showed that heavy-element ions have comparable masses in the cold and warm plasmas. To further probe hydrogen depletion in the cold phase, we also estimated the H$^{+}$ mass ratios between the two components. Following \citet{2022MNRAS.510.5444G}, we determined  the fractional contribution of H$\beta$ emission from the cold plasma, denoted as $\omega$. The intensity ratio of a given line to H$\beta$ can be expressed using Equation (3) of \citet{2022MNRAS.510.5444G}. However, $\omega$ differs from the H$^+$ mass fraction in cold plasma which we define as $\omega^{\prime}$. Their relation is given by:
\begin{equation}\label{eq:9}
{\omega \over 1-\omega} = {\omega^{\prime}\times j(H\beta,T{\rm_e^c}) \over (1-\omega^{\prime})\times j(H\beta,T{\rm_e^w})}
\end{equation}
Since the H$\beta$ emissivity decreases with increasing electron temperature, $\omega^{\prime}$ is expected to be smaller than $\omega$.

Inspired by \citet{2020MNRAS.497.3363G}, \citet{2022MNRAS.510.5444G} used the Paschen jump temperature to determine the weights. We adopted this method, and additionally incorporated Balmer jump temperatures. Figure \ref{fig:H_weight} shows the Balmer and Paschen discontinuity temperatures as a function the cold plasma temperature and $\omega^{\prime}$, assuming a warm plasma temperature of 9000 K. The positions of our targets in the T$^e_c-\omega^{\prime}$ plane are also shown in Figure \ref{fig:H_weight}, and their corresponding H$^+$ weights in the cold plasma are listed in Table \ref{tab:H_weight}. The uncertainties in the H$^+$ fractions were conservatively estimated, considering only the uncertainties in the Balmer or Paschen temperatures, excluding those associated with the cold and warm plasma temperatures. The lower Paschen temperatures in our targets led to higher $\omega^{\prime}$ values.

The corresponding weights of H$\beta$ emission from the cold plasmas, $\omega$, derived from Equation \ref{eq:9}, are also presented in Table \ref{tab:H_weight}. Despite the relatively low mass fractions of H$^+$ in the cold gas, this ion contributes a significant portion of the H~{\sc i} emission. Specifically, for results based on Balmer jump temperatures, the hydrogen ions in the cold plasmas of the three sources account for only about 17\%, 11\%, and 8\% of the total H$^+$ content, yet they contribute approximately 65\%, 33\%, and 20\% of the H$\beta$ emission, respectively. This significant contribution implies that a small fraction of H$^+$ embedded in the cold gas is sufficient to produce the observed low H~{\sc i} temperatures..

The ionic abundances in cold and warm plasmas can be corrected using the weights of H$\beta$ or H$^+$. For 1D spectrum-based ionic abundance calculations, the atomic data for H$^+$ and X$^{i+}$ were assumed to be at the same temperature. However, H$^+$ recombination coefficients at different temperatures were used for ORL and CEL abundances (Section \ref{sec:abundance}). Therefore, correcting the H$\beta$ fluxes using $\omega$ is sufficient in these cases. If the H$^+$ parameters at the same temperature are used for both ORL and CEL abundance determinations (as in Section \ref{sec:AD}), $\omega^{\prime}$ will be adopted instead. 

\citet{2020MNRAS.497.3363G} introduced the abundance contrast factor (ACF), defined as the ratio of the true ionic abundance in metal-rich region to that in the main nebula. The relationship between ADF from 1D spectra and ACF is:
\begin{equation}\label{eq:10}
ACF = ADF\times {1-\omega \over \omega},
\end{equation}
where $\omega$ needs to be replaced with $\omega^{\prime}$ when the ADFs in Section \ref{sec:AD} were used. 

Table \ref{tab:H_weight} also lists the logarithmic ACF values calculated using the H$\beta$ weights derived from Balmer jumps and the ADFs in Table \ref{tab:ion_adf}. Only Hf\,2-2 shows ACF lower than its corresponding ADF. The ACFs of O$^+$ and O$^{2+}$ for NGC\,6153 reported by \citet{2024AA...689A.228G} are approximately 2.5 and 1.7, respectively, consistent with our results. The ACF(O$^{2+}$) values reported by \citet{2022MNRAS.510.5444G} are broadly consistent with ours within the large uncertainties, though some differences remain. Even when adopting similar cold plasma temperatures, their values tend to be higher. 

\citet{2022AJ....164..243R} determined H$\beta$ and H$^+$ weights for NGC\,6153 using PV diagrams from the same dataset. At a kinematic temperature of 8000 K, which homogenizes the H$\beta$/[O~{\sc iii}] PV map, they subtracted their assumed H$\beta$ weights from the cold plasma and selected the value that yielded the flattest ratio. Their estimated cold plasma weights are about 10\% for H$\beta$ emission and 3-4\% for H$^+$ mass, which are broadly consistent with our results. We did not apply this method to the other two sources due to their lower spectral quality and the subjectivity of the procedure. Moreover, spatial variations in uneven [O~{\sc iii}] temperatures can introduce significant inconsistencies in the H$\beta$/[O~{\sc iii}] ratio, which is particularly problematic for Hf\,2-2, where [O~{\sc iii}] temperatures vary significantly.

\section{Summary}
\label{sec:summary}

We carefully reprocessed the VLT/UVES data of Hf\,2-2, M\,1-42 and NGC\,6153 with meticulous absolute flux calibration. One-dimensional spectra were extracted from the spatial regions jointly covered by all slits. We measured the wavelengths and fluxes of 417, 674, and 773 emission lines from the spectra of Hf\,2-2, M\,1-42 and NGC\,6153, respectively.  Line identification was conducted with the help of PyEMILI \citep{2025ApJS..277...13T}. The extinction coefficients of the three PNe, as well as the flux scaling between different wavelength intervals of the NGC\,6153 spectra \citet{2022AJ....164..243R} were estimated and corrected. The c(H$\beta$) of NGC\,6153 is approximately 1.13, which is larger than that reported by \citet{2022AJ....164..243R} but consistent with other observations.  

We analyzed the 1D spectra carefully due to known issues in \citet{2016MNRAS.461.2818M} -- specifically a mismatch between blue- and red-arm extraction regions and problems with absolute flux calibration.  Plasma diagnostics and CEL-based abundance calculations were carried out using \texttt{PyNeb}, while ORL analyse used newly calculated atomic data. Heavy-element ORLs yield the lowest temperatures, followed by H~{\sc i} recombination temperatures, with CEL temperatures being the highest.  ORL-based electron densities are higher than those derived from CELs. We obtained both CEL and ORL abundances for O$^+$, O$^{2+}$ and Ne$^{2+}$, finding that CEL abundances are systematically lower. Ne$^{2+}$ ORL abundances from 3$-$3 transitions differ significantly  from those from 3d$-$4f transitions, highlighting the need to recalculate Ne~{\sc ii} atomic parameters. The ADFs derived for these ions confirm that these PNe are high-ADF objects, aligning with previous studies. Elemental abundances calculated via ICFs are also consistent with previous studies.

We constructed Wilson diagrams for Hf\,2-2 and M\,1-42, finding that the kinematics of CEL- and heavy-element ORL-emitting regions are distinct. The velocity splitting of CELs decreases with increasing ionization potential, while the those of ORLs show no significant trend. To explore the spatial distribution and kinematic properties of various physical parameters, we also constructed PV diagrams for Hf\,2-2 and M\,1-42 based on the 2D spectra.  The morphologies of the ORL PV maps are clearly distinct from those of CELs, and are essentially identical for different species, from C~{\sc ii} to Ne~{\sc ii}.  The PV maps of electron temperature and density indicate that these parameters are nearly constant within the main shells of each PN.  Kinematic temperatures were also estimated, and they lie between the H\,{\sc i} discontinuity temperature and CEL temperature.  The PV diagrams of ADF(O$^{2+}$) show that the ADF value peaks in the nebular centre and decreases outward.  Our results are supported by the recent work of \citet{10.1093/mnras/staf2088}. 

We also calculated the mass ratios of N$^{2+}$, O$^{2+}$ and Ne$^{2+}$ between cold and warm plasmas, as well as the weights of the H$\beta$ emission and H$^+$ ions in the cold gas. The results show that the masses of the heavy-element ions in the cold plasma are of the same order of magnitude as those in the warm plasma. In contrast, the fraction of H$^+$ in the cold plasma is much lower than in warm component. This suggests that the ORL-emitting region is hydrogen-deficient. Although there is little H$^+$ in the cold gas, the lower electron temperature leads to an increased emissivity, resulting in a large percentage of the H~{\sc i} emission originating from cold gas, which explains the lower Balmer and Paschen jump temperatures. 

The VLT/UVES data we analysed were observed more than two decades ago. Advances in observational techniques, especially the availability of high spectral resolution IFUs, now enable the acquisition of three-dimensional spectra that simultaneously probe the 2D spatial distribution and radial velocity of emission lines. This provides a promising avenue to study the differences in the dynamics of different plasma components in PNe and thus understand the origin of the metal-rich region responsible for high ADFs. We subsequently intend to use MEGARA \citep{2020MNRAS.493..871G}, mounted on 10.4m GTC to study high ADF PNe.

\begin{acknowledgments}
We are grateful to the anonymous referee, whose excellent comments and suggestions greatly improved this article.  This paper is Based on observations collected at the European Southern Observatory under ESO programme ID 69.D-0174A.  This work was supported by the New Cornerstone Science Foundation through the New Cornerstone Investigator Program, the National Key R\&D Program of China (Grant No. 2023YFA1607902), and China Manned Space Program with grant No.\ CMS-CSST-2025-A14.  J.G.-R. acknowledges financial support from grant PID-2022136653NA-I00 (DOI:10.13039/501100011033) funded by the Ministerio de Ciencia, Innovaci\'{o}n y Universidades (MCIU/AEI) and by ERDF ``A way of making Europe'' of the European Union.  X.F.\ acknowledges support from the Youth Talent Program (2021) from the Chinese Academy of Sciences (CAS, Beijing) and the ``Tianchi Talents'' Program (2023) of the Xinjiang Autonomous Region, China. 
\end{acknowledgments}

\section*{Data Availability}

Plots of the fully reduced and calibrated 2D VLT/UVES echelle spectra, as well as the extracted 1D spectra, of the three PNe (Hf\,2-2, M\,1-42 and NGC\,6153) will be publicly available via an online repository (our data collection, organization and deposition have been underway).  Readers can also separately contact the authors of this article for the fully calibrated spectra, both 2D and 1D. 

We have also created and analyzed the P-V diagrams of NGC\,6153 using our fully reduced VLT/UVES echelle spectra.  However, the P-V diagrams of NGC\,6153 are not presented in this article, given that they have already been reported in \citet{2022AJ....164..243R}, but will be uploaded, together with those of the other two PNe, to an online repository.  Readers can contact the authors of this article for these data.

\appendix

\restartappendixnumbering

\section{Emission line lists} \label{appendix:linelist}

Measurements and identification of the emission lines detected in the VLT/UVES spectra of Hf\,2-2, M\,1-42 and NGC\,6153 are compiled in Tables\,\ref{Hf22_linelist}, \ref{M142_linelist} and \ref{NGC_linelist}, respectively.  For each PN, only a section of the table is presented here; the complete tables are available in the machine-readable format.  The columns from the left to the right are observed wavelengths, observed and dereddened fluxes (and uncertainties, normalized to H$\beta=100$), identifications, laboratory wavelengths, radial velocities (in unit of km s$^{-1}$), and transitions (lower$-$upper transition terms). The lines with uncertain identifications are marked with ``?" behind the identification species. The unidentified lines are marked with ``??" in the latter 4 columns, while the ``$\ast$" in the first 3 columns means the corresponding line is blended with its former line and contributes the flux.

\begin{table*}
\begin{center}											
\caption{Line List of Hf\,2-2}
\label{Hf22_linelist}						
\begin{tabular}{ccclccc}						
\hline\hline						
$\rm{\lambda_{obs}\ (\AA)}$ & $F$($\lambda$) & $I$($\lambda$) & Ion & $\rm{\lambda_{lab}\ (\AA)}$ & V$\rm{_{rad}\ (km s^{-1})}$ & Transition \\
\hline						
3133.28 & 1.502$\pm$: & 2.455$\pm$: & O~{\sc iii}  & 3132.79 & 46.92 & \textrm{$3p\ ^3S_1-3d\ ^3P^{\rm{o}}_2$}  \\
3188.23 & 3.313$\pm$0.166 & 5.269$\pm$0.264 & He~{\sc i}  & 3187.74 & 46.11 & \textrm{$2s\ ^3S_1-4p\ ^3P^{\rm{o}}$}  \\
3203.61 & 0.847$\pm$0.083 & 1.338$\pm$0.131 & He~{\sc ii}  & 3203.17 & 41.21 & \textrm{$3d\ ^2D-5f\ ^2F^{\rm{o}}$}  \\
3244.61 & 0.169$\pm$0.094 & 0.262$\pm$0.146 & Ne~{\sc ii}  & 3244.10 & 47.16 & \textrm{$3p\ ^4D^{\rm{o}}_{5/2}-3d\ ^2F_{7/2}$}  \\
3312.86 & 0.140$\pm$0.116 & 0.212$\pm$0.175 & O~{\sc iii}  & 3312.32 & 48.91 & \textrm{$3s\ ^3P^{\rm{o}}_1-3p\ ^3S_1$}   \\
3335.35 & 0.570$\pm$0.103 & 0.855$\pm$0.154 & Ne~{\sc ii}  & 3334.84 & 45.88 & \textrm{$3s\ ^4P_{5/2}-3p\ ^4D^{\rm{o}}_{7/2}$}  \\
3341.29 & 0.212$\pm$0.065 & 0.317$\pm$0.097 & O~{\sc iii}  & 3340.76 & 47.59 & \textrm{$3s\ ^3P^{\rm{o}}_2-3p\ ^3S_1$}  \\
3355.54 & 0.383$\pm$0.073 & 0.571$\pm$0.109 & Ne~{\sc ii}  & 3355.02 & 46.50 & \textrm{$3s\ ^4P_{3/2}-3p\ ^4D^{\rm{o}}_{5/2}$}  \\
3418.19 & 0.425$\pm$0.157 & 0.622$\pm$0.230 & Ne~{\sc ii}  & 3417.69 & 43.89 & \textrm{$3p\ ^2D^{\rm{o}}_{5/2}-3d\ ^2F_{7/2}$}  \\
3429.18 & 0.071$\pm$0.038 & 0.104$\pm$0.056 & O~{\sc iii}  & 3428.62 & 49.00 & \textrm{$3p\ ^3P_1-3d\ ^3P^{\rm{o}}_2$}  \\
$\cdots$ & $\cdots$ & $\cdots$ & $\cdots$ & $\cdots$ & $\cdots$ & $\cdots$ \\
\hline
\end{tabular}
\begin{description}
Table \ref{Hf22_linelist} is published in its entirety in the machine-readable format. A portion is shown here for guidance regarding its form and content.
\end{description}
\end{center}
\end{table*}

\begin{table*}
\begin{center}
\caption{Line List of M\,1-42}
\label{M142_linelist}
\begin{tabular}{ccclccc}
\hline\hline
$\rm{\lambda_{obs}\ (\AA)}$ & $F$($\lambda$) & $I$($\lambda$) & Ion & $\rm{\lambda_{lab}\ (\AA)}$ & V$\rm{_{rad}\ (km s^{-1})}$ & Transition \\
\hline 
 3120.69 & 0.257$\pm$0.070 & 0.603$\pm$0.164 & O~{\sc iii} & 3121.64 & -91.30 & $\rm{3p\ ^3S_1-3d\ ^3P^{\rm{o}}_1}$ \\
 3131.83 & 12.540$\pm$0.105 & 29.138$\pm$0.244 & O~{\sc iii} & 3132.79 & -91.93 & $\rm{3p\ ^3S_1-3d\ ^3P^{\rm{o}}_2}$ \\
 3186.74 & 2.224$\pm$0.078 & 4.932$\pm$0.173 & He~{\sc i} & 3187.74 & -93.68 & $\rm{2s\ ^3S_1-4p\ ^3P^{\rm{o}}}$ \\
 3197.66 & 0.090$\pm$0.028 & 0.198$\pm$0.062 & Ne~{\sc ii} & 3198.59 & -87.23 & $\rm{3p\ ^4D^{\rm{o}}_{5/2}-3d\ ^4F_{7/2}}$ \\
 3202.11 & 3.398$\pm$0.071 & 7.442$\pm$0.155 & He~{\sc ii} & 3203.17 & -98.85 & $\rm{3d\ ^2D-5f\ ^2F^{\rm{o}}}$ \\
 3217.19 & 0.329$\pm$0.038 & 0.712$\pm$0.082 & Ne~{\sc ii} & 3218.19 & -93.22 & $\rm{3p\ ^4D^{\rm{o}}_{7/2}-3d\ ^4F_{9/2}}$ \\
 3240.66 & 0.083$\pm$0.042 & 0.177$\pm$0.089 & Si~{\sc iii}? & 3241.62 & -88.84 & $\rm{4p\ ^3P^{\rm{o}}_2-5s\ ^3S_1}$ \\
 3243.04 & 0.145$\pm$0.047 & 0.308$\pm$0.100 & Ne~{\sc ii} & 3244.10 & -98.02 & $\rm{3p\ ^4D^{\rm{o}}_{5/2}-3d\ ^2F_{7/2}}$ \\
 3259.84 & 0.128$\pm$0.030 & 0.269$\pm$0.063 & O~{\sc iii} & 3260.86 & -93.84 & $\rm{3p\ ^3D_2-3d\ ^3F^{\rm{o}}_3}$ \\
 3264.32 & 0.142$\pm$0.024 & 0.297$\pm$0.050 & O~{\sc iii} & 3265.33 & -92.79 & $\rm{3p\ ^3D_3-3d\ ^3F^{\rm{o}}_4}$ \\
 $\cdots$ & $\cdots$ & $\cdots$ & $\cdots$ & $\cdots$ & $\cdots$ & $\cdots$ \\
\hline
\end{tabular}
\begin{description}
Table \ref{M142_linelist} is published in its entirety in the machine-readable format. A portion is shown here for guidance regarding its form and content.
\end{description}
\end{center}
\end{table*}

\begin{table*}
\begin{center}
\caption{Line List of NGC\,6153}
\label{NGC_linelist}
\begin{tabular}{ccclccc}
\hline\hline
$\rm{\lambda_{obs}\ (\AA)}$ & $F$($\lambda$) & $I$($\lambda$) & Ion & $\rm{\lambda_{lab}\ (\AA)}$ & V$\rm{_{rad}\ (km s^{-1})}$ & Transition \\
\hline
 3122.00 & 0.229$\pm$0.036 & 0.851$\pm$0.134 & O~{\sc iii} & 3121.64 & 35.01 & $\rm{3p\ ^3S_1-3d\ ^3P^{\rm{o}}_1}$ \\
 3133.20 & 8.317$\pm$0.086 & 30.449$\pm$0.315 & O~{\sc iii} & 3132.79 & 39.26 & $\rm{3p\ ^3S_1-3d\ ^3P^{\rm{o}}_2}$ \\
 3188.18 & 1.026$\pm$0.034 & 3.495$\pm$0.116 & He~{\sc i} & 3187.74 & 41.06 & $\rm{2s\ ^3S_1-4p\ ^3P^{\rm{o}}}$ \\
 3198.95 & 0.031$\pm$0.016 & 0.104$\pm$0.054 & Ne~{\sc ii} & 3198.59 & 33.76 & $\rm{3p\ ^4D^{\rm{o}}_{5/2}-3d\ ^4F_{7/2}}$ \\
 3203.50 & 1.783$\pm$0.032 & 5.960$\pm$0.107 & He~{\sc ii} & 3203.17 & 30.58 & $\rm{3d\ ^2D-5f\ ^2F^{\rm{o}}}$ \\
 3218.62 & 0.181$\pm$0.021 & 0.594$\pm$0.069 & Ne~{\sc ii} & 3218.19 & 40.08 & $\rm{3p\ ^4D^{\rm{o}}_{7/2}-3d\ ^4F_{9/2}}$ \\
 3230.43 & 0.050$\pm$0.021 & 0.162$\pm$0.068 & Ne~{\sc ii} & 3230.07 & 33.44 & $\rm{3s\ ^2D_{5/2}-3p\ ^2D^{\rm{o}}_{5/2}}$ \\
 3244.51 & 0.066$\pm$0.015 & 0.210$\pm$0.048 & Ne~{\sc ii} & 3244.10 & 37.91 & $\rm{3p\ ^4D^{\rm{o}}_{5/2}-3d\ ^2F_{7/2}}$\\
 3261.38 & 0.039$\pm$0.010 & 0.122$\pm$0.031 & O~{\sc iii} & 3260.86 & 46.00 & $\rm{3p\ ^3D_2-3d\ ^3F^{\rm{o}}_3}$ \\
 3265.72 & 0.053$\pm$0.016 & 0.165$\pm$0.050 & O~{\sc iii} & 3265.33 & 35.83 & $\rm{3p\ ^3D_3-3d\ ^3F^{\rm{o}}_4}$ \\
 $\cdots$ & $\cdots$ & $\cdots$ & $\cdots$ & $\cdots$ & $\cdots$ & $\cdots$ \\
\hline
\end{tabular}
\begin{description}
Table \ref{NGC_linelist} is published in its entirety in the machine-readable format. A portion is shown here for guidance regarding its form and content.
\end{description}
\end{center}
\end{table*}

\section{Telluric Lines} \label{appendix:telluric}

\begin{figure*}[ht!]
\begin{center}
\includegraphics[width=18 cm,angle=0]{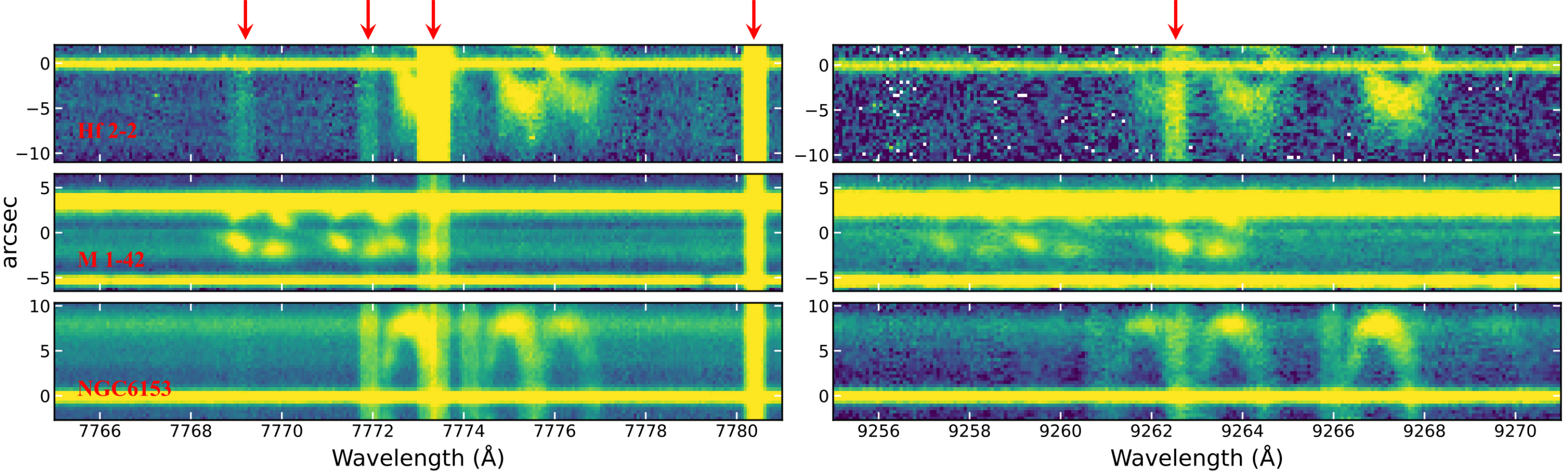}
\caption{Two-dimensional VLT/UVES echelle spectra of Hf\,2-2 (top), M\,1-42 (middle) and NGC\,6153 (bottom), showing the O~{\sc i} recombination lines and the adjacent telluric emission.  The vertical bright features, as marked by red arrows on top, are the telluric lines whose emission fills the whole UVES slit, while the arc-like features, due to nebular expansion revealed in high-dispersion spectroscopy, are the nebular O~{\sc i} ORLs.  In the left panels, the three nebular lines are O~{\sc i} $\lambda\lambda$7771.94, 7774.17, 7775.39; in the right panels, the nebular lines are O~{\sc i} $\lambda\lambda$9260.84, 9262.67, 9265.94.  The two sets of O\,{\sc i} nebular lines are affected by telluric emission at different levels among the three PNe.} 
\label{fig:telluric} 
\end{center}
\end{figure*}

Figure\,\ref{fig:telluric} shows two sections of the 2D VLT/UVES echelle spectra of Hf\,2-2, M\,1-42 and NGC\,6153, where the O\,{\sc i} nebular lines from the M1 ($\lambda\lambda$7771.94,\,7774.17,\,7775.39) and M8 ($\lambda\lambda$9260.84,\,9262.67,\,9265.94) multiplets are located along with adjacent bright telluric emission.

\bibliography{reference}{}

\bibliographystyle{aasjournal}

\end{document}